\pdfoutput=1
\documentclass[prl,reprint, epsfig,amssymb,longbibliography,amsmath,superscriptaddress,twocolumn,nofootinbib]{revtex4-2}

\usepackage{blkarray}
\usepackage[dvipsnames]{xcolor}

\usepackage{amssymb,amsmath,bm,epsfig,graphicx,subfigure,hyperref}

\usepackage[vietnamese, english]{babel}
\usepackage{chngcntr}
\usepackage{xcolor}
\usepackage{braket}
\usepackage{physics}
\usepackage{sidecap}
\usepackage{lipsum}
\usepackage{amsfonts}
\usepackage{bbold}
\usepackage{tensor}
\usepackage{makecell}
\usepackage{multirow}
\usepackage{rotating}
\usepackage{float}
\usepackage{amstext}

\begin{document}

\title{\textcolor{Blue}{Coulomb Interaction-Stabilized Isolated Narrow Bands with Chern Numbers $\mathcal{C} > 1$\\ in Twisted Rhombohedral Trilayer-Bilayer Graphene   }}
\author{\foreignlanguage{vietnamese}{Võ Tiến Phong}}
\email{vophong@magnet.fsu.edu}
\affiliation{Department of Physics, Florida State University, Tallahassee, FL, 32306, U.S.A.}
\affiliation{National High Magnetic Field Laboratory, Tallahassee, FL, 32310, U.S.A.}
\author{Cyprian Lewandowski}
\email{clewandowski@magnet.fsu.edu}
\affiliation{Department of Physics, Florida State University, Tallahassee, FL, 32306, U.S.A.}
\affiliation{National High Magnetic Field Laboratory, Tallahassee, FL, 32310, U.S.A.}

\date{\today}
\begin{abstract} 
Recently, fractional quantum anomalous Hall effects have been discovered in two-dimensional moir\'{e} materials when a topologically nontrivial band with Chern number $\mathcal{C}=1$ is partially doped. Remarkably, superlattice Bloch bands can carry higher Chern numbers that defy the Landau-level paradigm and may even host exotic fractionalized states with non-Abelian quasiparticles. Inspired by this exciting possibility, we propose twisted \textit{rhombohedral} trilayer-bilayer graphene at  $\theta \sim 1.2^\circ$ as a field-tunable quantum anomalous Chern insulator that features spectrally-isolated, kinetically-quenched, and topologically-nontrivial bands with $\mathcal{C} = 2,3$ favorable for fractional phases once fractionally doped, as characterized by their quantum geometry. Based on extensive self-consistent mean-field calculations, we show that these phases are stabilized by Coulomb interactions and are robust against variations in dielectric environment, tight-binding hopping parameters, and lattice relaxation. 
\end{abstract}

\maketitle

Fractionalized phases that emerge from doping a quantum anomalous Chern insulator are highly-sought after because they may host quasiparticles with non-Abelian exchange statistics, importantly in lattice systems without a magnetic field \cite{Neupert2011Fractional,Tang2011High,Sun2011Nearly,Regnault2011Fractional,Wu2012Zoology,Parameswaran2012Fractional,Behrmann2016Model}. These exotic quasiparticles are fascinating on their own right \cite{greiter2024fractional}, but much of the excitement surrounding them concerns their potential application in topological quantum computation \cite{Nayak2008Non-Abelian}. Within the past two years, these fractional Chern insulating (FCI) states have finally been experimentally measured in two classes of materials: twisted MoTe$_2$ bilayers without displacement field \cite{cai2023signatures,zeng2023thermodynamic,Xu2023Observation,redekop2024direct,ji2024local,xu2025signatures,park2025observation,wang2025hidden,chang2025emergent} and rhombohedral graphene multilayers aligned with hexagonal boron nitride (hBN) with displacement field \cite{lu2024fractional,xie2025tunable,lu2025extended}. In twisted MoTe$_2,$ theoretical calculations suggest that the fractional states emerge from partial filling of a valence band with $|\mathcal{C}| = 1$ that exists already in non-interacting band structure modeling \cite{Reddy2023Toward,Reddy2023Fractional,Wang2024Fractional,Morales2024Magic,Shi2024Adiabatic,Jia2024Moir,Yu2024Fractional,xu2024maximally,Li2025Variational,chen2025fractional}. In rhombohedral graphene-hBN heterostructures, fractional states have only been observed at displacement fields $\sim 1$ V/nm where the conduction electrons are pushed \textit{away} from the substrate \cite{lu2024fractional,xie2025tunable,lu2025extended}. In this regime, the non-interacting  conduction bands are essentially those of pristine rhombohedral graphene multilayer with minimal influence from the hBN substrate \cite{Dong2024Anomalous,Zhou2024Fractional,Dong2024Theory,Dong2024Stability,kwan2023moir,Herzog2024Moir,Guo2024Fractional,Huang2024Self,Soejima2024Anomalous,Huang2025Fractional,Tan2024Parent,li2025multi}. That is, the conduction band where electrons would populate is nearly dispersionless but it has no discernible gap to the next higher-energy band \cite{Min2008Chiral,Koshino2009Trigonal,Zhang2010Band}. However, upon including electron-electron interactions, at least at the mean-field level, the narrow band decouples from the other energy bands, generating a mean-field gap $\sim 10-20$ meV, and carrying a Chern number $|\mathcal{C}| = 1$ \cite{Dong2024Anomalous,Zhou2024Fractional,Dong2024Theory,Dong2024Stability,kwan2023moir,Herzog2024Moir,Guo2024Fractional,Huang2024Self,Soejima2024Anomalous,Huang2025Fractional}. This band remains more or less flat at the mean-field level, and is thus, conducive to further strong interaction effects at partial filling. These graphene-based platforms offer the insight that non-interacting band structures can only predict a subset of possible topological materials that can fractionalize, i.e. they can predict twisted MoTe$_2$ but not graphene-hBN moir\'{e} systems.

Despite their important differences, both aforementioned systems fractionalize from doping $|\mathcal{C}|=1$ bands. In principle, superlattice systems can support bands with any integer Chern number \cite{Hasan2010Colloquium,Qi2011Topological}. Theories predict that these states are qualitatively much richer than their Landau-level analogues \cite{Wang2012Fractional,Barkeshli2012Topological,Liu2012Fractional,Sterdyniak2013Sterdyniak,Liu2013Fractional,Moller2015Fractional,Andrews2021Stability}. Higher-Chern bands can stabilize fractional states at generalized Jain bosonic and fermionic fillings; these states can be both Abelian and non-Abelian. Therefore, superlattice materials present the exciting possibility of fractionalizing higher-Chern bands without any Landau-level counterpart. Many moir\'{e} materials have been proposed to host higher-Chern bands on the basis of non-interacting band structure modeling, including, just to name a few, twisted double bilayer graphene \cite{Chebrolu2019Flat,Koshino2019Band,Liu2021Gate,perea2024quantum}, twisted AB-ABC graphene \cite{Zhang2019Nearly}, helical trilayer graphene and analogues \cite{devakul2023magic, Kwan2024Strong, Datta2024Helical,Makov2024Flat}, buckled bilayer graphene \cite{Wan2023Nearly,Fujimoto2025Higher}, periodically gated multilayer graphene \cite{Ghorashi2023Topological,Ghorashi2023Multilayer,Seleznev2024Inducing}, and others \cite{Zhou2022Moir,wang2024fractional2,Wang2024Electrically}. Various higher-Chern bands have been observed experimentally as well, which however host other (non-FCI) correlated states at fractional fillings of the corresponding higher-Chern bands \cite{chen2020tunable, chen2021electrically, xia2023helical,peng2024abundant}. Despite extensive effort, the experimental observation of fractionalized states emerging from higher-Chern bands remains an unfulfilled aspiration to date.

In this work, we take a step towards making realistic predictions for a material platform that hosts higher-Chern bands endowed with favorable quantum geometry for FCI states. Our system is constructed from rhombohedral graphene multilayers whose conduction bands can be kinetically quenched by a moderate displacement field when the number of layers is large\footnote{It is a bit subjective what ``large" here means. Essentially, what we need is a system that can be made to have a nearly dispersionless conduction band minimum at experimentally-accessible displacement fields. For concreteness, let us just say that large is anything above two layers.}. By bringing a rhombohedral trilayer and bilayer stack together with a relative twist angle, henceforth referred to as twisted rhombohedral multilayer graphene ($\mathrm{TRMG}_{3+2}$), we show this system is a suitable candidate for higher-Chern physics. $\mathrm{TRMG}_{3+2}$ is one member of a broader family of $\mathrm{TRMG}_{M+N}$ that has been considered in several earlier theoretical works \cite{Zhang2019Nearly,Liu2019Quantum,Ledwith2022Family,Wang2022Hierarchy,Dong2023Many,yang2023flat,Zhang2022Spin}, wherein it has been shown that the non-interacting band structures contain higher-Chern narrow bands. On the basis of the chiral model that retains only a minimal set of known hoppings \cite{San2012Non,Tarnopolsky2019Origin}, Refs. \cite{Ledwith2022Family,Wang2022Hierarchy,Dong2023Many} argue that these bands are ideal Chern bands where the Berry curvature and quantum metric exactly satisfy proportionality \cite{Roy2014Band,Claassen2015Position,Ledwith2020Fractional,Wang2021Exact}. Building on these works with the insight from graphene-hBN FCI's deterministic role of Coulomb interactions, we analyze the formation of narrow Chern bands with $|\mathcal{C}| = 2,3$ after Hartree-Fock (HF) renormalization. We find that Coulomb interactions are simultaneously crucial to the spectral isolation of these bands from their proximal partners and to the stabilization of their favorable quantum geometry. In the interest of realistic modeling, we include all known hoppings of rhombohedral graphite, account for electrostatic screening from both metallic gates and charge redistribution among the different layers of $\mathrm{TRMG}_{3+2}$, and allow for some layer buckling across the twist interface. We find that these post-HF Chern bands are stable in prominent regions of phase space and are robust against changes in dielectric environment and hopping parameters. Therefore, on the basis of our extensive mean-field calculations, we propose $\mathrm{TRMG}_{3+2}$ as a suitable candidate for the exploration of possible FCI states emerging from a higher-Chern narrow band.

\begin{figure}
    \centering
    \includegraphics[width=1\linewidth]{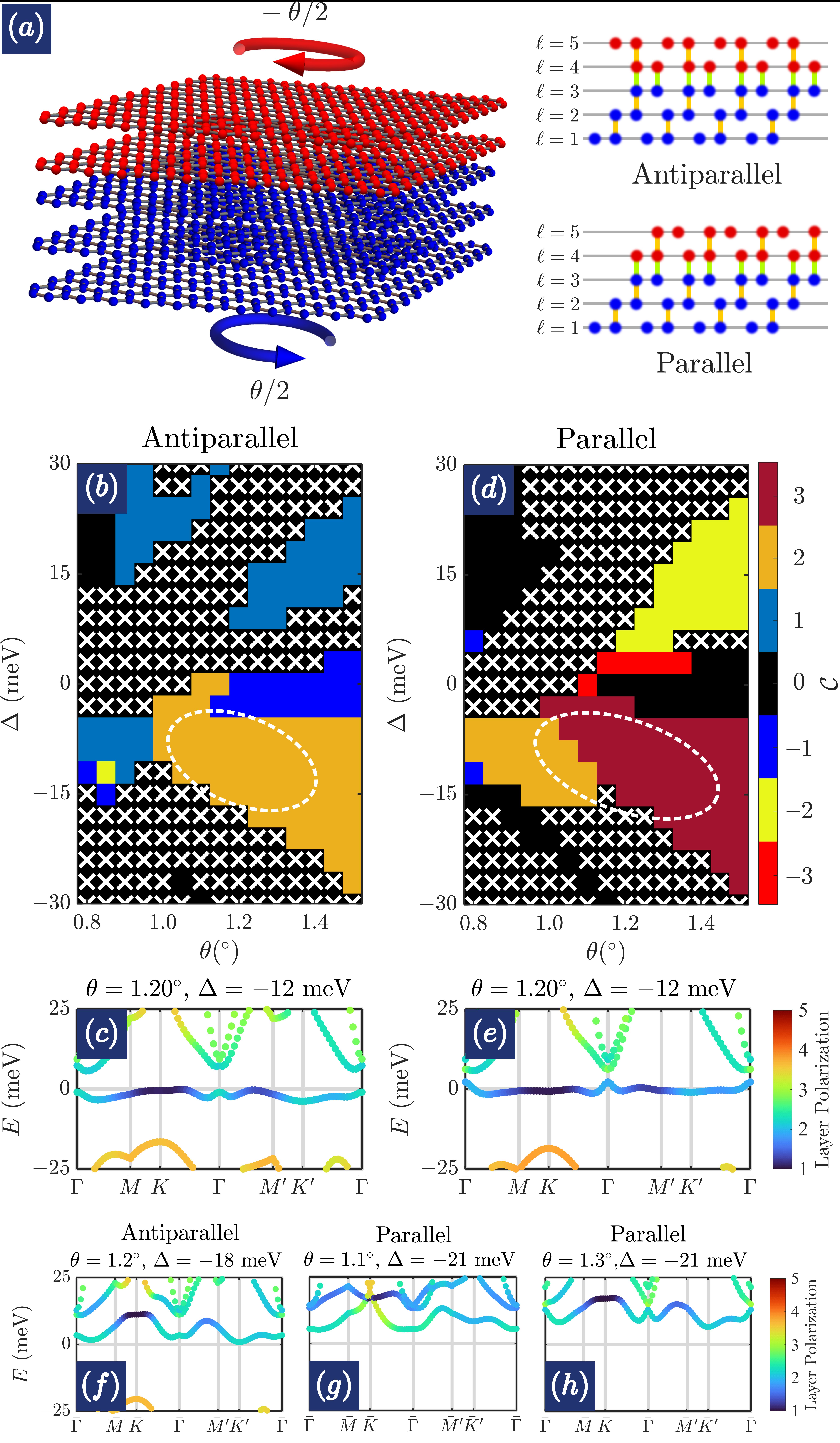}
    \caption{\textbf{Moir\'{e} heterostructure schematic and topological characterization of the non-interacting band structures.} (a) Construction of a twisted $3+2$ rhombohedral stack by rotating the bottom trilayer (top bilayer) by $+\theta/2$ ($-\theta/2$). There are two inequivalent configurations, antiparallel and parallel, depending on the relative stacking orders of the two subsystems. (b) Single-particle (non-interacting) Chern-number characterization of first band above charge neutrality of the antiparallel configuration for values of $\theta$ and $\Delta$ where said band is globally isolated. $\times$ indicates where that band is \textit{not} isolated from the other bands. A dashed line encircles the region of interest with $\mathcal{C}=2$, reasonably large band gaps, relatively small bandwidths, and small trace condition violations (see discussion in the main text).(c) An example band structure in that region is color coded by layer polarization. (d) Single-particle (non-interacting), Chern-number characterization for the parallel configuration showing a contiguous region featuring $\mathcal{C} = 2$ and $\mathcal{C} = 3$. A $\mathcal{C} = -2$ phase is also present in this phase diagram, but it does not have the desired spectral properties and quantum geometry. (e) A layer-resolved band structure is shown for a representative point in the dashed circle in (d). (f,g,h) Band structures for values of $(\theta,\Delta)$ where no clear isolated bands are observed. They are included here for later comparison to HF bands.}
    \label{fig:fig1_main}
\end{figure}

To begin, we construct $\mathrm{TRMG}_{3+2}$ by stacking Bernal bilayer graphene on top of rhombohedral trilayer graphene with a relative twist angle $\theta.$ We focus on this combination because both the trilayer and bilayer substacks can be locally gapped by an interlayer displacement field and made to feature spectral regions with high density of states. This feature is crucial for our study as it allows us to tune into different regions in phase space where the interplay between topology and electron-electron interactions is important. $\mathrm{TRMG}_{3+2}$ is chosen to comprise of different layer numbers so that the competition between disparate layer numbers can be ascertained\footnote{Said differently, our choice of $3+2$ is informed by the following logic. We wish to keep the layer number relatively low for computational speed for this work. So, we fix one of the substacks to be a trilayer because it is the lowest layer number where the rhombohedral nature is evident. The remaining substack can be a monolayer, bilayer, or trilayer. We wish to observe the competition between the two substacks to determine which one, if any, dominates the relevant physics. Therefore, we eliminate the trilayer option. Between the monolayer and bilayer options, we choose the latter since bilayer graphene also locally gaps out in a displacement field.}. Two inequivalent stacking orders are possible depending on whether the \textit{pre-twist} alignment of the trilayer and bilayer are parallel or antiparallel\footnote{Concretely, let us label layers $\ell = 1,2,3$ those of the trilayer and layers $\ell = 4,5$ those of the bilayer. The twist interface occurs between layers $\ell = 3, 4.$ The trilayer is oriented in such a way that layer $\ell +1$ is shifted by $+a/\sqrt{3}\hat{y}$ relative to layer $\ell.$ If the bilayer is oriented the same way, we refer to the stack as parallel. If instead, layer 5 is shifted by $-a/\sqrt{3}\hat{y}$ relative to layer 4, then we refer to the stack as antiparallel. After alignment, we twist the trilayer by $+\theta/2$ and the bilayer by $-\theta/2.$ }. In the parallel (antiparallel) stacking order, the zone-corner degeneracies carry net chiralities of the same (opposite) sign. The two stacking orders are related to each other by a $C_{2z}$ rotation of the bilayer. The schematic construction of our system is shown in Fig. \ref{fig:fig1_main}(a). As we will show, these two stacking orders have significantly different mean-field phase diagrams, illustrating the importance of alignment precision. We model the band structure of $\mathrm{TRMG}_{3+2}$ using  a conventional valley-projected continuum  approach appropriate for small twist angles \cite{bistritzer2011moire,Lopes2012Continuum}:
\begin{equation}
\begin{split}
\label{eq: kinetic energy - main}
    \mathbb{K}_{3+2,\nu}  = \begin{pmatrix}
        \mathbb{K}_{3,\nu}(-i\nabla_{\mathbb{R}_-\mathbf{r}}) & \mathbb{T}_\nu(\mathbf{r}) \\
        \mathbb{T}_\nu^\dagger(\mathbf{r}) & \mathbb{K}_{2,\nu}(-i\nabla_{\mathbb{R}_+\mathbf{r}})
    \end{pmatrix} + \mathbb{D},
\end{split}
\end{equation}
where $\mathbb{K}_{3,\nu}$ and $\mathbb{K}_{2,\nu}$ are the kinetic energy for the rhombohedral trilayer and bilayer at valley $\nu$ respectively, $\mathbb{T}$ is the tunneling interaction between the two interfacial layers,  $\mathbb{D} = \Delta\mathrm{diag}\left(-2,-1,0,1,2\right)_\ell\otimes \mathbb{1}_\sigma$ is the \textit{external} interlayer displacement field that is diagonal in layer space $\ell$ and is proportional to the identity matrix in sublattice space $\sigma$, and $\nabla_{\mathbb{R}_\pm \mathbf{r}}$ is a shorthand notation  for rotating the gradient operator by $\pm \theta/2$. For accurate modeling, we include all known hoppings $\lbrace \gamma_0,\gamma_1,\gamma_2,\gamma_3,\gamma_4 \rbrace = \lbrace -3.1,0.38,-0.015,0.29,0.141 \rbrace$ eV \cite{dresselhaus1981intercalation,shi2020electronic,zhou2021half} in $\mathbb{K}_3$ and $\mathbb{K}_2$ but we neglect the dimerization energy on eclipsed sites. In particular, the $\gamma_3$ and $\gamma_4$ terms that are responsible for trigonal warping can significantly deform the spectrum near charge neutrality.  The tunneling matrix depends on two hopping parameters at the $AA$ and $AB$ regions of the twist interface $\gamma_\mathrm{AA}$ and $\gamma_\mathrm{AB}.$ Symmetry allows these two parameters to be different, and adjusting their relative amplitudes accounts for some lattice relaxation at the interface. Results in the main text are obtained using $\gamma_\mathrm{AA}/\gamma_\mathrm{AB} = 0.1\text{ }\mathrm{eV}/0.12 \text{ }\mathrm{eV}$ \cite{Koshino2018Maximally,lisi2021observation, Carr2019Exact,Leconte2022Relaxation}. In Ref. \cite{SI}, we show that qualitative conclusions are robust against varying this ratio. That said, a comprehensive analysis of lattice relaxation is beyond the scope of the present work and is relegated to future studies.

$\mathrm{TRMG}_{3+2}$ does not respect many point symmetries. Since rhombohedral multilayers of graphene do not respect $C_{2z}$ symmetry, the twisted stack cannot either. Due to different layer numbers in the two substacks, symmetries that exchange layers across the twist interface like $C_{2x}$ and $C_{2y}$ rotations are automatically broken. However, $C_{3z}$ symmetry is respected by each layer individually and hence is preserved in the twisted stack. Also, when both valleys are considered,  time-reversal symmetry $\mathcal{T}$ is respected. In our numerical calculations, we explicitly confirm that $C_{3z}$ and $\mathcal{T}$ symmetries are preserved in all cases.

At the non-interacting level, the band structure of TRMG$_{3+2}$ at zero displacement field for $\theta \sim 1^\circ$ for both configurations features a pair of bands at charge neutrality per isospin flavor\footnote{For the non-interacting band structures, valley-spin states are degenerate; so we shall restrict our discussion to just the $\lbrace \nu = +1,s=\uparrow\rbrace$ flavor until interactions are considered where this degeneracy is broken.}. These bands are relatively narrow in spectral width (for instance, at $\theta = 1.25^\circ,$ the bandwidth is about $25$ meV), but they are \textit{not} well isolated from the other higher energy bands. The suppressed bandwidth here mirrors the bandwidth of the active bands in twisted bilayer graphene, but in the latter scenario, the bands are also well isolated from the rest of the band structure. Because of this spectral complexity, at zero displacement field, we heuristically do not expect that electron-electron interactions would isolate a single band with nontrivial topology. Indeed, extensive calculations that include Coulomb interactions confirm with this heuristic argument: the sought-after physics of a topologically nontrivial singlet band has to be found at non-zero displacement field.

The application of finite $\Delta \neq 0$ has two qualitative effects on the band structure: (1) the shifting of the local spectra of the trilayer and bilayer subsystems relative to each other, and (2) the genesis of local gaps in each substack due to layer inequivalence (essentially, this is the effect of a $\sigma_z$ mass). Interestingly, the formation of \textit{local} gaps is \textit{not} enough to generate a global gap in this system. To see this, let us consider the limit where the hoppings across the twist interface are switched off (i.e. the moir\'{e} structure can be gauged away) so that the subsystems can be regarded as independent. Placing the energy origin at the nodes of the multilayer spectra when $\Delta = 0,$ the trilayer spectrum is centered at $-\Delta$ with a gap of $2|\Delta|$ and disperses away its extrema as $\pm |\mathbf{k}|^3,$ while the bilayer spectrum is centered at $3\Delta/2$ with a gap of $|\Delta|$ and disperses away from its extrema as $\pm |\mathbf{k}|^2.$ Importantly, for $\Delta > 0,$ the conduction band minimum of the trilayer is located at $E = 0$ while the valence band maximum of the bilayer is located at $E = \Delta.$ Therefore, the two spectra completely overlap. For $\Delta < 0,$ the valence band maximum of the trilayer is at $E = 0,$ while the conduction band minimum of the bilayer is at $E = \Delta.$ Thus, the same conclusion applies that the two independent spectra completely overlap. This simple analysis immediately traces the origin of any robust global gap in this system (which does indeed exist) to the existence of the moir\'{e} structure, elevating its important status in the subsequent analysis. For comparison, we recall that in rhombohedral graphene multilayers aligned with hBN, the role of the substrate in band-structure modeling is entirely negligible in the large $\Delta$ limit where the conduction electrons are pushed to the layer farthest away from the substrate\footnote{The role of the substrate on stabilizing the observed anomalous Hall effects is much more subtle and continues to generate debate and interest to the present \cite{Herzog2024Moir,aronson2024displacement}.}; the giant gap at charge neutrality is just the gap of the graphene stack itself without any discernible contribution from the substrate \cite{Zhou2024Fractional,Dong2024Anomalous}.

For $\Delta < 0$, the two active bands, initially entangled, are decoupled and pushed away from each other with increasing $\Delta$. Focusing on the conduction band, we find its states are predominantly localized on layer 1, suggesting that for negative $\Delta,$ the trilayer plays a more dominant role than the bilayer. Not surprisingly then, we find that the conduction band can become very flat ($< 5$ meV) because the $\sqrt{\Delta^2+|\mathbf{k}|^6}$ dispersion of the trilayer can be made narrow for reasonable values of $\Delta.$ Now, band suppression does not always occur simultaneously with spectral isolation from the other bands. So, even though the conduction band can be made quite flat, it is often close by to, even if technically not overlapping, the next higher energy band. In contrast, when $\Delta$ is increased in the positive direction, the conduction band is localized primarily on the bilayer sector. In this case, we seldom find bands as flat as in the negative $\Delta$ direction. Of course, this is because the bilayer dispersion requires higher $\Delta$ to flatten. As a result, all of the interesting correlated physics reported in this work is confined to the negative $\Delta$ region of parameter space.

Spectrally, the band structures for both the antiparallel and parallel configurations are qualitatively similar. However, their differences are encoded in the quantum geometry of the Bloch states. For cases where the conduction band is fully gapped\footnote{Fully gapped here means that it is gapped both above and below.}, we calculate its Chern number and associated quantum metric violation defined as $\lambda = (2\pi)^{-1}\int d^2\mathbf{k} \left[\mathrm{Tr}g_{\mu\nu}(\mathbf{k}) - |\mathcal{B}_z(\mathbf{k})| \right],$ where $g_{\mu\nu}(\mathbf{k})$ is the quantum metric and $\mathcal{B}_z(\mathbf{k})$ is the Berry curvature \cite{Wilczek1984Appearance, pati1991relation, fukui2005chern, cheng2010quantum,Mera2021Engineering}. The $\lambda$ quantity is a proxy for the propensity of a band to become an FCI once doped. It is based on analogy to the quantum geometry of the zeroth Landau level \cite{Ledwith2022Family,Wang2022Hierarchy,Dong2023Many}. We shall use $\lambda$ as a possible FCI diagnostic. The smaller it is, the better for FCI's. For comparison, the trace condition violations calculated for systems that have been experimentally shown to host FCI states are comparable to our values. For graphene pentalayer on hBN, $\lambda \sim 0.6-1.45$ \cite{Dong2024Theory,Dong2024Anomalous,Herzog2024Moir,Zhou2024Fractional}; for twisted MoTe$_2,$ $\lambda \sim 0.1-0.7$ \cite{mao2024transfer,Reddy2023Toward,xu2024maximally}. The Chern characterization of the antiparallel and parallel configurations in $(\theta,\Delta)$ space is shown in Fig. \ref{fig:fig1_main}(b,d). We find that the antiparallel configuration hosts a large region where $\mathcal{C} = +2.$ In this region, the smallest trace condition violation is $\lambda \approx 1.7.$ This coincides with the where the bandwidths are comparatively small ($\sim 3$ meV) and band gaps relatively large $\sim 6$ meV\footnote{There are also regions where $|\mathcal{C}| = 1.$ Since we are interested in higher Chern numbers in this work, we do not analyze these regions here. However, it is perhaps worth mentioning that $\lambda >2$ for these regions. The bandwidth of the $\mathcal{C}=-1$ region is relatively small ($\sim 2.5$ meV), but the bandwidth of the $\mathcal{C} = 1$ region is quite large ($>30$ meV).}. The parallel configuration is even more exotic in that it hosts both a $\mathcal{C}=2$ and a $\mathcal{C}=3$ region. The $\mathcal{C} = 2$ region here features a slightly smaller trace condition violation, with $\lambda \approx1.3$ at best, compared to the antiparallel configuration. On the other hand, the $\mathcal{C}=3$ region has quite small $\lambda < 1.$ However, the bandwidths of the $\mathcal{C}=3$ phase are generally larger ($\sim 4.5$ meV) than the $\mathcal{C}=2$ phase ($\sim 1.5$ meV). All in all, we find that for both the parallel and antiparallel configurations, there are regions with nonzero Chern numbers where relatively small bandwidths, large band gaps, and small $\lambda$ coexist. We mark these regions with dashed lines in Fig. \ref{fig:fig1_main}(b,d). From each region, we display a representative band structure in Fig. \ref{fig:fig1_main}(c,e) showing the narrow band of interest polarized primarily to layer 1. We also show three band structures in Fig. \ref{fig:fig1_main}(f,g,h) that seemingly do not look especially conducive to interaction effects, but yet they will turn out to be important once interactions are included.

\begin{figure*}[t!]
    \centering
    \includegraphics[width=1\linewidth]{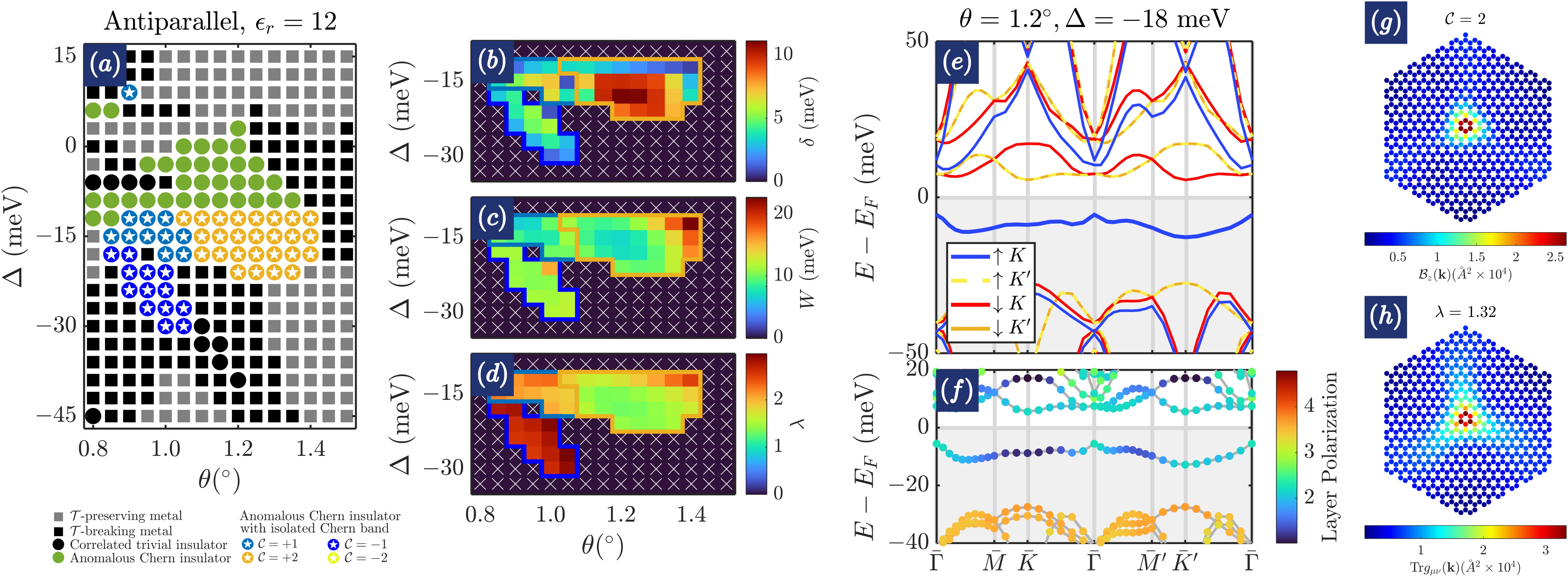}
    \caption{\textbf{Self-consistent mean field phase diagram of the antiparallel configuration and characterization of the $\mathcal{C}=2$ phase.} (a) Phase diagram in $(\theta,\Delta)$ space showing several distinct metallic and insulating phases. Of particular interest is the presence of a $\mathcal{C}=2$ phase at negative $\Delta$ and $\theta \in[1.1^\circ,1.4^\circ]$ featuring a spectrally-isolated, isospin-polarized, topologically-nontrivial band right below the Fermi level. The band gap, bandwidth, and trace condition violation of that band are shown in (b-d). A representative band structure from that phase color coded by isospin and by layer polarization is shown in (e) and (f) respectively. Here, the existence of an isolated narrow band with significant electron localization to the outermost layers of the trilayer substack is apparent. The distribution of the Berry curvature and quantum metric of that band in momentum space is plotted in (g) and (h) respectively. Here, $\epsilon_r = 12$ and $D=40$ nm. Energies are measured relative to Fermi level, $E_F.$}
    \label{fig:fig2_main}
\end{figure*}

A narrow band implies a large density of states, which, in turn, suggests that electron-electron interactions should significantly modify these bands, both in terms of their energetic and topological properties. Therefore, the preceding analysis based on non-interacting band structures is only suggestive of a rough region in phase space where topologically nontrivial bands may give rise to higher-Chern anomalous quantum Hall phases. It is entirely possible that the inclusion of interactions may trivialize non-interacting Chern bands or topologicalize non-interacting trivial ones. As such, we need to account for Coulomb interactions in order to make reliable predictions. To this end, we perform self-consistent mean-field HF calculations. The mean-field Hamiltonian is written in band basis as
\begin{equation}
\begin{split}
    \hat{\mathcal{H}} &= \sum_{\substack{\mathbf{k}\in\mathrm{mBZ},\\n,n',\xi,\xi'}} \hat{\Psi}^\dagger_{n,\xi,\mathbf{k}} \left[  \mathbb{K}^{n,n'}_{\xi,\xi',\mathbf{k}} + \mathbb{H}^{n,n'}_{\xi,\xi',\mathbf{k}}+ \mathbb{F}^{n,n'}_{\xi,\xi',\mathbf{k}} \right]\hat{\Psi}_{n',\xi',\mathbf{k}},
\end{split}
\end{equation}
where $\hat{\Psi}^\dagger_{n,\xi,\mathbf{k}}$ is a fermionic creation operator for a Bloch state at momentum $\mathbf{k},$ $\lbrace\xi,\xi',n,n'\rbrace$ label isospins and bands respectively, and the kinetic energy (same as Eq. \eqref{eq: kinetic energy - main}), Hartree energy, and Fock exchange energy in the band basis are
\begin{subequations}
\label{eq: HF equations - main}
    \begin{align}
            \mathbb{K}^{n,n'}_{\xi,\xi',\mathbf{k}} &= \delta_{\xi,\xi'} \delta_{n,n'} E_{n,\xi,\mathbf{k}} , \\
    \mathbb{H}^{n,n'}_{\xi,\xi',\mathbf{k}} &= +\delta_{\xi,\xi'} \sum_{\substack{\mathbf{p} \in \mathrm{mBZ}\\ n_1, n_3, \xi_1}} \mathbb{V}^{n_1,n,n_3,n'}_{\xi_1,\xi,\mathbf{p},\mathbf{k},\mathbf{0}} \mathbb{D}^{n_1,n_3}_{\xi_1,\xi_1,\mathbf{p}}, \\
     \mathbb{F}^{n,n'}_{\xi,\xi',\mathbf{k}} &= -\sum_{\substack{\mathbf{p} \in \mathrm{mBZ}\\ n_1, n_4}} \mathbb{V}^{n_1,n,n',n_4}_{\xi',\xi,\mathbf{k},\mathbf{p},\mathbf{p}-\mathbf{k}}\mathbb{D}_{\xi',\xi,\mathbf{p}}^{n_1,n_4}.
    \end{align}
\end{subequations}
In Eq. \eqref{eq: HF equations - main}, $E_{n,\xi,\mathbf{k}}$ are the non-interacting band eigenvalues, and the Coulomb matrix elements are calculated from layer-resolved form factors $\mathbb{\Lambda}^{n,n'}_{\xi,\ell,\mathbf{k},\mathbf{q}+\mathbf{Q}} = \bra{u_{n,\xi,\mathbf{k}+\mathbf{q}}} e^{i \mathbf{Q} \cdot \hat{\mathbf{r}}} \mathbb{P}_\ell \ket{u_{n',\xi,\mathbf{k}}},$ where $\mathbb{P}_\ell$ is the projection to layer $\ell$ and $\ket{u_{n,\xi,\mathbf{k}}}$ are the periodic part of Bloch wavefunctions, as follows \cite{Huang2023Spin,Kol2023Electrostatic,Kol2025SingleGate}
\begin{subequations}
    \begin{align}
        \mathbb{V}^{n_1,n_2,n_3,n_4}_{\xi_1,\xi_2,\mathbf{k},\mathbf{p},\mathbf{q}} &= \sum_{\mathbf{Q},\ell_1,\ell_2} \mathbb{\Lambda}^{n_1,n_3}_{\xi_1,\ell_1,\mathbf{k},\mathbf{q}+\mathbf{Q}} \mathcal{V}^{\ell_1,\ell_2}_{\mathbf{q}+\mathbf{Q}} \mathbb{\Lambda}^{n_2,n_4}_{\xi_2,\ell_2,\mathbf{p},-\mathbf{q}-\mathbf{Q}}, \label{eq: Coulomb matrix elements - main} \\
        \mathcal{V}^{\ell,\ell'}_{\mathbf{k}} &= \frac{e^2\csch(k D)}{2\epsilon_0\epsilon_r kV_\mathrm{sys}} \left[ \cosh (k \left[D-|z_{\ell}-z_{\ell'}|\right]) \right. \nonumber \\
        &\left. -  \cosh (k \left[D-|z_{\ell}+z_{\ell'}|\right]) \right], \label{eq: layer dependent - main}
    \end{align}
\end{subequations}
where $D$ is the distance between top and bottom gates, $\epsilon_0$ and $\epsilon_r$ are the permitivity of free space and dielectric constant respectively, $V_\mathrm{sys}$ is the system volume, and $z_\ell$ is the vertical position of layer $\ell.$ Throughout, lower-case momenta $\mathbf{k},$ $\mathbf{p},$ $\mathbf{q}$ are confined within a moir\'{e} Brillouin zone (mBZ) while capitalized momenta $\mathbf{Q}$ are reciprocal lattice vectors. We emphasize that we are using a layer-dependent Coulomb potential, Eq. \eqref{eq: layer dependent - main}, that not only accounts for screening effects from the dual gates  but also takes into consideration charge redistribution among the five layers to screen the applied displacement field \cite{Kol2023Electrostatic,Kol2025SingleGate}\footnote{The simple form of the Coulomb potential assumes that the bottom gate is located at $z = 0$ and the top gate is located at $z = D.$ We provide the full derivation of this Coulomb potential using Green's functions in Ref. \cite{SI}. There, we also compare this functional form of the Coulomb potential at $\mathbf{k}=\mathbf{0}$ to the classical capacitor model and find exact agreement (see also Refs. \cite{Kol2023Electrostatic,Kol2025SingleGate}). As a quick consistency check here, if we take $z_\ell = z_{\ell'}=D/2,$ we obtain exactly the usual dual-gate potential: $\mathcal{V}_\mathbf{k} = e^2\tanh\left(kD/2\right)/2\epsilon_0\epsilon_rkV_\mathrm{sys}$.}. For simplicity, we assume that the graphene stack is located exactly half-way between the two gates. Unless specified otherwise, $D = 40$ nm. In Eq. \eqref{eq: HF equations - main}, $ \mathbb{D}^{n,n'}_{\xi,\xi',\mathbf{p}} = \langle \hat{\Psi}_{n, \xi, \mathbf{p}}^\dagger \hat{\Psi}_{n', \xi', \mathbf{p}} \rangle-\frac{1}{2} \delta_{n,n'}\delta_{\xi,\xi'}$ is the density matrix measured from a neutral uniform background at half filling in order to avoid double-counting Coulomb energy contributions already accounted for in \textit{ab initio} calculations \cite{Xie2020Nature,Bernevig2021Twisted}.

In our iterative HF calculations, we only consider filling factor of one electron per moir\'{e} cell. Since we are only interested in valley-polarized anomalous quantum Hall phases, in analogy to the MoTe$_2$ and the rhombohedral pentalayer system, we seed our calculations with initial ansatz that preferentially fill one of the isospin flavors with the extra electron per unit cell\footnote{Searching for more general ground states such as valley-coherent states or nematic states would form an interesting further study. }. To determine the nature of the ground state, the energy of the converged solution from this symmetry-broken ansatz is compared to the solution of the corresponding symmetric state. As mentioned before, the gaps in the non-interacting model are not enormous (on the order of a few tens of meV at best, not hundreds) in the relevant regime of $\Delta$. Therefore, we need to include a decently large number of bands in order to accurately capture Coulomb renormalization. In particular, there is no apparent sensible way to discard the valence bands as was customary in studies of pentalayer graphene on hBN \cite{Dong2024Theory,Soejima2024Anomalous,Dong2024Anomalous, Dong2024Stability, Zhou2024Fractional}. For the majority of the results in the main text, we include 8 (4 valence and 4 conduction) bands per isospin flavor. The grid size is always $18\times18$ $\mathbf{k}$ points.

We first present the mean-field results for the antiparallel configuration in Fig. \ref{fig:fig2_main}. We identify the following phases: (1) a $\mathcal{T}$-preserving metal resulting from the symmetric ansatz with total energy lower or very close to the solution of the symmetry-broken ansatz, (2) a $\mathcal{T}$-breaking metal that unequally occupies the isospin flavors and contains a Fermi surface, (3) a correlated trivial insulator that breaks $\mathcal{T}$-symmetry and is gapped phase with zero Hall conductivity,  (4) an anomalous Chern insulator that features quantized Hall conductivity but \textit{without} an isolated Chern band right below chemical potential, and (5) an anomalous Chern insulator with isolated Chern band and quantized Hall conductivity whose first valence band is spectrally-isolated, topological band. The last phase is what we are after in this work since its existence suggests the possibility of exotic fractionalized states upon partial doping. We label this state by the Chern number of its first valence band (which is often, but not always, the same as the Hall conductivity of the corresponding gap at Fermi level), which we will use to refer to this state henceforth. The phase diagram simulated with $\epsilon_r = 12$ is shown in Fig.  \ref{fig:fig2_main}(a).  As is evident, the phase diagram features considerable complexity owning to the complex evolution of the band structure as a function of $\theta$ and $\Delta.$ All five phases aforementioned are represented in this phase diagram. Most importantly, we find the $\mathcal{C}=2$ phase spanning an angle range $\theta \in [1.1^\circ, 1.4^\circ]$ for $\Delta \in [-21,-12]$ meV. We also find a region where $|\mathcal{C}| = 1$ at smaller angles.

Focusing on the $\mathcal{C}=2$ phase, we further characterize its spectral and quantum geometric properties in Fig. \ref{fig:fig1_main}(b-d). We quantify an ``optimal" combination of $(\theta,\Delta)$ within this phase with a figure of merit defined simply as $\mathfrak{f.o.m.} = \delta/W\lambda,$ where $\delta$ is the minimum of the gaps above and below the band of interest and $W$ is its bandwidth. A large $\mathfrak{f.o.m.} $ means that the band is susceptible to further renormalization by residual Coulomb interactions once partially doped and that the band has a small trace condition violation. In other words, this band is a candidate for hosting a fractional quantum hall phase that requires strong Coulomb repulsion to stabilize. A similar, but not the same, $\mathfrak{f.o.m.}$ is used in Ref. \cite{Zhou2024Fractional,waters2024topological}. For the $\mathcal{C}=2$ phase, the optimal $\mathfrak{f.o.m.}$ occurs at $\theta = 1.20^\circ$ and $\Delta = -18$ meV where $W = 7.5$ meV, $\delta = 11.1$ meV, $\lambda = 1.32.$ Referring back to the non-interacting band structures in Fig. \ref{fig:fig1_main}(f), we find that the band is not even fully gapped without interactions at this $(\theta,\Delta)$. Therefore, we attribute the origins of this isolated topological band to Coulomb interactions. The HF-renormalized bands color coded by isospin polarization and valley polarization are shown in Fig. \ref{fig:fig2_main}(e) and (f) respectively. The Berry curvature and quantum metric of this band, shown in Fig. \ref{fig:fig2_main}(g,h), are concentrated primarily at the $\bar{\Gamma}$ point. Remarkably, we find that interactions reduce the optimal values of $\lambda,$ $\lambda = 1.7\rightarrow1.32.$ As is clear from Fig. \ref{fig:fig2_main}(f), the electrons in this band are localized mostly in layer 1, validating a prior assentation that the trilayer substack plays the dominant role in the observed physics. Once the charge density is summed over all states from that band, we find that the total charge density has $\gtrsim 70 \%$ of its share on layer 1, where the charge density also fluctuates laterally with a higher concentration at one of the three Wyckoff positions in the moir\'{e} unit cell.

\begin{figure}[h!]
    \centering
    \includegraphics[width=1\linewidth]{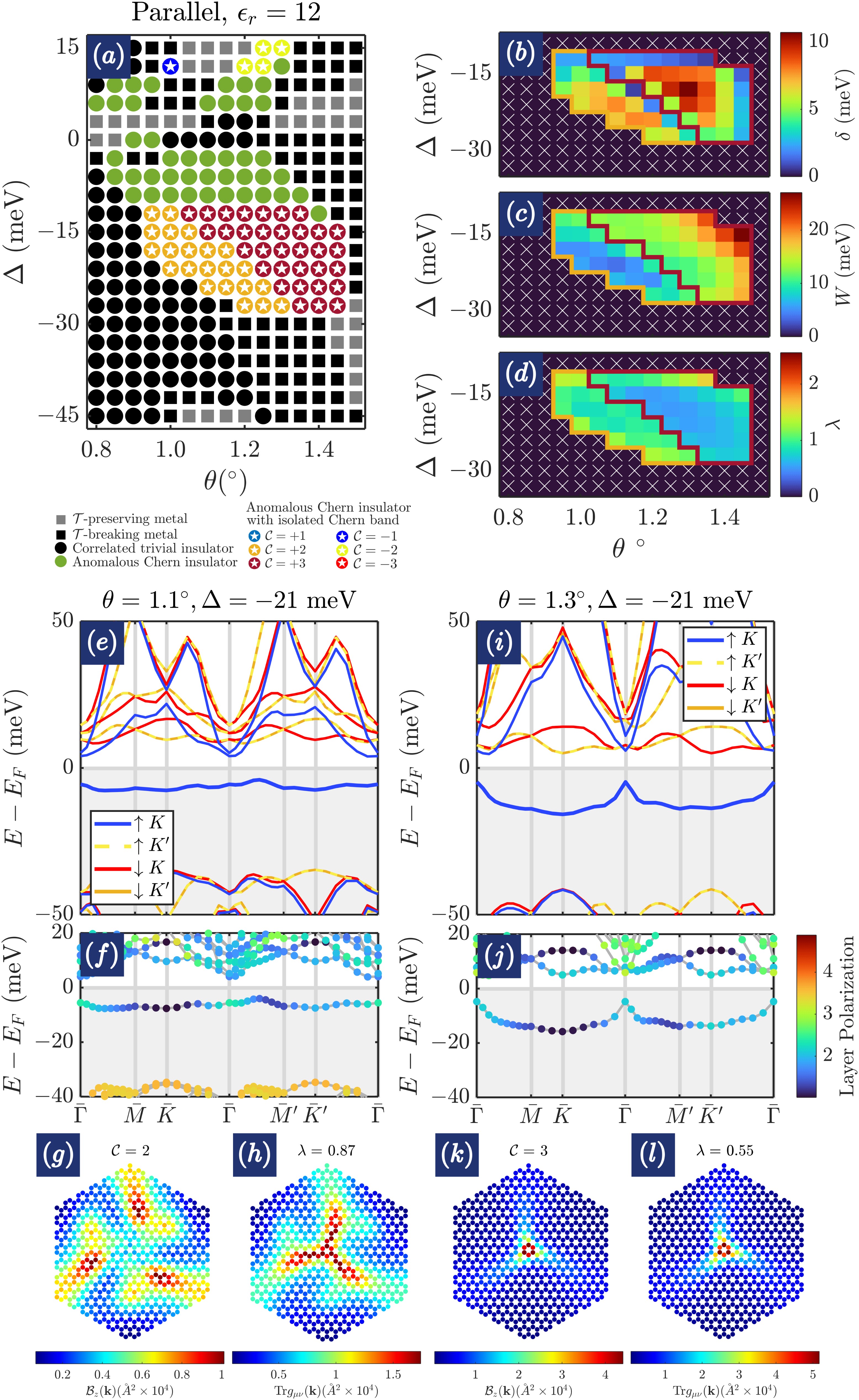}
    \caption{\textbf{Self-consistent mean field phase diagram of the parallel configuration and characterization of the $\mathcal{C}=2$ and $\mathcal{C}=3$ phases.} (a) Phase diagram in $(\theta,\Delta)$ space showing several distinct metallic and insulating phases. Of importance is the presence of a contiguous region showing $\mathcal{C}=2$ and $\mathcal{C}=3$ phases at negative $\Delta$ and $\theta \in[0.95^\circ,1.45^\circ]$ featuring a spectrally-isolated, isospin-polarized, topologically-nontrivial bands right below the Fermi level. The band gap, bandwidth, and trace condition violation of that band are shown in (b-d). A representative band structure for the $\mathcal{C}=2$ phase color coded by isospin and by layer polarization is shown in (e) and (f) respectively. The band of interest is exceptionally narrow with large gap below and a smaller gap above. A representative band structure for the $\mathcal{C}=3$ phase color coded by isospin and by layer polarization is shown in (i) and (j) respectively. The bandwidth of this phase is not as small as the $\mathcal{C}=2$ phase. The distribution of the Berry curvature and quantum metric is shown in (g,h,k,l). Here, $\epsilon_r = 12$ and $D=40$ nm. Energies are measured relative to Fermi level, $E_F.$} 
    \label{fig:fig3_main}
\end{figure}

Next, we present the mean-field results for the parallel configuration in Fig. \ref{fig:fig3_main}. We again find in Fig. \ref{fig:fig3_main}(a) a plethora of metallic and insulating phases. Crucially, we find a prominent contiguous region that starts with $\mathcal{C} = 2$ at lower angles and crosses over to $\mathcal{C} = 3$ at larger angles. Again, while these phases no doubt are related to the $\mathcal{C}=2$ and $\mathcal{C}=3$ bands in the non-interacting limit, we emphasize that they occur here in the HF phase diagram at values of $(\theta,\Delta)$ where the non-interacting band structures do not feature a fully isolated Chern band near charge neutrality. Thus, comparing to Fig. \ref{fig:fig1_main}(g,h) in the non-interacting limit, these fully-isolated bands must be stabilized by Coulomb interactions. For the $\mathcal{C} = 2$ phase, optimal $\mathfrak{f.o.m.}$ occurs at $\theta = 1.1^\circ$ and $\Delta = -21$ meV where $\delta = 7.9$ meV, $W =3.8$ meV, and $\lambda = 0.87.$ The band structure at this value of $(\theta,\Delta)$ is shown in Fig. \ref{fig:fig3_main}(e,f). We notice that the band is exceptionally narrow with a smaller gap above and a much larger gap below. The Berry curvature is more delocalized in this case, as shown in Fig. \ref{fig:fig3_main}(g), compared to the antiparallel situation. As before, we find that Coulomb interactions reduce $\lambda.$ After HF renormalization, optimal $\lambda \approx0.9$ is smaller than the non-interacting values around $\lambda \approx 1.3.$ Moving onto the $\mathcal{C}=3$ phase, we find that optimal $\mathfrak{f.o.m.}$ occurs at a higher angle $\theta = 1.3^\circ$ but same $\Delta = -21$ meV. Here, $\delta = 9.5$ meV while $W = 11$ meV. Generally, we find that the $\mathcal{C}=3$ phase features large bandwidths compared to the $\mathcal{C}=2$ phases both in the parallel and antiparallel configurations. This is perhaps a bit surprising considering the energy spectra for both Chern phases are similar in the non-interacting limit, highlighting the role of wavefunctions (form factors) in controlling the interaction renormalization of electronic properties. The trace condition violation here is remarkably low $\lambda = 0.55$; it is already low in the non-interacting limit, $\lambda \approx 0.75$, but HF renormalization reduces it even more. Thus, we find in all cases that Coulomb interactions have the tendency to lower the trace condition violation\footnote{By this, we mean that in both configurations and for both $\mathcal{C}=2$ and $\mathcal{C}=3,$ the optimal values of $\lambda$ are reduced after including HF interactions. We note, however, that the optimal values of $\lambda$ may occur at different points in $(\theta,\Delta)$ space. We have not done a systematic comparison of $\lambda$ before and after HF renormalization for the same values of $(\theta,\Delta)$ because it is not always sensible to do so since there are many instances where the non-interacting bands are not gapped but the HF bands are gapped and vice versa.}. The Berry curvature and quantum metric for the $\mathcal{C} = 3$ phase are both concentrated at the $\bar{\Gamma}$ point, as shown in Fig. \ref{fig:fig3_main}(k,l). The charge densities for both the $\mathcal{C}=2$ and $\mathcal{C}=3$ phases are localized primarily on layer 1 ($\gtrsim 75 \%$). On that layer, the charge density fluctuates laterally as well, with more concentration at one of the three Wyckoff positions in the moir\'{e} unit cell. However, the $\mathcal{C}=2$ phase chooses a different Wyckoff position to localize charge compared to the $\mathcal{C}=3$ phase.

In both stacking configurations, we have established that electron-electron interactions at $\epsilon_r =12$ and $D=40$ nm are favorable to the formation of topologically-nontrivial bands that give rise to anomalous quantum Hall phases at an integer filling and are possible candidates for fractional quantum Hall phases at partial fillings. Coulomb interactions generate fully-isolated, relatively-narrow bands for values of $(\theta,\Delta)$ where these bands are entangled with other bands in the non-interacting limit; Coulomb interactions also reduce the quantum metric violation in these phases compared to their non-interacting counterparts. From the presented phase diagrams, it is also clear that these phases are robust with variations in $\Delta$ and $\theta$ because the phase regions span prominent domains in phase space. In other words, no fine tuning of $\theta$ or $\Delta$ is necessary.

\begin{figure}[t!]
    \centering
    \includegraphics[width=1\linewidth]{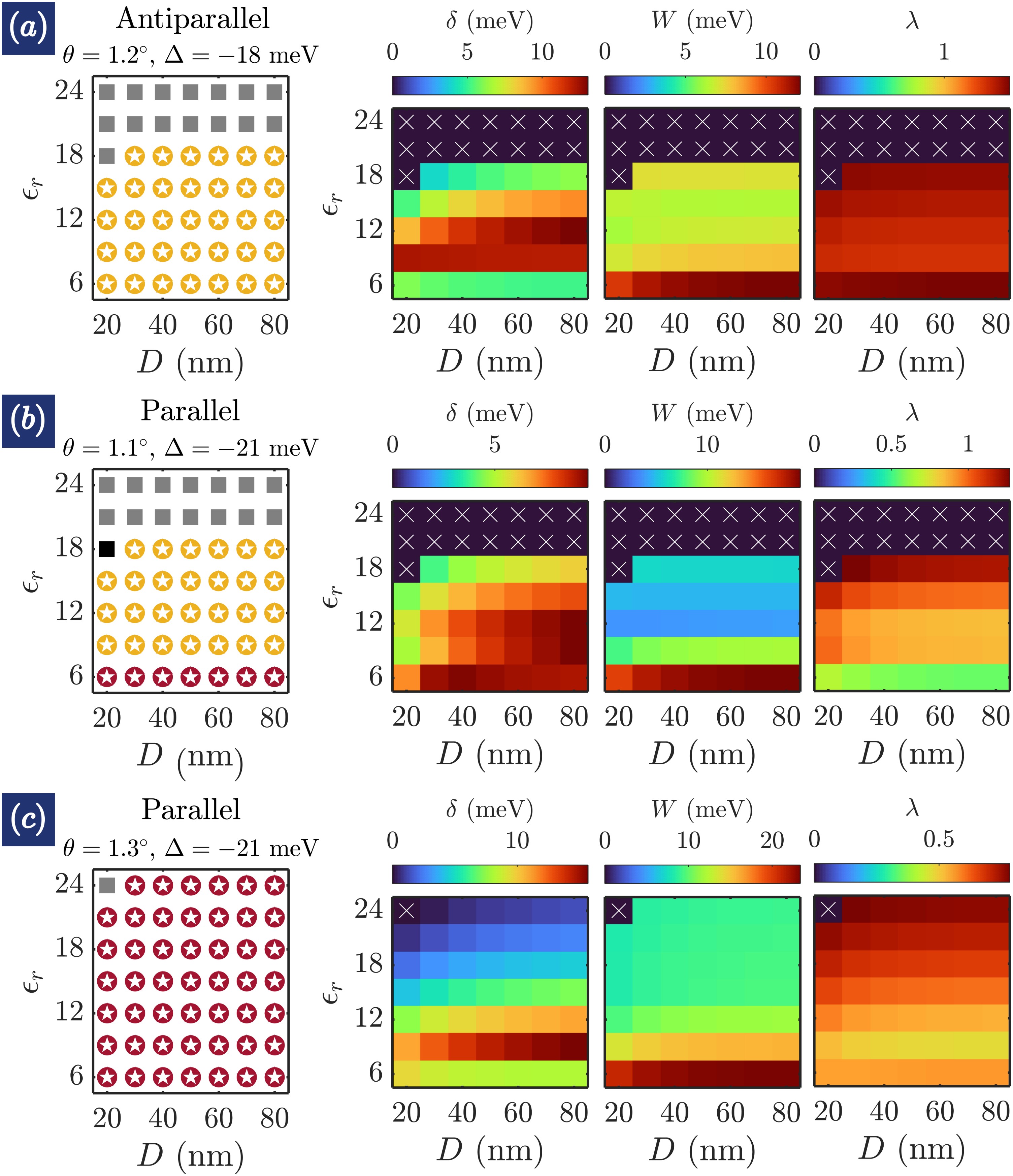}
    \caption{\textbf{Stability of topologically nontrivial phases.} We show the evolution of states at fixed $(\theta,\Delta)$ as a function of $(D,\epsilon_r).$ In the first column, we classify the states using the same method as in Figs. \ref{fig:fig2_main} and \ref{fig:fig3_main}. In the second to fourth columns, we characterize the band gap, bandwidth, and trace condition violation of the band of interest for values of $(D,\epsilon_r)$ where it is isolated. For these simulations, we include $6$ bands per isospin.}
    \label{fig:fig4_main}
\end{figure}

Next, we analyze the stability of these phases as the Coulomb interaction is modified. This is an important consideration because the strength of electron-electron interactions can be experimentally controlled by dielectric engineering (changing $\epsilon_r$) and gate distance (changing $D$). For this purpose, we park ourselves at a particular $(\theta,\Delta)$ and sweep $\left(D,\epsilon_r\right).$ For demonstration, we choose the optimal combinations of $(\theta,\Delta)$ as diagnosed by $\mathfrak{f.o.m.}$ in the preceding analysis. The results are shown in Fig. \ref{fig:fig4_main}. In general, all of these states seem quite insensitive to changing the gate distance $D.$ The dependence on $\epsilon_r$ is more intricate. In Fig. \ref{fig:fig4_main}(a), we assess the stability of the antiparallel configuration for a $\mathcal{C}=2$ state. As is clear, this state is robust for $\epsilon_r < 21.$ At the highest $\epsilon_r >21$ (i.e. the weakest Coulomb strength), this state is out-competed by the symmetric metallic state. The value of $\epsilon_r \sim 9-15$ appears optimal for this state and lies in the range of $\epsilon_r$ that are typically used \cite{laturia2018dielectric}. $\lambda$ remains relatively unchanged as $\epsilon_r$ and $D$ are varied. For a $\mathcal{C} = 2$ state in the parallel configuration, we again observe that there is a finite range of $\epsilon_r\sim 9-18$ where this phase exists. For weaker Coulomb interactions (larger $\epsilon_r$), this state is outcompeted by the symmetric metallic phase too. For smaller $\epsilon_r \sim 6$, this state merges into a $\mathcal{C} = 3$ state. The $\mathcal{C}=3$ state in the parallel configuration is much more robust than the other two states. It persists essentially in the entire domain of $(D,\epsilon_r)$ for our scan. The optimal dielectric constant occurs around $\epsilon_r =9.$ We also show extensive phase diagrams for $\epsilon_r = 6$ in Ref. \cite{SI}.

In Ref. \cite{SI}, we consider many other different perturbations to assess the robustness of our predictions. In particular, we study the absence of lattice relaxation, i.e. $\gamma_\mathrm{AA}/\gamma_{\mathrm{AB}} = 1,$ the role of reduced $\gamma_0=2.7$ eV\footnote{It appears that there is no consensus on the value of $\gamma_0$ in the literature. Values extracted from experiments tend to be at the higher end $3.1-3.2$ eV \cite{dresselhaus1981intercalation,shi2020electronic,zhou2021half}, while values calculated from density functional theory tend to be at the lower end $2.6-2.7$ eV \cite{Malard2007Probing,Jung2014Accurate}.  We take the perspective that experimentally-extracted values are more representative of reality. In the interest of making realistic predictions, all results presented in the main text are simulated with a higher value of $\gamma_0.$ However, to their assess robustness, we also show results simulated with a smaller $\gamma_0$ in Ref. \cite{SI}. } \cite{Malard2007Probing,Jung2014Accurate}, and the neglect of layer dependence on the Coulomb potential energy, i.e. $\mathcal{V}_{\mathbf{k}} \propto \tanh(kD/2)/k.$ We find that the $\mathcal{C}=2$ phase in the antiparallel configuration is robust in all the cases examined. The parallel configuration is more sensitive to changes in parameters. While both Chern $\mathcal{C} = 2$ and $\mathcal{C} = 3$ phases always exist, we find that their areas in phase space can be significantly enlarged or diminished depending on simulation parameters. Nevertheless, we emphasize again that we always find that either $\mathcal{C}=2$, $\mathcal{C}=3$, or both phases exist. We provide further discussion of the robustness of these phases in Ref. \cite{SI}.

We end by addressing a question that naturally follows the preceding discussion: to which of the two known experimental FCI systems is our platform similar? The quick answer is that our system shares similarities with both, but also exhibits some distinct properties. Here, we set aside the obvious difference that our system features $|\mathcal{C}|>1$ unlike the two experimental systems that have $|\mathcal{C}|=1$. On one hand, the isolated nontrivial bands already exist in the non-interacting limit, like in twisted MoTe$_2$. On the other, optimal values of $(\theta,\Delta)$ where isolated bands with favorable quantum geometry exist after HF renormalization are where non-interacting bands are not isolated at all, like in graphene-hBN heterostructures. Furthermore, a majority of the electron population for the active band resides on the outer layer of the trilayer substack, bearing resemblance to the electronic occupation in the conduction band of graphene-hBN heterostructures under a displacement field. However, the role of the moir\'{e} structure (e.g. the twist interface) is important in our system as it would not be gapped otherwise even with a displacement field, unlike in graphene-hBN heterostructures where the large gap below the flat band is field-induced\footnote{An interesting investigation for future work might be to carefully explore if electron-electron interactions alone would be enough to generate the isolated narrow Chern bands even when one completely turns off the hoppings across the twist interface. That is, when the moir\'{e} structure is non-existent, what happens to the correlated phases? However the answer to this interesting question turns out to be, it shall have no material effect on the conclusions of the present work. The fact that we observe qualitatively different phase diagrams depending on alignment configuration already points to the importance of the twist interface in our system. }. Therefore, from our perspective, it is fairest to say that our system is a paradigm of its own, with important similarities and differences to both of the known FCI systems.

In this work, we have demonstrated that $\mathrm{TRMG}_{3+2}$ is likely an anomalous Chern insulator whose valence bands immediately below the Fermi level have $|\mathcal{C}|>1.$ These phases are stabilized by Coulomb interactions. Most interestingly, the isolated, narrow, HF-renormalized bands only mildly violate the trace condition. Therefore, we are optimistic that our proposed platform might also become an FCI once partially doped. In the present work, we have not explored the different possible FCI states that can form in this system \cite{Wang2012Fractional,Barkeshli2012Topological,Liu2012Fractional,Sterdyniak2013Sterdyniak,Liu2013Fractional,Moller2015Fractional,Andrews2021Stability}. In the near future, we shall report results from ongoing work devoted to exploring such possibilities using numerical techniques suitable for calculating strongly-correlated ground states (such as many-body exact diagonalization).  $\mathrm{TRMG}_{3+2}$ is just one example in a family of twisted materials constructed from an $M$-layer rhombohedral stack rotationally misaligned on top of an $N$-layer rhombohedral stack, $\mathrm{TRMG}_{M+N}.$ We expect these related systems to contain similar phases, perhaps even with higher Chern numbers for isolated narrow bands. Our hypothesis is supported by the fact that all of the relevant high-Chern phases in $\mathrm{TRMG}_{3+2}$  occur in regions of displacement field where the electron occupation is peaked on the trilayer substack (rather than the bilayer substack), suggesting that higher layer number associated with flat-band formation may have an outsized influence on the observed phenomenology. 
Our results demonstrate that layer number adds yet another ingredient to an already-versatile set of tunable knobs (relative stacking order, displacement field, twist angle, etc...) endowed to this family of materials. Therefore, we expect continued investigation into $\mathrm{TRMG}_{M+N}$ to yield a rich understanding of correlated and fractionalized physics.

\textit{Note added:} After this work had been posted to the arXiv, Ref. \cite{dong2025observation} reported experimental observations of $\mathcal{C}=3$ integer Chern insulating states in $\mathrm{TRMG}_{3+2}$ at $\theta \approx 1.5^\circ$ for displacement fields that drive conduction electrons towards the outer layer of the trilayer substack. These observations are entirely consistent with our predictions of stable $\mathcal{C} = 3$ bands in the parallel configuration for negative $\Delta$ and angles $\theta \in [ 1.05^\circ,1.45^\circ].$

We thank Eugene J. Mele, Francisco Guinea, Nicholas Bonesteel, Mitali Banerjee,  Matthew Yankowitz,  Xirui Wang, and Jose Angel Silva Guill\'{e}n for fruitful discussions. We are grateful to Hitesh Changlani for sharing computational resources. Numerical calculations are done using the High Performance Compute Cluster  of the Research Computing Center (RCC) at Florida State University. This work was supported by  start-up funds from Florida State University and the National High Magnetic Field Laboratory and also by an intramural award from the Florida State University's Provost’s Office and the Office of Postdoctoral Affairs. The National High Magnetic Field Laboratory is supported by the National Science Foundation through NSF/DMR-2128556 and the State of Florida.

\newpage

\setcounter{equation}{0}
\setcounter{figure}{0}
\renewcommand{\theequation}{S\arabic{equation}}
\renewcommand{\thefigure}{S\arabic{figure}}
\renewcommand{\bibnumfmt}[1]{[#1]}
\renewcommand{\citenumfont}[1]{#1}

\onecolumngrid

\begin{large}
\begin{center}
\textbf{Supplementary Material} \textit{for}\\
Coulomb Interaction-Stabilized Isolated Narrow Bands with Chern Numbers $\mathcal{C} > 1$\\ in Twisted Rhombohedral Trilayer-Bilayer Graphene
\end{center}    
\end{large}

\begin{center}
\begin{normalsize}
\foreignlanguage{vietnamese}{Võ Tiến Phong}$^{1,2}$ and Cyprian Lewandowski$^{1,2}$\\
\vspace{5pt}
\end{normalsize}
\begin{small}
$^{1}$\textit{Department of Physics, Florida State University, Tallahassee, FL, 32306, U.S.A.}\\
$^{2}$\textit{National High Magnetic Field Laboratory, Tallahassee, FL, 32310, U.S.A.}\\
\vspace{5pt}
\today
\end{small}
\end{center}

\tableofcontents
\listoffigures
\listoftables

\section{Continuum Hamiltonian}
\label{sec: Continuum Hamiltonian}

We construct twisted rhombohedral multilayer graphene ($\mathrm{TRMG}_{3+2}$) in the following way. We start with a rhombohedral trilayer whose valley-projected Hamiltonian at valley $\nu = \pm$ is given by 
\begin{subequations}
\begin{align}
\mathbb{K}_{3,\nu}(\mathbf{k}) &= \begin{pmatrix}
    \mathbb{K}_{1,\nu}(\mathbf{k}) & \mathbb{U}_\nu(\mathbf{k}) & \mathbb{W} \\
    \mathbb{U}_\nu^\dagger(\mathbf{k}) & \mathbb{K}_{1,\nu}(\mathbf{k}) & \mathbb{U}_\nu(\mathbf{k}) \\
    \mathbb{W}^\dagger & \mathbb{U}_\nu^\dagger(\mathbf{k})  & \mathbb{K}_{1,\nu}(\mathbf{k}) 
    \end{pmatrix}  \\
\mathbb{K}_{1,\nu}(\mathbf{k}) &= \begin{pmatrix}
    0 & \hbar v_0 \Pi_\nu(\mathbf{k}) \\ 
    \hbar v_0 \Pi^\dagger_\nu(\mathbf{k}) & 0
\end{pmatrix},\\
\mathbb{U}_\nu(\mathbf{k}) & = \begin{pmatrix}
    \hbar v_4 \Pi_\nu(\mathbf{k}) & \hbar v_3 \Pi^\dagger_\nu(\mathbf{k}) \\ 
    \gamma_1 & \hbar v_4 \Pi_\nu(\mathbf{k})
\end{pmatrix}, \\
\mathbb{W} &= \begin{pmatrix}
    0 & \gamma_2/2\\
    0 & 0
\end{pmatrix},
\end{align}
\end{subequations}
where $\mathbf{k}$ is measured from the center of the microscopic Brillouin zone, $\Pi_\nu(\mathbf{k}) = \nu k_x - i k_y - 4\pi/3a,$ $a = 2.46$
\AA \vspace{1pt} is lattice constant of graphene, $\hbar v_i = -\sqrt{3} \gamma_i a/2,$ and $\gamma_i$ are the hopping parameters. We ignore dimerization energies on eclipsed sites in this work. The parameters used in this work are listed in Table \ref{tab:hopping parameters}. Next, we stack Bernal bilayer graphene on top of the trilayer in perfect crystallographic alignment. If the orientation of the bilayer is \textit{parallel} to the trilayer, then the bilayer Hamiltonian is 
\begin{equation}
\mathbb{K}_{2,\nu}(\mathbf{k}) = \begin{pmatrix}
    \mathbb{K}_{1,\nu}(\mathbf{k}) & \mathbb{U}_\nu(\mathbf{k}) \\
    \mathbb{U}_\nu^\dagger(\mathbf{k}) & \mathbb{K}_{1,\nu}(\mathbf{k})
    \end{pmatrix};
\end{equation}
otherwise, if the orientation of the bilayer is \textit{antiparallel} to the trilayer, then the bilayer Hamiltonian is 
\begin{equation}
\mathbb{K}_{2,\nu}(\mathbf{k}) = \begin{pmatrix}
    \mathbb{K}_{1,\nu}(\mathbf{k}) & \mathbb{U}^\dagger_\nu(\mathbf{k}) \\
    \mathbb{U}_\nu(\mathbf{k}) & \mathbb{K}_{1,\nu}(\mathbf{k})
    \end{pmatrix}.
\end{equation}
The two inequivalent stacking orders differ by taking the Hermitian adjoint of the interlayer hopping matrix $\mathbb{U}_\nu(\mathbf{k})$ in the bilayer sector. One stacking order is related to the other by a $C_{2z}$ rotation of the bilayer. Next, we rotate the trilayer by $+\theta/2$ and the bilayer by $-\theta/2$ around a common vertical axis. By continuing to measure momentum using a common (non-rotated) coordinate axis for both stacks, the momentum in the trilayer gets rotated  by $-\theta/2$ while the momentum in the bilayer gets rotated by $+\theta/2.$ The rotation in reciprocal space is opposite to the rotation in real space to compensate for the use of a common axis. Thus, the Hamiltonian of a twisted trilayer-bilayer stack is simply 
\begin{equation}
    \mathbb{K}_{3+2,\nu}(\mathbf{k}) = \begin{pmatrix}
        \mathbb{K}_{3,\nu}(\mathbb{R}[-\theta/2]\mathbf{k}) & * \\
        * & \mathbb{K}_{2,\nu}(\mathbb{R}[+\theta/2]\mathbf{k})
    \end{pmatrix},
\end{equation}
where $\mathbb{R}[\theta]$ is the rotation matrix that implements counterclockwise rotation by angle $\theta.$

\begin{figure}
    \centering
    \includegraphics[width=1\linewidth]{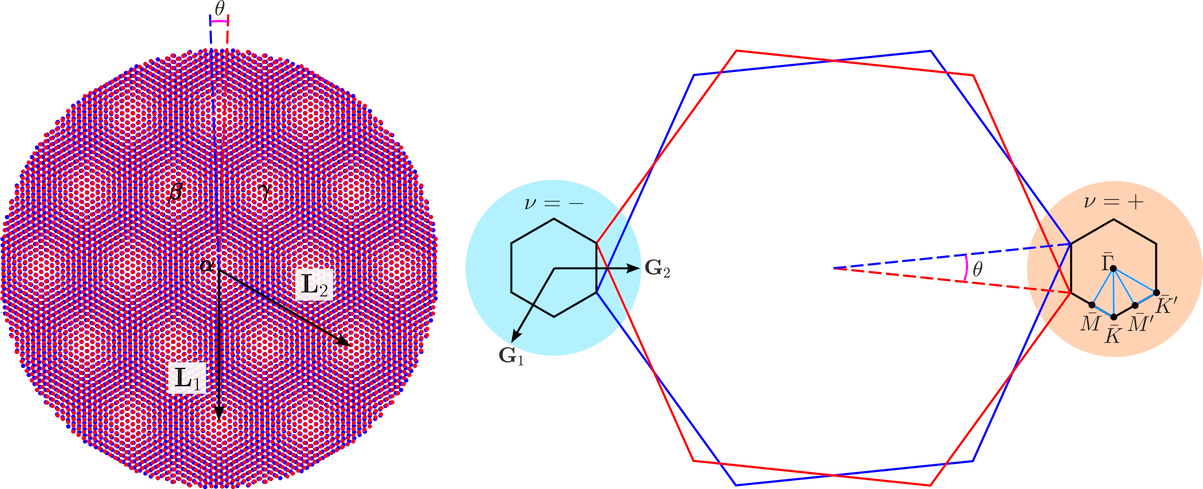}
    \caption{\textbf{Lattice representation of twisted rhombohedral trilayer-bilayer graphene.} (Left) Moir\'{e} pattern of rhombohedral bilayer grahene, whose atoms are colored red, stacked on top of rhombohedral trilayer grahene, whose atoms are colored blue, with a relative twist angle $\theta$. Regions of high-symmetry stacking order are labeled $\alpha,$ $\beta,$ $\gamma.$ (Right) Reconstructed moir\'{e} Brillouin zone shown by the black hexagons, where high-symmetry points are marked. The larger blue and red  hexagons are the rotated Brillouin zones of the graphene multilayer subsystems before hybridization.}
    \label{fig:Lattice structure}
\end{figure}

At the interface between the trilayer and bilayer substacks, a long-wavelength moir\'{e} pattern is formed; this pattern modulates the hopping amplitude between regions of different local atomic stacking registry. Denoting the microscopic primitive translation vectors and primitive reciprocal lattice vectors as 
\begin{equation}
    \mathbf{a}_1 = a \left( 1, 0 \right), \quad \mathbf{a}_2 = a \left( \frac{1}{2}, \frac{\sqrt{3}}{2} \right), \quad \mathbf{b}_1 = \frac{4\pi}{\sqrt{3}a} \left( \frac{\sqrt{3}}{2},-\frac{1}{2}\right), \quad \text{and} \quad \mathbf{b}_2 = \frac{4\pi}{\sqrt{3}a} \left( 0,1\right),
\end{equation}
the moir\'{e} reciprocal lattice vectors are defined as $\mathbf{G}_i = \left(\mathbb{R}[-\theta/2]-\mathbb{R}[+\theta/2] \right)\mathbf{b}_i$:
\begin{equation}
    \mathbf{G}_1 = \frac{4\pi}{\sqrt{3}L} \left(-\frac{1}{2}, - \frac{\sqrt{3}}{2}\right) \quad \text{and} \quad \mathbf{G}_2 = \frac{4\pi}{\sqrt{3}L} \left(1,0\right),
\end{equation}
where $L = a/2\sin(\theta/2) \approx a/\theta$ is the moir\'{e} period. The corresponding moir\'{e} lattice vectors $\mathbf{L}_1$ and $\mathbf{L}_2$ are shown in Fig. \ref{fig:Lattice structure}. This period diverges as $\theta \rightarrow 0.$ In terms of these moir\'{e} reciprocal lattice vectors, the tunneling matrix between the trilayer and bilayer substacks is given in real space by 
\begin{equation}
\label{eq: hopping matrix}
    \mathbb{T}_\nu(\mathbf{r}) =   \begin{pmatrix}
        \gamma_\mathrm{AA} & \gamma_\mathrm{AB} \\
        \gamma_\mathrm{AB} & \gamma_\mathrm{AA}
    \end{pmatrix}  + \begin{pmatrix}
        \gamma_\mathrm{AA} & \gamma_\mathrm{AB}\omega^{-\nu} \\
        \gamma_\mathrm{AB}\omega^{\nu} & \gamma_\mathrm{AA}
    \end{pmatrix}  e^{-i\nu \mathbf{G}_1 \cdot \mathbf{r}}     + \begin{pmatrix}
        \gamma_\mathrm{AA} & \gamma_\mathrm{AB}\omega^{\nu} \\
        \gamma_\mathrm{AB}\omega^{-\nu} & \gamma_\mathrm{AA}
    \end{pmatrix}  e^{-i\nu (\mathbf{G}_1 +\mathbf{G}_2)\cdot \mathbf{r}},
\end{equation}
where $\omega = e^{2\pi i /3}$ and $\gamma_\mathrm{AA}$ and $\gamma_\mathrm{AB}$ are hopping amplitudes at $AA$ and $AB$ regions. Because of the possibility of lattice relaxation, $|\gamma_\mathrm{AA}| \leq |\gamma_\mathrm{AB}|$. In twisted bilayer graphene, $\gamma_\mathrm{AA}/\gamma_\mathrm{AB} = 0.8$ in a typical experimental setup. Presumably, with more layers, it is more difficult to fully relax. We do not scrutinize lattice relaxation in this work. Instead, we leave the ratio $\gamma_\mathrm{AA}/\gamma_\mathrm{AB}$ as an adjustable parameter in the modeling. The effect of lattice relaxation on the band structure can be an interesting future study. The parameters we use in our study are listed in Table \ref{tab:hopping parameters}. For {\fontfamily{cmtt}\selectfont Param\_one} and {\fontfamily{cmtt}\selectfont Param\_two}, we include the parameters from Ref. \cite{zhou2021half}, which were deduced experimentally. They are close to the values from this experiment \cite{shi2020electronic}. Therefore, we believe these two sets to represent parameters somewhat closer to experiments. For $\gamma_\mathrm{AB},$ we take it to be roughly one-third of $\gamma_1$ since they are essentially the same term if the twist were not there. In experimental Ref. \cite{lisi2021observation} and theoretical Refs. \cite{Carr2019Exact,Leconte2022Relaxation}, $\gamma_\mathrm{AB} = 110$ meV, which is roughly what we use $\gamma_\mathrm{AB} = 120$ meV. For {\fontfamily{cmtt}\selectfont Param\_three} and {\fontfamily{cmtt}\selectfont Param\_three}, we reduce $\gamma_0 = -2700$ meV and $\gamma_\mathrm{AB} = 100$ meV to values closer to what density functional theory calculations predict \cite{Jung2014Accurate, Koshino2018Maximally,Leconte2022Relaxation}. It is worth mentioning parenthetically that we include $\gamma_2$, hopping between next-nearest layers, in the rhombohedral trilayer stack, but we do not include such a hopping across the twist interface. This is because we assume that the misalignment due to twist diminishes that already-weak tunneling significantly. By testing parameters from both DFT calculations and experiments, we demonstrate that our principle conclusions are robust against these deviations in parameters. Now, including only nearest-layer hopping across the twist interface, the Hamiltonian in real space now becomes
\begin{equation}
    \mathbb{K}_{3+2,\nu}(\mathbf{r}) = \begin{pmatrix}
        \mathbb{K}_{3,\nu}(-i\mathbb{R}[-\theta/2]\nabla_\mathbf{r}) & \mathbb{T}_\nu(\mathbf{r}) \\
        \mathbb{T}^\dagger_\nu(\mathbf{r}) & \mathbb{K}_{2,\nu}(-i\mathbb{R}[+\theta/2]\nabla_\mathbf{r})
    \end{pmatrix}.
\end{equation}
Finally, we add to the Hamiltonian a vertical interlayer electrostatic potential (corresponding to the external displacement field):
\begin{equation}
\mathbb{D} = \Delta \begin{pmatrix}
    -2 & 0 & 0 & 0 & 0 \\
    0 & -1 & 0 & 0 & 0 \\
    0 & 0 & 0 & 0 & 0 \\
    0 & 0 & 0 & +1 & 0 \\
    0 & 0 & 0 & 0 & +2 \\
\end{pmatrix}_\ell\otimes \mathbb{1}_\sigma,
\end{equation}
where $\sigma$ and $\ell$ denote sublattice and layer degrees of freedom respectively and $\Delta$ is the energy difference between adjacent layers. The energy difference between the outermost layers in this five-layer stack is $4|\Delta|.$ This vertical displacement is an essential feature of our study since it functions as a tunable knob with which to control band flatness and isolation.

By themselves, rhombohedral graphene multilayers already do not respect $C_{2z}$ symmetry. So, of course, $\mathrm{TRMG}_{3+2}$ also does not respect $C_{2z}$ symmetry. Because the top and bottom stacks do not have the same number of layers, any possible symmetry exchanging layers from the trilayer with those of the bilayer, such as $C_{2x}$ and $C_{2y},$ is violated, even when $\Delta = 0.$ There are also no vertical mirror planes in this system. However, $C_{3z}$ and time-reversal $\mathcal{T}$ symmetries are respected when both valleys are considered. All in all, this system has only a few symmetries. In our simulations, we ensure that both $\mathcal{T}$ and $\mathcal{C}_{3z}$ symmetries are numerically respected in every converged solution.

Throughout this work, we shall restrict ourselves to  $\theta > 0$ without loss of generality. For $\theta < 0,$ we can map it to a value in $\theta>0$ using the following transformation $\lbrace \theta, \nu, k_x,x\rbrace \mapsto - \lbrace \theta, \nu, k_x,x\rbrace.$ This is $M_x$ mirror reflection. One can see that this is a good transformation in the following way. First, by noting that $\theta \mapsto - \theta$ leads to $\mathbf{G}_i \mapsto -\mathbf{G}_i,$ we immediately find that $\lbrace\theta,\nu\rbrace \mapsto - \lbrace \theta,\nu\rbrace$ leaves $\nu\mathbf{G}_i$ invariant. However, $\nu \mapsto -\nu$ flips the two matrices involving $\nu$ in the hopping matrix of Eq. \eqref{eq: hopping matrix}. To switch them back, we implement $x \mapsto -x$ to exchange $\mathbf{G}_1$ and $\mathbf{G}_1 + \mathbf{G}_2.$ This proves that the hopping matrix is invariant under this transformation. The diagonal elements are also invariant since $\left(\nu k_x - i k_y\right) e^{\pm i \nu \theta/2}$ is manifestly invariant under this transformation. Thus, every state in valley $\nu$ at momentum $(k_x,k_y)$ and twist angle $\theta$ has a partner in valley $-\nu$ at momentum $(-k_x,k_y)$ and twist angle $-\theta.$ This immediately implies that the bands at $-\theta$ have the same spectral and topological properties as those at $\theta.$ So, we do not need to assess $-\theta$ separately.

\begin{table}[]
    \centering
    \begin{tabular}{||c||c|c|c|c| c|c| c|  c ||}
        \hline\hline
         Parameter set name & $\gamma_0$  & $\gamma_1$ & $\gamma_2$ & $\gamma_3$ & $\gamma_4$ & $\gamma_\mathrm{AA}$& $\gamma_\mathrm{AB}$ & Description\\
         \hline
         \hline
          {\fontfamily{cmtt}\selectfont Param\_one} & $-3100$ & $380$ & $-15$ & $290$ & $141$ & $100$ & $120$  & Includes some lattice relaxation\\
          \hline
          {\fontfamily{cmtt}\selectfont Param\_two}  & $-3100$ & $380$ & $-15$ & $290$ & $141$ & $120$ & $120$ & No  lattice relaxation\\
          \hline
          {\fontfamily{cmtt}\selectfont Param\_three} & $-2700$ & $380$ & $-15$ & $290$ & $141$ & $80$ & $100$ & Includes some lattice relaxation, reduced Fermi velocity\\
          \hline
          {\fontfamily{cmtt}\selectfont Param\_four}  & $-2700$ & $380$ & $-15$ & $290$ & $141$ & $100$ & $100$& No  lattice relaxation, reduced Fermi velocity\\
          \hline\hline
    \end{tabular}
    \caption{\textbf{Hopping parameters for rhombohedral graphene multilayers.} All values are quoted in millielectronvolts. $\gamma_0,$ $\gamma_1,$ $\gamma_2,$ $\gamma_3,$ and $\gamma_4$ in {\fontfamily{cmtt}\selectfont Param\_one} and {\fontfamily{cmtt}\selectfont Param\_two} are taken from Ref. \cite{zhou2021half}. Results presented in the main text come from {\fontfamily{cmtt}\selectfont Param\_one} because we believe it contains values closest to those found in experiments.}
    \label{tab:hopping parameters}
\end{table}

\section{Non-Interacting Phase Diagrams}
\label{sec:Non-Interacting Phase Diagrams}

\subsection{Band Alignment Considerations}

\begin{figure}
    \centering
    \includegraphics[width=0.8\linewidth]{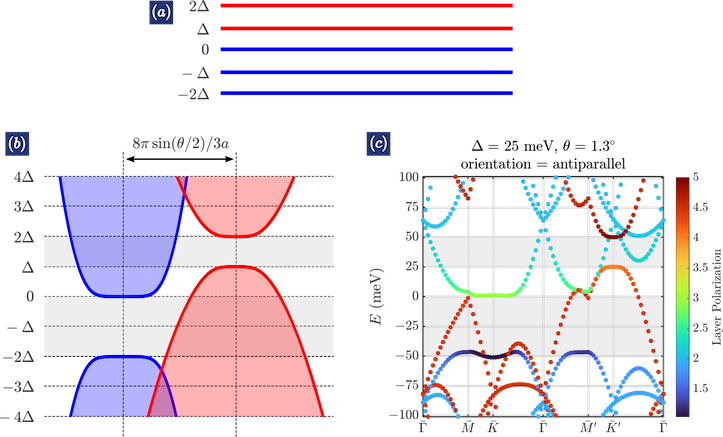}
    \caption{\textbf{Band alignment of a twisted stack without hopping across the twist interface.} (a) Real-space distribution of the on-site energies induced by an interlayer displacement field. (b) Schematic representation of the band alignment of the trilayer and bilayer spectra showing the absence of a global energy gap even the local spectrum of each substack is individually gapped. (c) Band structure with trigonal warping terms ($\gamma_i$ from {\fontfamily{cmtt}\selectfont Param\_one}) but without hopping across the twist interface ($\gamma_\mathrm{AA} = \gamma_\mathrm{AB} = 0$). Here, states belonging to one substack as opposed to another are identified by their layer polarization. States belonging to the trilayer have layer polarization between 1 and 3, while those of the bilayer have layer polarization between 4 and 5.}
    \label{fig:bandalignment}
\end{figure}

In this section, we study the phase diagrams calculated in the non-interacting limit. Without interactions, the four isospin flavors are necessarily degenerate; thus, it is sufficient to inspect just the $\lbrace \nu = +, s = \uparrow \rbrace $ flavor. To begin, it is instructive to examine the band alignment of the two substacks in a minimal model that neglects all trigonal warping terms as well as hopping across the twist interface. In this case, the two substacks are completely independent from each other. At this level of simplicity, the discussion is valid for both the antiparallel and parallel configurations. Because of the relative twist, the band structures of the substacks are shifted away from each other by $8\pi\sin(\theta/2)/3a$ in momentum space. Near their respective rotated zone corners, the energy spectra of the two substacks are \cite{Koshino2009Trigonal}
\begin{subequations}
\begin{align}
    \varepsilon_3(\mathbf{k}) &= -\Delta \pm \sqrt{\Delta^2 + \frac{(\hbar v_0 |\mathbf{k}|)^6}{\gamma_1^4}} ,    \\
    \varepsilon_2(\mathbf{k}) &= \frac{3}{2}\Delta \pm \sqrt{ \frac{\Delta^2}{4} +\frac{(\hbar v_0 |\mathbf{k}|)^4}{\gamma_1^2}}.
\end{align}
\end{subequations}
The trilayer spectrum is centered at $-\Delta$ with a gap of $2|\Delta|$ and disperses away its extrema as $\pm |\mathbf{k}|^3,$ while the bilayer spectrum is centered at $3\Delta/2$ with a gap of $|\Delta|$ and disperses away from its extrema as $\pm |\mathbf{k}|^2.$ Importantly, for $\Delta > 0,$ the conduction band minimum of the trilayer is located at $E = 0$ while the valence band maximum of the bilayer is located at $E = \Delta.$ Therefore, the two spectra overlap, and there are no global gaps that result from local gap openings due to $\Delta$ because of the spectral shifts of the two substacks. For $\Delta < 0,$ the valence band maximum of the trilayer is at $E = 0,$ while the conduction band minimum is at $E = \Delta.$ Thus, the same conclusion applies that there are no resulting global gaps. This is illustrated schematically in Fig. \ref{fig:bandalignment}(b). This basic observation survives even if the trigonal warping terms are switched back on, as shown in Fig. \ref{fig:bandalignment}(c). In this band structure, we identify a state $\ket{\Psi_\mathbf{k}}$ as belonging to one substack as opposed to another according to its layer polarization 
\begin{equation}
\label{eq: layer polarization}
    \chi_\ell = \sum_{\ell = 1}^5 \ell \bra{\Psi_\mathbf{k}}\mathbb{P}_\ell \ket{\Psi_\mathbf{k}},
\end{equation} 
 where $\mathbb{P}_\ell$ is the layer projector matrix. If $1 \leq \chi_\ell \leq 3,$ the state belongs to the trilayer sector;  if $4 \leq \chi_\ell \leq 5,$ the state belongs to the bilayer sector. From this simple exercise, we observe that the displacement field plays a minimal role in generating robust gaps in our system of twisted rhombohedral multilayers despite the fact that the displacement field generates significant gaps in each of the substack individually. Consequently, global gaps in our system are produced predominantly by hybridization across the twist interface, demonstrating the importance of the moir\'{e} potential in our work.

We now turn on the interlayer hopping across the twist interface. This generically produces anticrossings where the bands approach each other in energy unless otherwise prohibited by a symmetry. Because our system belongs to a small symmetry group, there are not many, if any, protected band crossings. This is the origin of many gaps, sometimes global, in the band structure. We are interested in the formation of \textit{global} gaps at charge neutrality and above the first conduction band. Wherever these two gaps are significantly larger than the bandwidth of the first conduction band, that band can be treated to a good approximation as an isolated band wherein physics driven purely by electron-electron interactions may be qualitatively important. Even if there is no global gap at charge neutrality, having a gap above the first conduction is sometimes enough to stabilize an anomalous quantum Hall effect at filling of one electron per moir\'{e} unit cell. Thus, we are also interested in situations where there are only global gaps above the first conduction band. However, in these cases, it is not expected that interaction-driven physics can be ascertained by projecting to just a single isolated conduction band. Nonetheless, by considering a subset of nearby bands, we can still ascertain whether this situation can give rise to a symmetry-broken phase in the presence of Coulomb interactions. Given these considerations, we first inspect the spectral gap formation in the band structures. For our system of interest, twist angle $\theta$ and interlayer displacement field $\Delta$ are the two experimentally tunable parameters. So, we scan the phase space of $\theta \text{ vs } \Delta$ looking for these gaps.

\subsection{Antiparallel Stacking Order}

\subsubsection{With Some Lattice Relaxation}
\begin{figure}
    \centering
    \includegraphics[width=1\linewidth]{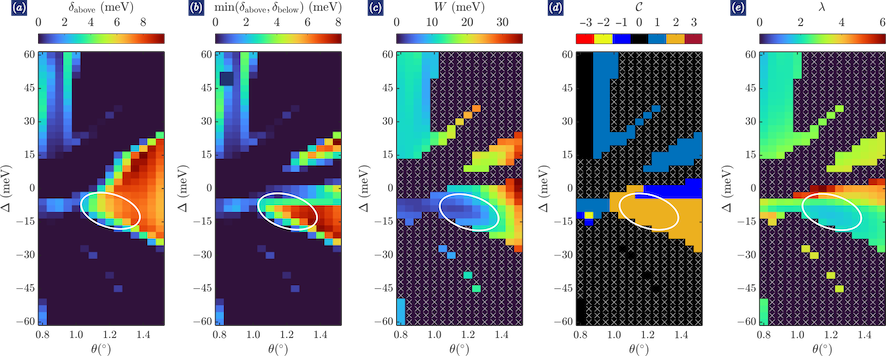}
    \caption{\textbf{Non-interacting phase diagram of the antiparallel stacking order with lattice relaxation.} (a) Value of the global gap above the first conduction band. (b) Global gap of the first conduction band defined as the minimum of the gap below and above that band. (c) Bandwidth of the first conduction band in cases where it is isolated, $i.e. \min \left( \delta_\mathrm{above},\delta_\mathrm{below}\right) > 0.$  (d)-(e) Chern number and trace condition of isolated first conduction bands.  $\times$ denotes points where the gap is zero, for which the task of defining an associated bandwidth, Chern number, or trace condition violation either makes no sense or is useless. {\fontfamily{cmtt}\selectfont Param\_one} is used for the results reported here.}
    \label{noninteractingphase1_antiparallel}
\end{figure}

For the antiparallel stacking order simulated using {\fontfamily{cmtt}\selectfont Param\_one}, which includes some lattice relaxation controlled by a non-unity ratio $\gamma_\mathrm{AA}/\gamma_\mathrm{AB},$ the phase diagram results are shown in Fig. \ref{noninteractingphase1_antiparallel}. Here, $\delta_\mathrm{above}$ is the global gap between the first conduction band and second conduction band while $\delta_\mathrm{below}$ is the global gap between the first conduction band and the first valence band. $\delta_\mathrm{above}$ is plotted in Fig. \ref{noninteractingphase1_antiparallel}(a) where it is shown that for small $\theta < 1^\circ,$ there is a  region in parameter space at large positive $\Delta > 15$ meV where there exists small gaps $ \lesssim 4 $ meV above the first conduction band. More importantly, at larger $\theta > 1^\circ,$ there is a region surrounding $\Delta = 0$ meV where $\delta_\mathrm{above}$ can be as large as $9$ meV. This region grows in area as $\theta$ increases. To search for an anomalous Hall phase, $\delta_\mathrm{above}$ is a useful metric to suggest the possibility of symmetry breaking once interactions are accounted for. To find isolated bands, we further need $\delta_\mathrm{below}$ to be non-zero. The gap associated with a band is defined as the lesser of $\delta_\mathrm{above}$ and $\delta_\mathrm{below}.$ This is plotted in Fig. \ref{noninteractingphase1_antiparallel}(b). As demonstrated here, only a subset of the nonzero region in Fig. \ref{noninteractingphase1_antiparallel}(a) remains nonzero in Fig. \ref{noninteractingphase1_antiparallel}(a) since the complement has a vanishing $\delta_\mathrm{below}.$ We find the region of the largest $\min(\delta_\mathrm{above},\delta_\mathrm{below})$ occurs around $\Delta = 0$, starts from $\theta \sim 1.0^\circ$, and enlarges as $\theta$ increases. An additional indicator of the tendency to stabilize interaction-driven phases is the bandwidth of the band nearest to chemical potential. In Fig. \ref{noninteractingphase1_antiparallel}(c), we plot the bandwidth $W$ of the first conduction band only for values of $(\Delta,\theta)$ where $\min(\delta_\mathrm{above},\delta_\mathrm{below}) >0.$ Where $\min(\delta_\mathrm{above},\delta_\mathrm{below}) = 0,$ we simply identify that point in parameter space as having $W = 0.$ The smallest bandwidths for isolated bands occur near $\Delta = 0$ for $\theta < 1.3^\circ.$ This overlaps partially with the region of largest $\min(\delta_\mathrm{above},\delta_\mathrm{below}).$ Therefore, we identify this region, marked by a while oval in Fig. \ref{noninteractingphase1_antiparallel}, as the most conducive to symmetry-broken phases based on non-interacting band structures. Next, we characterize the topology of the isolated bands. The results are in shown Fig. \ref{noninteractingphase1_antiparallel}(d). We find a large region where the $\mathcal{C} = 2$ as well as regions where the $\mathcal{C} = \pm 1.$ The $\mathcal{C} = 2$ region is particularly interesting because it overlaps in phase space with regions of small bandwidths and large band gaps. In addition to the Chern number, we also characterize the isolated bands by the trace condition, which is a dimensionless quantity defined as 
\begin{equation}
    \lambda = \frac{1}{2\pi}\int d^2\mathbf{k} \left[ \mathrm{Tr}g_{\mu\nu}(\mathbf{k}) - |\mathcal{B}_z(\mathbf{k})| \right],
\end{equation}
$g_{\mu\nu}(\mathbf{k})$ and $\mathcal{B}_z(\mathbf{k})$ are the quantum metric and Berry curvature as defined in Sec. \ref{sec: Numerical Evaluation of Quantum Geometric Tensor}. This trace condition measures how ``close" a band is to the zeroth Landau level in terms its quantum geometry in $\mathbf{k}$-space. A small value of $\lambda$ means that the trace of the quantum metric nearly saturates the bound $g_{\mu\nu}(\mathbf{k}) \geq |\mathcal{B}_z(\mathbf{k})|.$ The trace condition is smallest also in the white oval region of Fig. \ref{noninteractingphase1_antiparallel}(e). The smallest values are $\lambda \approx 1.7$ in that region. All in all, from a non-interacting phase diagram, we identify the white oval region as most promising for stabilizing interaction-driven symmetry-broken phases.

\begin{figure}
    \centering
    \includegraphics[width=\linewidth]{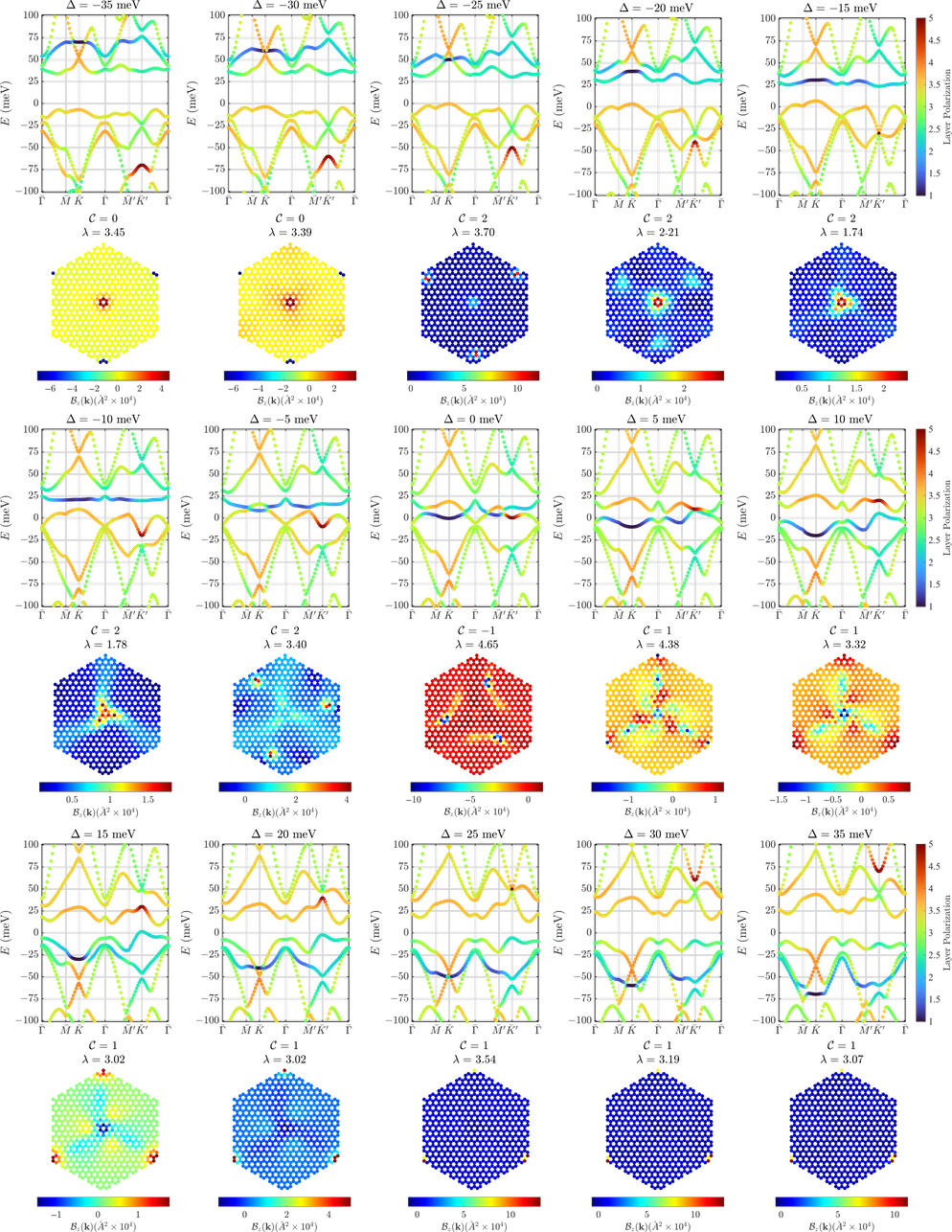}
    \caption{\textbf{Evolution of band structures as a function of $\Delta$ at $\theta = 1.25^\circ$ for the antiparallel stacking order with lattice relaxation.} For each Bloch state, we color code it by its layer polarization. Below each band structure is shown the corresponding Berry curvature distribution of its first conduction band. The Chern number and trace condition violation for that band are also displayed. {\fontfamily{cmtt}\selectfont Param\_one} is used for the results reported here.}
    \label{fig:bandstructure1_antiparallel}
\end{figure}

\begin{sidewaysfigure}
    \centering
    \includegraphics[width=1\linewidth]{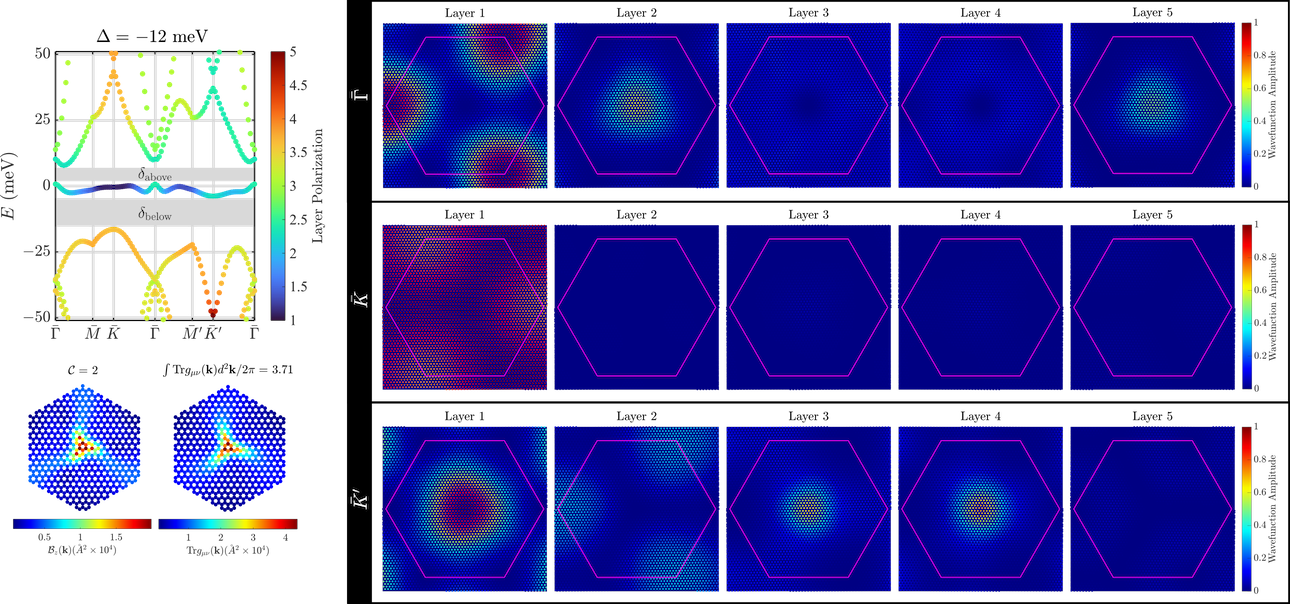}
    \caption{\textbf{Charge density of $\mathcal{C} = 2$ band in the non-interacting limit with some lattice relaxation  for the antiparallel configuration.} We show the charge density in each layer for states at the high-symmetry points $\bar{\Gamma},$ $\bar{K},$ and $\bar{K}'$ for the first conduction band that has Chern number $\mathcal{C} = 2.$ We also compute the $C_{3z}$ symmetry eigenvalues: $\omega(\bar{\Gamma}) = 1$ and $\omega(\bar{K}) = \omega(\bar{K}') = \exp \left(-2\pi i/3\right).$ This confirms that the Chern number is indeed two. The quantum metric and Berry curvature are shown below the band structure. {\fontfamily{cmtt}\selectfont Param\_one} is used for the results reported here.}
    \label{fig:chargedensity1_antiparallel}
\end{sidewaysfigure}

To gain insight about whether interactions would significantly alter the non-interacting phase diagram, we now investigate the charge distribution of the first conduction band. For concreteness, let us take $\theta = 1.25^\circ$ since this angle contains many of the important qualitative features for the other angles as well. In Fig. \ref{fig:bandstructure1_antiparallel}, we plot the evolution of the band structures as $\Delta$ varies for $\theta = 1.25^\circ.$ At $\Delta = 0$ meV, there are two bands near charge neutrality isolated from the rest of the band structure: one band is occupied (valence band) and one band is empty (conduction band) at neutrality. These two bands have mixed layer character, meaning that there are states with $\chi_\ell$ in the trilayer substack and those with $\chi_\ell$ in the bilayer substack coexisting in the same band. As $\Delta$ is increased in the negative direction, these two bands separate, and the resulting conduction band acquires a Chern number of $\mathcal{C} = 2.$ Recall that negative $\Delta$ is the regime where in the trilayer substack is shifted higher in energy than the bilayer substack in the absence of hopping across the twist interface. Therefore, for small negative $\Delta,$ the first conduction band is essentially polarized to the trilayer substack. This band initially becomes narrower in bandwidth as $\Delta$ grows in magnitude before the trend reverses as the the state at $\bar{K}$ is pushed upwards in energy to band invert with the second conduction band at larger $\Delta,$ causing the bandwidth to increase and the global gap surrounding the first conduction band to collapse. In the regime where $\mathcal{C} =  2$ and there is global gap surrounding the first conduction band, the Berry curvature peaks near the $\bar{\Gamma}$ point, but it also has non-negligible density spreading across the moire Brillouin zone as well. Going in the opposite direction to increase positive $\Delta,$ we also observe the first conduction getting pushed to higher energy. Sometimes, this conduction band is isolated global gaps and has a Chern number of $\mathcal{C} = +1.$ However, there is no value of positive $\Delta$ for which its bandwidth is narrow. This is because this band primarily polarizes to the bilayer substack for which band flatness over a large region of momentum space emerges at  higher values of $\Delta.$

The above analysis reveals that in the $\mathcal{C} = 2$ phase, the average density is mostly in the trilayer substack. However, this is only part of the picture as variations of the charge density are not captured in the calculation of their average. To gain more insight, we now inspect the real-space charge density for $\Delta = -12$ meV and $\theta = 1.25^\circ$ as an illustration. In Fig. \ref{fig:chargedensity1_antiparallel}, we map the charge density for each layer for states at the high-symmetry points $\bar{\Gamma},$ $\bar{K},$ and $\bar{K}'.$ We observe significant charge redistribution moving from one $\mathbf{k}$ point to the next. At $\bar{\Gamma},$ most of the charge is concentrated at $\gamma$ regions on layer 1 with some residual charge located on layers 2 and 5 at  $\alpha$ regions. At $\bar{K}$, nearly all of the charge is located on layer 1 at $\gamma$ regions. There is virtually no charge on the remaining four layers. At $\bar{K}'$, most of the charge is localized in $\alpha$ regions divided up among layers 1,3, and 4. In layer 2, there is some charge density in $\gamma$ regions. Even if all the layers were treated as one, there would still be significant charge reshuffling laterally as one moves across the mBZ. Therefore, we expect Hartree-Fock (HF) renormalization to change the shape of this band. Moreover, there is also charge redistribution across the layers. Thus, we expect out-of-plane Coulomb interactions to contribute as well. As such, in order to accurately predict symmetry breaking in the presence of Coulomb interactions between charges, we take into consideration whether the charges are of the same layer or different layers, as detailed later.

Equipped with the real-space representation of the wave functions, it is straightforward to calculate the $C_3$ symmetry eigenvalues $\omega(\mathbf{k}_*)$ of wavefunctions:
\begin{equation}
    \hat{C}_3\ket{\Psi_{\mathbf{k}_*}} = \omega(\mathbf{k}_*) \ket{\Psi_{\mathbf{k}_*}},
\end{equation}
where $\hat{C}_3$ is a three-fold rotation operator and $\mathbf{k}_*$ is a high-symmetry momentum that brings a wavefunction to itself (up to a phase) under $C_3$ rotation. Using the $C_3$ eigenvalues, we can verify the Chern number (modulo three) using the formula \cite{Fang2012Bulk}
\begin{equation}
    \omega(\bar{K})\omega(\bar{K}')\omega(\bar{\Gamma}) = \exp \left(- \frac{2 \pi i \mathcal{C}}{3} \right).
\end{equation}
This serves as a check of the Chern number calculated by numerically integrating the local Berry curvature. As an example, for the states in Fig. \ref{fig:chargedensity1_antiparallel}, $\omega(\bar{\Gamma}) = 1$ and $\omega(\bar{K}) = \omega(\bar{K}') = \exp \left(-2\pi i/3 \right),$ confirming indeed that $\mathcal{C} = 2.$

\subsubsection{Without  Lattice Relaxation}

\begin{figure}
    \centering
    \includegraphics[width=1\linewidth]{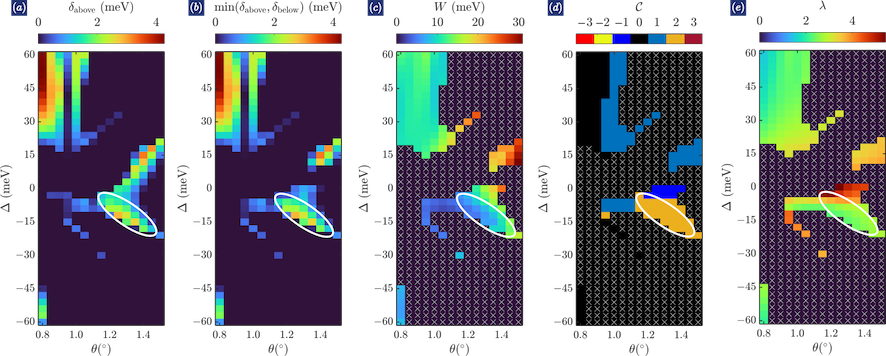}
    \caption{\textbf{Non-interacting phase diagram of the antiparallel stacking order without lattice relaxation.} (a) Value of the global gap above the first conduction band. (b) Global gap of the first conduction band defined as the minimum of the gap below and above that band. (c) Bandwidth of the first conduction band in cases where it is isolated, $i.e. \min \left( \delta_\mathrm{above},\delta_\mathrm{below}\right) > 0.$  (d)-(e) Chern number and trace condition of isolated first conduction bands.  $\times$ denotes points where the gap is zero, for which the task of defining an associated bandwidth, Chern number, or trace condition violation either makes no sense or is useless. {\fontfamily{cmtt}\selectfont Param\_two} is used for the results reported here. Like the results reported in Fig. \ref{noninteractingphase1_antiparallel}, the phase diagram here features a $\mathcal{C} = 2$ phase. However, unlike in the case with lattice relaxation, the band gaps here are much smaller.   }
    \label{noninteractingphase2_antiparallel}
\end{figure}

We now analyze the non-interacting phase diagram simulated using parameters in {\fontfamily{cmtt}\selectfont Param\_two}. This set of parameters includes no lattice relaxation across the twist interface since $\gamma_\mathrm{AA} = \gamma_\mathrm{AB} = 120$ meV. The phase diagram is shown in Fig. \ref{noninteractingphase2_antiparallel}. Qualitatively, this phase diagram is generally the same as the phase diagram in Fig. \ref{noninteractingphase1_antiparallel} in the following ways. The two feature a $\mathcal{C} = 2$ phase starting around $\theta > 1.1^\circ.$ There are also $\mathcal{C} = \pm 1$ phases, but they appear at different regions in the two phase diagrams. On the other hand, the two phase diagrams differ in important ways as well. First, the Chern-nontrivial regions in parameter space are smaller in Fig. \ref{noninteractingphase2_antiparallel} compared to Fig. \ref{noninteractingphase1_antiparallel}. Second, the band gaps are much smaller without lattice relaxation. Lattice relaxation tends to enhance the direct gap at $\bar{\Gamma}$, and therefore is better at spectrally isolating the Chern-nontrivial phases ($\mathcal{C} = 2$ phase in particular). Without lattice relaxation, the largest gaps we obtain in the $\mathcal{C} = 2$ phase are about $2$ meV. Values like this are too small to well-isolate a band into which one can project the Coulomb interactions. Therefore, on the basis on non-interacting band structures, one might dismiss these systems without lattice relaxation as poor candidate for anomalous integer and fractional Chern insulating phases. However, as we shall show, non-interacting band structures are not necessarily a good predictor of the emergence of Chern phases stabilized by interactions. Therefore, we shall revisit this phase diagram after accounting for HF corrections in a later section.  In addition to the fact that the first conduction band is poorly isolated from the other bands in the region of interest in parameter space, that band also has a relatively larger trace condition violation compared to the results with lattice relaxation. For the $\mathcal{C} = 2$ phase, we find that the smallest violation is around $\lambda \approx 2.7$ (with lattice relaxation, this violation has smallest value around $\lambda \approx 1.7$). All in all, it does not appear from a non-interacting analysis that the bands calculated without lattice relaxation would be good candidates for anomalous Chern phases.

\begin{figure}
    \centering
    \includegraphics[width=\linewidth]{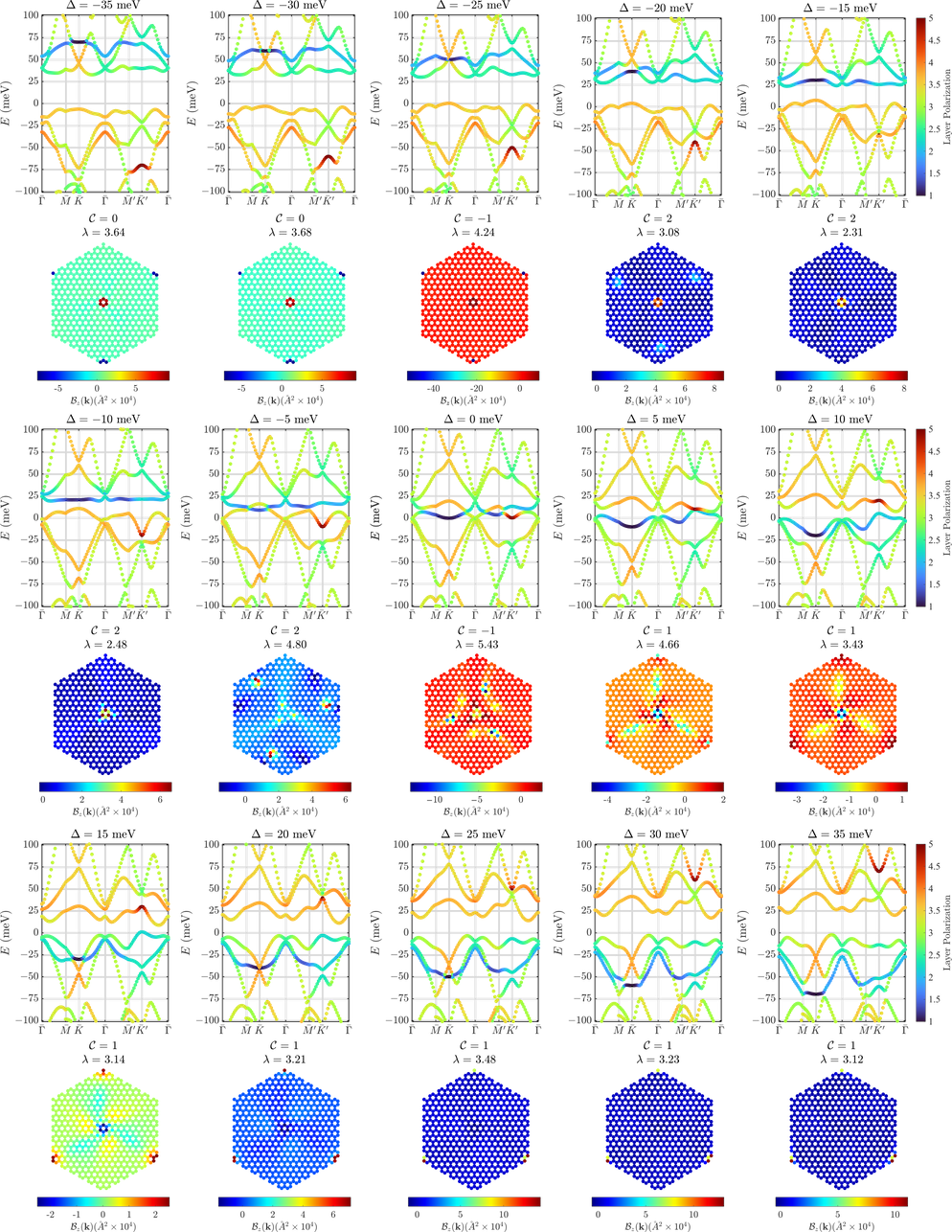}
    \caption{\textbf{Evolution of band structures as a function of $\Delta$ at $\theta = 1.25^\circ$ for the antiparallel stacking order without lattice relaxation.} For each Bloch state, we color code it by its layer polarization. Below each band structure is shown the corresponding Berry curvature distribution of its first conduction band. The Chern number and trace condition violation for that band are also displayed. {\fontfamily{cmtt}\selectfont Param\_two} is used for the results reported here.}
    \label{fig:bandstructure2_antiparallel}
\end{figure}

\begin{sidewaysfigure}
    \centering
    \includegraphics[width=1\linewidth]{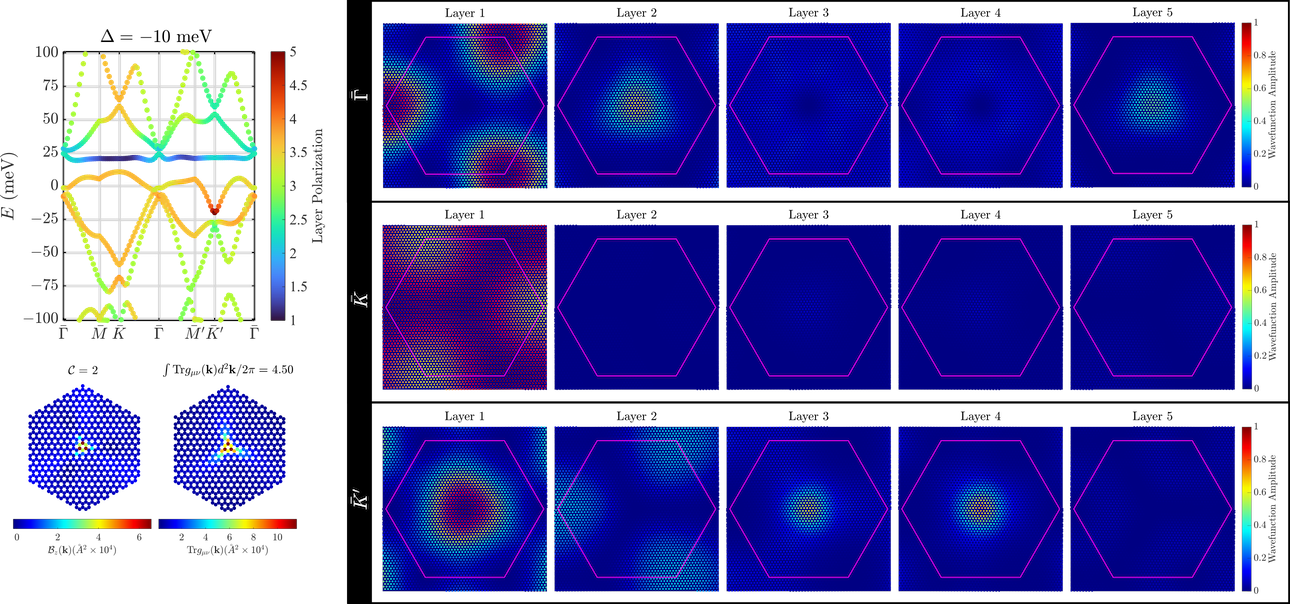}
    \caption{\textbf{Charge density of $\mathcal{C} = 2$ band in the non-interacting limit without lattice relaxation for the antiparallel configuration.} From the band structure, we observe that the first conduction band is not well isolated from its surrounding high-energy bands. We show the charge density in each layer for states at the high-symmetry points $\bar{\Gamma},$ $\bar{K},$ and $\bar{K}'$ for the first conduction band that has Chern number $\mathcal{C} = 2.$ We also compute the $C_{3z}$ symmetry eigenvalues: $\omega(\bar{\Gamma}) = 1$ and $\omega(\bar{K}) = \omega(\bar{K}') = \exp \left(-2\pi i/3\right).$ This confirms that the Chern number is indeed two. The quantum metric and Berry curvature are shown below the band structure. {\fontfamily{cmtt}\selectfont Param\_two} is used for the results reported here.}
    \label{fig:chargedensity2_antiparallel}
\end{sidewaysfigure}

To gain more insights into the small band gaps, we next inspect the band structures at a fixed $\theta=1.25^\circ.$ The results are shown in Fig. \ref{fig:bandstructure2_antiparallel}. Starting at $\Delta = 0$ meV, we already see that the bands are  not well separated from each other near charge neutrality, unlike in the situation with corrugation.  In particular, the $\bar{\Gamma}$ point is barely gapped. As we increase $\Delta$ in the negative direction, the two bands at charge neutrality spectrally disentangle. However, the first conduction band is never well separated from the higher-energy continuum, with its $\bar{\Gamma}$ point very close to the next energy band. This is in contrast to the situation with lattice relaxation where there is a region of negative $\Delta$ where this band is reasonably well isolated from the rest of the band structure. However, it is still true in the present case without lattice relaxation that this first conduction band is quite flat and is localized, on average, mostly in the trilayer substack. In some ways, this is reminiscent of the situation in pentalayer graphene aligned with hexagonal boron nitride where the narrow bottom of the conduction band is not separated from the rest of the energy spectrum. In the positive $\Delta$ region, the first band conduction is seldom flat or isolated but it does feature a Chern number of one. In this way, the positive $\Delta$ region here is qualitatively the same as the positive $\Delta$ region in the case with lattice relaxation. We have also inspected the charge density distribution across the layers. In Fig. \ref{fig:chargedensity2_antiparallel}, we show an example from $\theta = 1.25^\circ$ and $\Delta = -10$ meV. The general patterns look very much like those in Fig. \ref{fig:chargedensity1_antiparallel}. In particular, the symmetry eigenvalues at the high-symmetry points remain unchanged and they multiply to give the correct Chern number.

\subsubsection{With Reduced Fermi Velocity $\gamma_0$}

\begin{figure}[t!]
    \centering
    \includegraphics[width=1\linewidth]{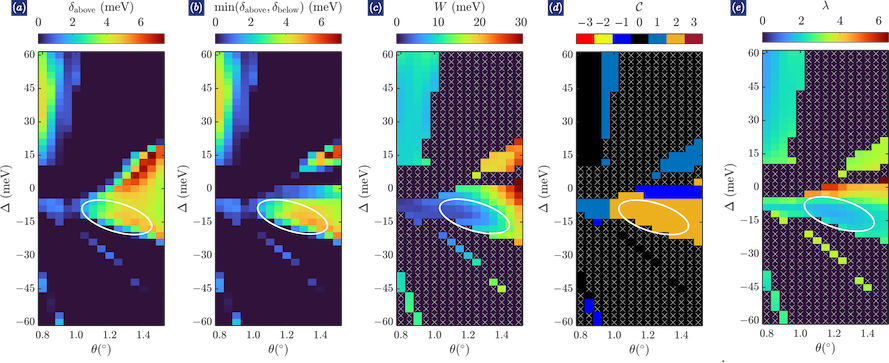}
    \caption{\textbf{Non-interacting phase diagram of the antiparallel stacking order with reduced $\gamma_0$ and with lattice relaxation.} (a) Value of the global gap above the first conduction band. (b) Global gap of the first conduction band defined as the minimum of the gap below and above that band. (c) Bandwidth of the first conduction band in cases where it is isolated, $i.e. \min \left( \delta_\mathrm{above},\delta_\mathrm{below}\right) > 0.$  (d)-(e) Chern number and trace condition of isolated first conduction bands.  $\times$ denotes points where the gap is zero, for which the task of defining an associated bandwidth, Chern number, or trace condition violation either makes no sense or is useless. {\fontfamily{cmtt}\selectfont Param\_three} is used for the results reported here. Like the results reported in Fig. \ref{noninteractingphase1_antiparallel}, the phase diagram here features a $\mathcal{C} = 2$ phase. However, unlike in the case with a larger $\gamma_0$ and with lattice relaxation, the band gaps here are smaller.  }
    \label{noninteractingphase3_antiparallel}
\end{figure}

\begin{figure}[t!]
    \centering
    \includegraphics[width=1\linewidth]{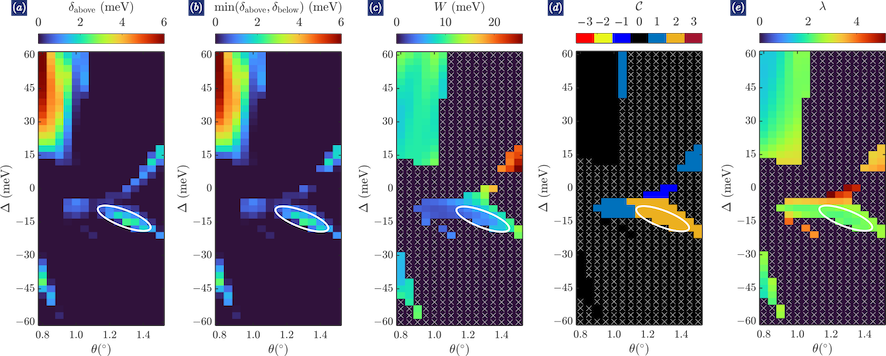}
    \caption{\textbf{Non-interacting phase diagram of the antiparallel stacking order with reduced $\gamma_0$ and without lattice relaxation.} (a) Value of the global gap above the first conduction band. (b) Global gap of the first conduction band defined as the minimum of the gap below and above that band. (c) Bandwidth of the first conduction band in cases where it is isolated, $i.e. \min \left( \delta_\mathrm{above},\delta_\mathrm{below}\right) > 0.$  (d)-(e) Chern number and trace condition of isolated first conduction bands.  $\times$ denotes points where the gap is zero, for which the task of defining an associated bandwidth, Chern number, or trace condition violation either makes no sense or is useless. {\fontfamily{cmtt}\selectfont Param\_four} is used for the results reported here. Like the results reported in Fig. \ref{noninteractingphase3_antiparallel}, the phase diagram here features a $\mathcal{C} = 2$ phase. However, unlike in the case with a larger $\gamma_0$ and with lattice relaxation, the band gaps here are smaller. Also, the region of $\mathcal{C} = 2$ is much smaller. This trend mirrors the trend seen in  Fig. \ref{noninteractingphase2_antiparallel} with a larger $\gamma_0.$ }
    \label{noninteractingphase4_antiparallel}
\end{figure}

To close our considerations of the non-interacting phase diagrams for the antiparallel configuration, we show in Figs. \ref{noninteractingphase3_antiparallel} and \ref{noninteractingphase4_antiparallel} the phase diagrams with a reduced $\gamma_0 = 2700$ meV with and without lattice relaxation respectively. Here, we use parameters from {\fontfamily{cmtt}\selectfont Param\_three} and {\fontfamily{cmtt}\selectfont Param\_four} respectively. Overall, there Chern-two phase remains robust on all cases. Compared to having lattice relaxation, the phase diagrams without lattice relaxation feature much smaller gaps and higher trace condition violation. This is true for $\gamma_0 = 3100$ meV and $\gamma_0 = 2700$ meV. The Chern-two phase of interest remains around $\theta = 1.25^\circ$ and $\Delta = -10$ meV with reduced $\gamma_0.$ This analysis inspires confidence that qualitative conclusions about the Chern-two phase are robust against changes in modeling parameters since using both parameters experimentally determined and calculated from density functional theory methods, we obtain basically the same results.

\subsection{Parallel Stacking Order}

\subsubsection{With Some Lattice Relaxation}

\begin{figure}
    \centering
    \includegraphics[width=1\linewidth]{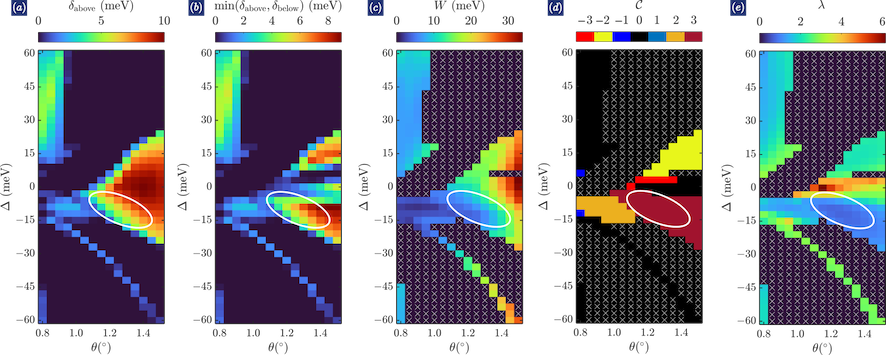}
    \caption{\textbf{Non-interacting phase diagram of the parallel stacking order with lattice relaxation.} (a) Value of the global gap above the first conduction band. (b) Global gap of the first conduction band defined as the minimum of the gap below and above that band. (c) Bandwidth of the first conduction band in cases where it is isolated, $i.e. \min \left( \delta_\mathrm{above},\delta_\mathrm{below}\right) > 0.$  (d)-(e) Chern number and trace condition of isolated first conduction bands. There is a large region with $\mathcal{C} = 3$ and small trace condition violation.  $\times$ denotes points where the gap is zero, for which the task of defining an associated bandwidth, Chern number, or trace condition violation either makes no sense or is useless. {\fontfamily{cmtt}\selectfont Param\_one} is used for the results reported here.}
    \label{noninteractingphase1_parallel}
\end{figure}

\begin{figure}
    \centering
    \includegraphics[width=\linewidth]{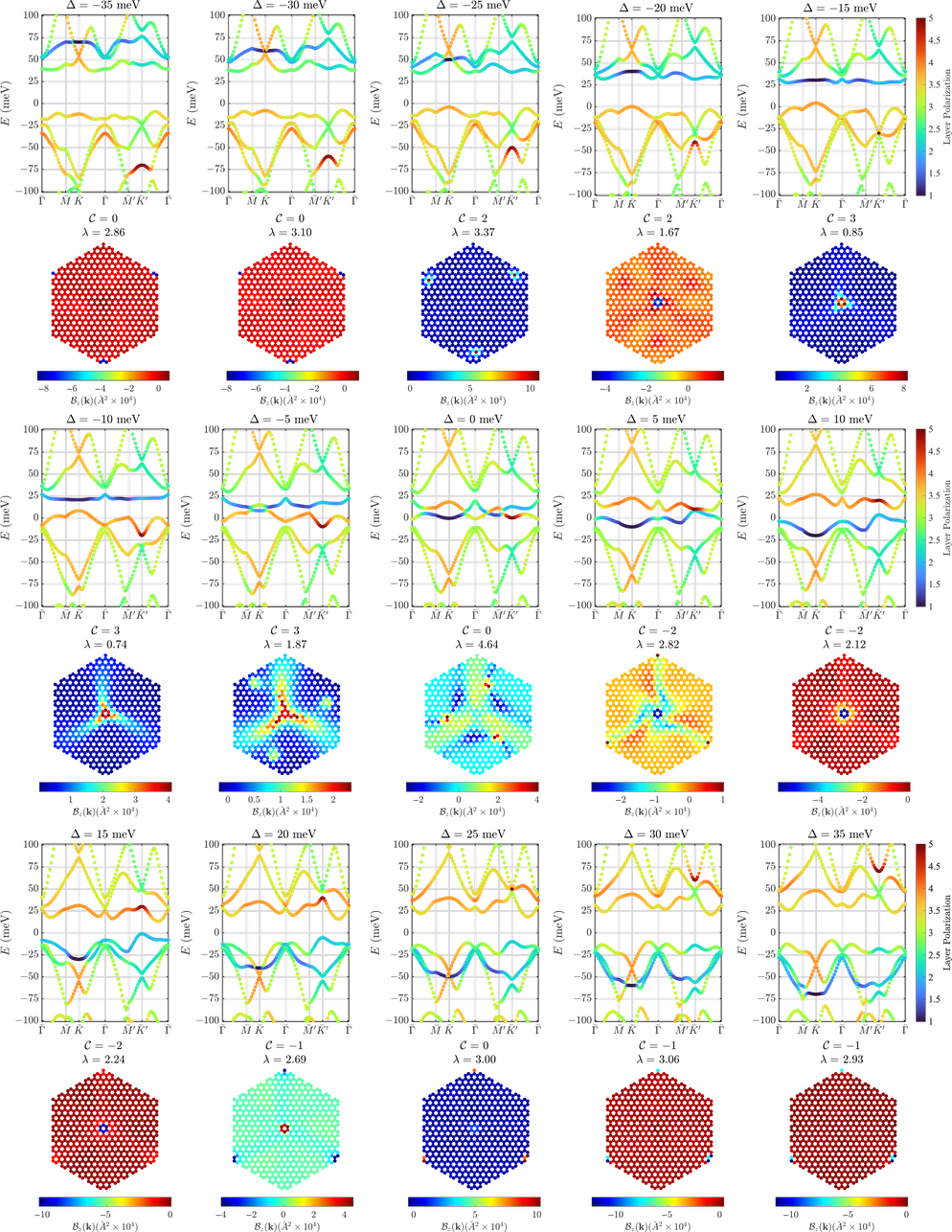}
    \caption{\textbf{Evolution of band structures as a function of $\Delta$ at $\theta = 1.25^\circ$ for the parallel stacking order.} For each Bloch state, we color code it by its layer polarization. Below each band structure is shown the corresponding Berry curvature distribution of its first conduction band. The Chern number and trace condition violation for that band are also displayed. {\fontfamily{cmtt}\selectfont Param\_one} is used for the results reported here.}
    \label{fig:bandstructure1_parallel}
\end{figure}

We now examine the phase diagram for the parallel configuration. To start, we consider the results simulated with {\fontfamily{cmtt}\selectfont Param\_one} shown in Fig. \ref{noninteractingphase1_parallel}. This phase diagram is richer than the antiparallel configuration's phase diagrams in the sense that it features isolated Chern bands with $|\mathcal{C}| = 1,2,3.$ Referring to Fig. \ref{noninteractingphase1_parallel}, we find that the $\mathcal{C} = 3$ and $\mathcal{C} = 2$ phases exist exclusively in the $\Delta < 0$ region. The latter develops prominently for $\theta > 1^\circ,$ while the latter exists around $\theta \approx 1^\circ.$ However, the $\mathcal{C} = 2$ phase features simultaneously small bandwidths and band gaps, on the order of $2$ meV or so. On the other hand, the $\mathcal{C}= 3$ phase exists with larger bandwidths and band gaps. Part of the reason is simply the fact that the $\mathcal{C} = 3$ generically exists at larger angles where the bands disperse more due to a larger mBZ (because of a smaller real-space unit cell). Since the $\mathcal{C} = 3$ phase is qualitatively different compared to the antiparallel configuration, we focus on this phase more carefully. The evolution of the band structures at $\theta = 1.25^\circ$ as $\Delta$ varies ins shown in Fig. \ref{fig:bandstructure1_parallel}. Based on these spectra alone, there does not appear to be much qualitative difference between the parallel configuration and the antiparallel configuration at the same values in parameter space. We obverse the following features that are also present in the corresponding antiparallel stacking order. As $\Delta$ is increased in the negative direction, the first conduction separates from the two-band manifold at charge neutrality and migrates upward in energy. Before it merges into the continuum, that band is spectrally isolated and has a Chern number of $\mathcal{C} = 3.$ The density of that band is, on average, localized on the trilayer substack. The Berry curvature peaks around the $\bar{\Gamma}$ point in the mBZ, but it has appreciable spread throughout the mBZ. It is worth emphasizing that the trace condition violation in this case is much smaller than in the antiparallel configuration. The best value is around $0.7,$ which is about half the trace condition violation of the antiparallel configuration. On the opposite side of $\Delta,$ the first conduction band is mostly localized on the bilayer substack and is never really well-isolated from the rest of the band structure or narrow in bandwidth.

\begin{sidewaysfigure}
    \centering
    \includegraphics[width=1\linewidth]{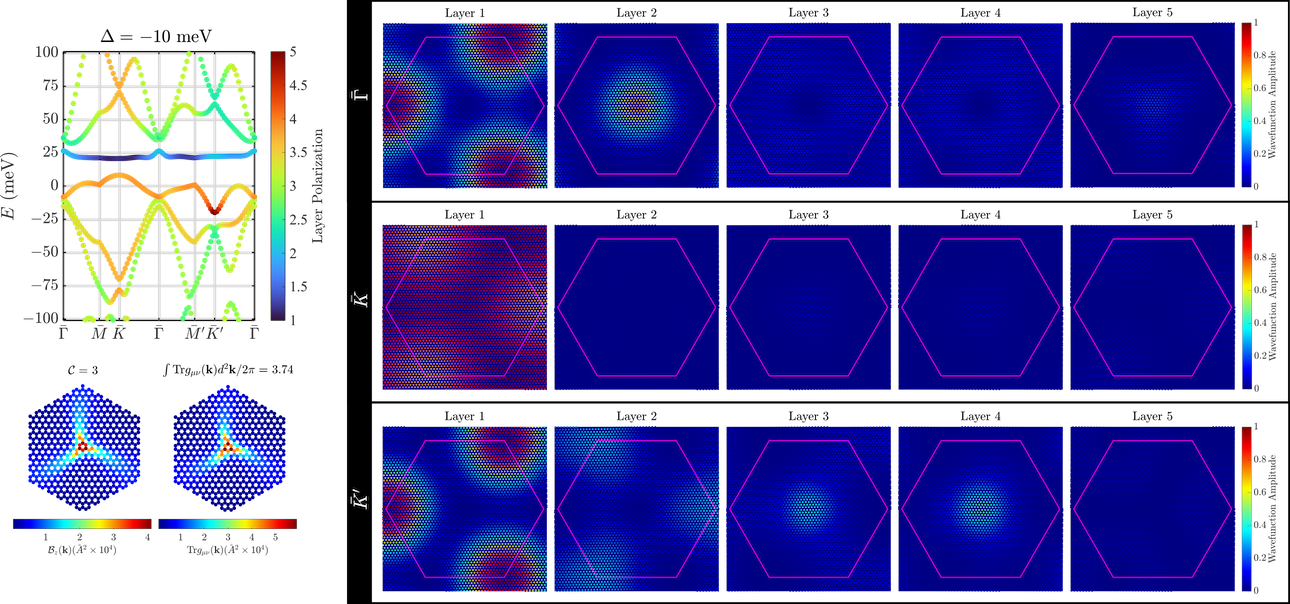}
    \caption{\textbf{Charge density of $\mathcal{C} = 3$ band in the non-interacting limit with some lattice relaxation for the parallel configuration.} We show the charge density in each layer for states at the high-symmetry points $\bar{\Gamma},$ $\bar{K},$ and $\bar{K}'$ for the first conduction band that has Chern number $\mathcal{C} = 3.$ We also compute the $C_{3z}$ symmetry eigenvalues: $\omega(\bar{\Gamma}) = 1,$ $\omega(\bar{K}) = \exp \left(-2\pi i/3\right),$ and $\omega(\bar{\bar{K}}) = \exp \left(+2\pi i/3\right).$ This confirms that the Chern number is indeed three. The quantum metric and Berry curvature are shown below the band structure. {\fontfamily{cmtt}\selectfont Param\_one} is used for the results reported here.}
    \label{fig:chargedensity1_parallel}
\end{sidewaysfigure}

Next, we inspect the real-space charge density of $\theta = 1.25^\circ$ and $\Delta = -10$ meV as an example. We map the charge density for each layer for the three $C_{3z}$-symmetric wavefunctions, shown in Fig. \ref{fig:chargedensity1_parallel}. For the $\bar{\Gamma}$ and $\bar{K}$ states, the charge density is basically the same as that in Fig. \ref{fig:chargedensity1_antiparallel}. For the $\bar{\Gamma}$ state, the charge is localized mostly at $\gamma$ regions  on layer 1 with a smaller concentration of charge on layer 2 at $\alpha$ regions. For the $\bar{K}$ state, almost all of the charge is located on layer 1, surrounding $\gamma$ regions. Both of these states transform in the same way as in the antiparallel configuration. In particular, $\omega(\bar{\Gamma}) = 1$ and $\omega(\bar{K}) = \exp \left(-2\pi i /3\right).$ On the contrary, the charge density for the $\bar{K}'$ state is completely different to that of the antiparallel case. In the present parallel case, the charge density is localized at $\gamma$ regions of layer 1, with residual charge density located at $\beta$ regions of layer 2 and $\alpha$ regions on layer 3 and layer 4. The symmetry eigenvalue of this state is $\omega(\bar{K}') = \exp \left(+2\pi i /3\right).$ This change in the charge distribution at the $\bar{K}'$ point is responsible for changing the Chern number from two to three.

\subsubsection{Without Lattice Relaxation}

\begin{figure}
    \centering
    \includegraphics[width=1\linewidth]{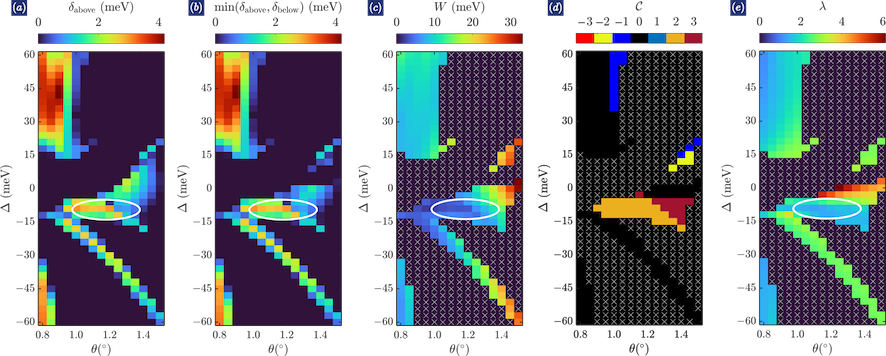}
    \caption{\textbf{Non-interacting phase diagram of the parallel stacking order without lattice relaxation.} (a) Value of the global gap above the first conduction band. (b) Global gap of the first conduction band defined as the minimum of the gap below and above that band. (c) Bandwidth of the first conduction band in cases where it is isolated, $i.e. \min \left( \delta_\mathrm{above},\delta_\mathrm{below}\right) > 0.$  (d)-(e) Chern number and trace condition of isolated first conduction bands. There is a small region with $\mathcal{C} = 3$ and relatively small trace condition violation.  $\times$ denotes points where the gap is zero, for which the task of defining an associated bandwidth, Chern number, or trace condition violation either makes no sense or is useless. {\fontfamily{cmtt}\selectfont Param\_two} is used for the results reported here.}
    \label{noninteractingphase2_parallel}
\end{figure}

\begin{figure}
    \centering
    \includegraphics[width=\linewidth]{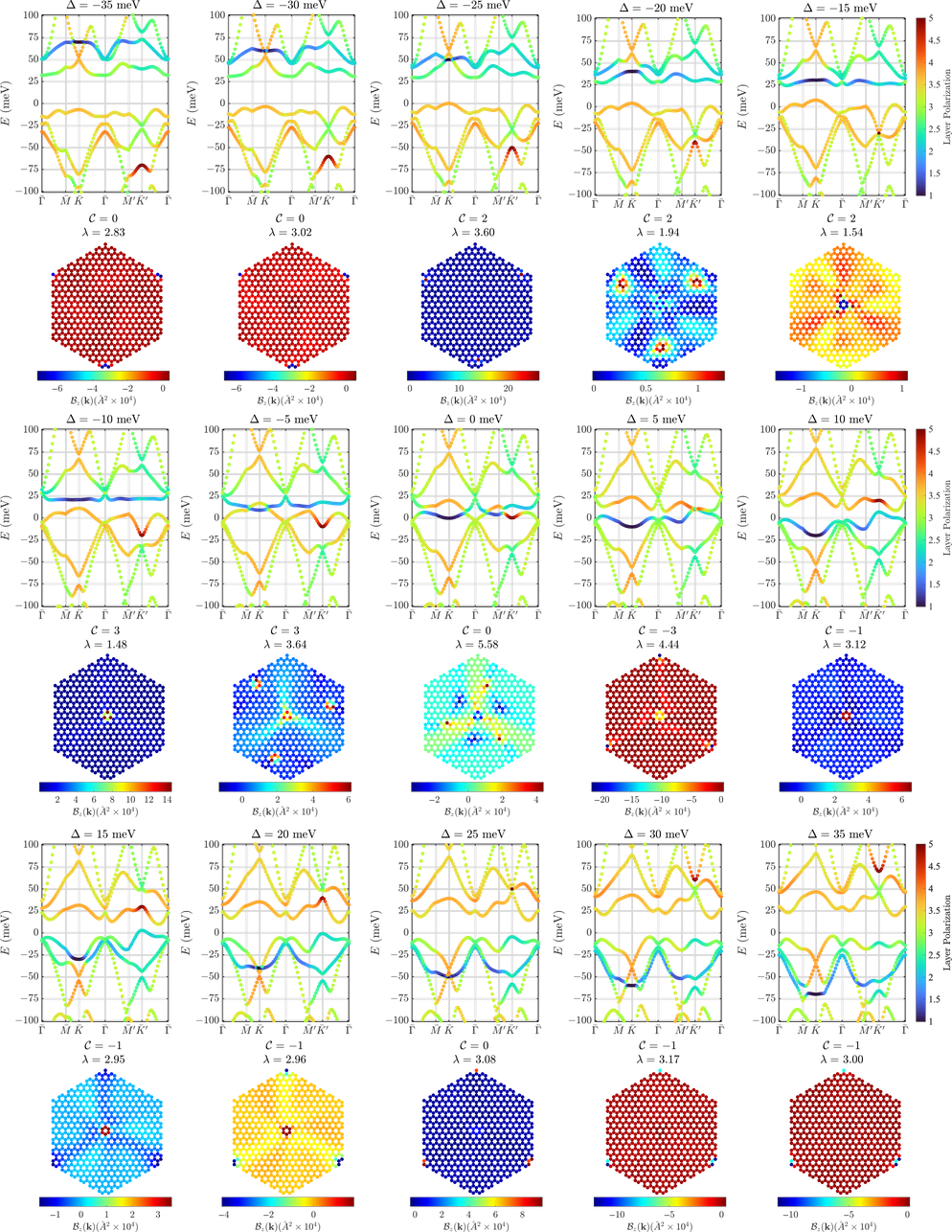}
    \caption{\textbf{Evolution of band structures as a function of $\Delta$ at $\theta = 1.30^\circ$ for the parallel stacking order.} For each Bloch state, we color code it by its layer polarization. Below each band structure is shown the corresponding Berry curvature distribution of its first conduction band. The Chern number and trace condition violation for that band are also displayed. {\fontfamily{cmtt}\selectfont Param\_two} is used for the results reported here.}
    \label{fig:bandstructure2_parallel}
\end{figure}

Without lattice relaxation, $\gamma_\mathrm{AA} = \gamma_\mathrm{AB} = 120$ meV, the phase diagram is shown in Fig. \ref{noninteractingphase2_parallel}. The first obvious difference here compared to Fig. \ref{noninteractingphase1_parallel} is the significant reduction in the $\mathcal{C} = 3$ region. While it is still there, it is now mostly succeeded by a $\mathcal{C} = 2$ phase as $\Delta$ is increased in the negative direction at a fixed $\theta.$ The band gaps in general are also smaller in Fig. \ref{noninteractingphase2_parallel} compared to Fig. \ref{noninteractingphase1_parallel}. Also, the trace condition violation is larger without lattice relaxation when compared to values obtained in the case with lattice relaxation. It is worth mentioning that the $\mathcal{C} = 2$ bands also exist with lattice relaxation too, as can be seen in Fig. \ref{fig:bandstructure1_parallel}, but they typically appear without a fully developed global gap above. So they do not show up in the phase diagram, which only displays a Chern phase if it has a clear gap above and below. All of these trends are also observed in the antiparallel configuration when comparing results with and without lattice relaxation.  Inspecting the band structures from Fig. \ref{fig:bandstructure2_parallel}, we observe the main difference here compared to Fig. \ref{fig:bandstructure1_parallel} is the fact that the first conduction band is barely isolated from the other bands for the values of $\Delta$ shown. Take $\Delta \approx -10$ meV as an example, we observe that the $\bar{\Gamma}$ point is almost touching the next higher-energy band. Because of this, the Berry curvature is almost singular at the $\bar{\Gamma}$ point. Again, this is also observed in the antiparallel configuration. So it appears that the general effects of reducing lattice relaxation are to diminish the regions of nontrivial Chern phases south of $\Delta = 0$ meV, decrease the bandwidths of the remaining topologically nontrivial regions, and exaggerates the corresponding trace condition violation. Nonetheless, it is worth emphasizing again that although the $\mathcal{C} = 3$ is significantly diminished in phase space in the absence of corrugation effects, it is still present. Once interactions are switched on at the mean-field level, it may be stabilized.

\subsubsection{With Reduced Fermi Velocity $\gamma_0$}

\begin{figure}
    \centering
    \includegraphics[width=1\linewidth]{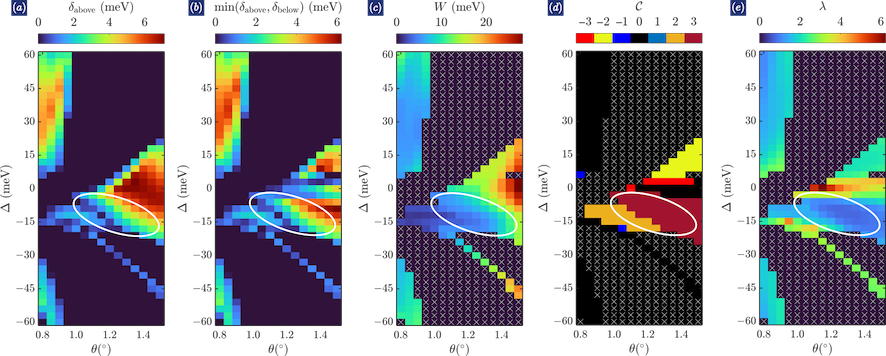}
    \caption{\textbf{Non-interacting phase diagram of the parallel stacking order with reduced $\gamma_0$ and with lattice relaxation.} (a) Value of the global gap above the first conduction band. (b) Global gap of the first conduction band defined as the minimum of the gap below and above that band. (c) Bandwidth of the first conduction band in cases where it is isolated, $i.e. \min \left( \delta_\mathrm{above},\delta_\mathrm{below}\right) > 0.$  (d)-(e) Chern number and trace condition of isolated first conduction bands. There is a large region with $\mathcal{C} = 3$ and relatively small trace condition violation. There is also a region with $\mathcal{C} = 2$ and small trace condition violation. $\times$ denotes points where the gap is zero, for which the task of defining an associated bandwidth, Chern number, or trace condition violation either makes no sense or is useless. {\fontfamily{cmtt}\selectfont Param\_three} is used for the results reported here.}
    \label{noninteractingphase3_parallel}
\end{figure}

\begin{figure}
    \centering
    \includegraphics[width=1\linewidth]{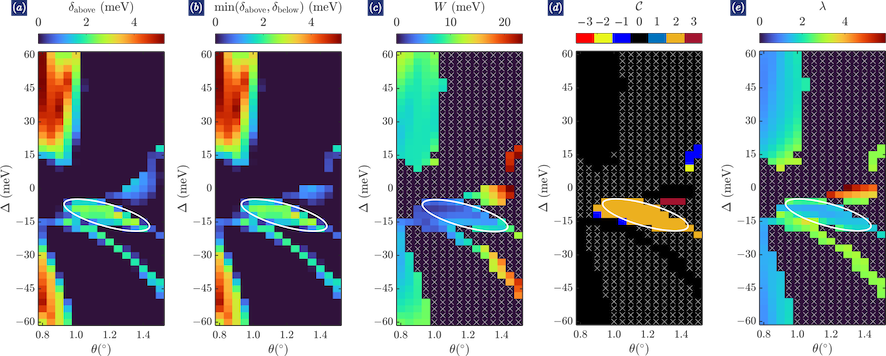}
    \caption{\textbf{Non-interacting phase diagram of the parallel stacking order with reduced $\gamma_0$ and without lattice relaxation.} (a) Value of the global gap above the first conduction band. (b) Global gap of the first conduction band defined as the minimum of the gap below and above that band. (c) Bandwidth of the first conduction band in cases where it is isolated, $i.e. \min \left( \delta_\mathrm{above},\delta_\mathrm{below}\right) > 0.$  (d)-(e) Chern number and trace condition of isolated first conduction bands. The $\mathcal{C} = 3$ is almost completely diminished. However, there is  a region with $\mathcal{C} = 2$ and small trace condition violation. $\times$ denotes points where the gap is zero, for which the task of defining an associated bandwidth, Chern number, or trace condition violation either makes no sense or is useless. {\fontfamily{cmtt}\selectfont Param\_four} is used for the results reported here.}
    \label{noninteractingphase4_parallel}
\end{figure}

Next, we present the phase diagrams for the cases with a reduced Fermi velocity $\gamma_0.$ The results with relaxation are reported in Fig. \ref{noninteractingphase3_parallel}, while the results without relaxation are reported in Fig. \ref{noninteractingphase4_parallel}. With relaxation, the phase diagram is similar to the results in Fig. \ref{noninteractingphase1_parallel}. There is a large region with $\mathcal{C} = 3$ and  quite small $\lambda.$ The smallest trace condition violation is around $0.75.$ However, the reduced $\gamma_0$ leads to relatively smaller global band gaps in this nontrivial phase. This is observed in the antiparallel configuration too where the reduction in the Fermi velocity generally leads to a decrease in the global band gaps. Moving onto the results without relaxation, we see here that the $\mathcal{C} = 3$ phase almost completely disappears, leaving only a robust $\mathcal{C} = 2$ phase that has a decently small trace condition violation $\lambda \approx 1.4.$ Recalling Fig. \ref{noninteractingphase2_parallel}, we see that the effect of removing lattice relaxation is to decrease the region in phase space where the first conduction band is an isolated, narrow, and $\mathcal{C} = 3$ band. That reduction is more dramatic here with $\gamma_0 = 2700$ meV than in the case where $\gamma_0 = 3100$ meV. In fact, the reduction in the $\gamma_0 = 2700$ meV case is so significant that we cannot conclude with confidence that the $\mathcal{C} = 3$ exists here at all. The phase area is so small that it could just be a numerical artifact due to fluctuation of the Chern number near a band inversion.

\subsection{Summary of Observations}

In the preceding sections, we have shown the phase diagrams for the antiparallel and parallel configurations using different sets of parameters that include or not include the effects of corrugation as well as having a larger or smaller $\gamma_0$. While there are some qualitatively important differences between the cases, there are clearly also some general conclusions that are insensitive to microscopic details. In this section, we summarize those robust conclusions. To aid in that effort, we recognize that the ample amount of data presented calls for the introduction of a figure of merit that can quantitatively capture the region in phase space most desirable. For the antiparallel configuration, let us focus only on the $\mathcal{C} = 2$ phase, while for the parallel configuration, let us focus only on the $\mathcal{C} = 2$ and $\mathcal{C} = 3$ phases. Then, within those regions, we define a dimensionless figure of merit as 
\begin{equation}
\label{eq: figure of merit}
    \mathfrak{f.o.m.} = \frac{\min \left( \delta_\mathrm{above}, \delta_\mathrm{below}\right)}{W\lambda}.
\end{equation}
The larger this figure of merit, naively, the likelier the band is at stabilizing a Chern phase in the presence of interactions.

\begin{table}[]
    \centering
    \begin{tabular}{||c||c|c|c|c|c|c| c|c||}
        \hline\hline
         Configuration & $\mathcal{C}$  & \# of points & $\theta$ (degrees)  &  $\Delta$ (meV) &$\delta_\mathrm{above}$ (meV) & $\delta_\mathrm{below}$ (meV) & $W$ (meV) & $\lambda$\\
         \hline
         \hline
         Antiparallel with relaxation & $2$  & $58$ & $1.15^\circ$  &  $-9$   &$5.63$   & $7.78$   & $2.75$   & $1.71$\\
         \hline
         Antiparallel without relaxation & $2$  & $27$ & $1.25^\circ$  &  $-12$   &$2.87$   & $11.51$   & $4.68$   & $2.32$\\
         \hline
         Antiparallel with reduced $\gamma_0$ and with relaxation & $2$  & $56$ & $1.20^\circ$  &  $-12$   &$4.21$   & $14.07$   & $2.99$   & $1.50$\\
         \hline
         Antiparallel with reduced $\gamma_0$ and without relaxation & $2$  & $26$ & $1.30^\circ$  &  $-15$   &$2.36$   & $19.16$   & $4.45$   & $2.16$\\
         \hline
         Parallel with relaxation & $2$  & $21$ & $0.90^\circ$  &  $-9$   &$2.33$   & $8.58$   & $1.47$   & $1.27$\\
         \hline
         Parallel without relaxation & $2$  & $26$ & $1.00^\circ$  &  $-9$   &$3.05$   & $5.17$   & $2.23$   & $1.50$\\
         \hline
         Parallel with reduced $\gamma_0$ and with relaxation & $2$  & $21$ & $0.95^\circ$  &  $-12$   &$1.81$   & $16.20$   & $1.63$   & $1.52$\\
         \hline
         Parallel with reduced $\gamma_0$ and without relaxation & $2$  & $34$ & $1.00^\circ$  &  $-9$   &$2.36$   & $5.95$   & $1.38$   & $1.39$\\
         \hline         
         Parallel with relaxation & $3$  & $51$ & $1.25^\circ$  &  $-12$   & $4.89$   & $17.24$   & $4.39$   & $0.74$\\
         \hline
         Parallel without relaxation & $3$  & $12$ & $1.30^\circ$  &  $-9$   &$0.88$   & $7.39$   & $8.45$   & $1.58$\\
         \hline
         Parallel with reduced $\gamma_0$ and with relaxation & $3$  & $48$ & $1.30^\circ$  &  $-12$   &$3.44$   & $18.63$   & $6.00$   & $0.76$\\
         \hline
         Parallel with reduced $\gamma_0$ and without relaxation & $3$  & $3$ & $1.35^\circ$  &  $-6$   &$0.34$   & $2.78$   & $12.90$   & $2.69$\\
         \hline          
         \hline
    \end{tabular}
    \caption{\textbf{Summary of observations from non-interacting band structures.} The first column lists the configuration and parameters used. The second column denotes the non-trivial Chern phase of interest. The third column provides the number of data points belonging to such phase and serves as a measure for the area of that phase within the phase diagram. For the data point within that area which minimizes the figure of merit in Eq. \ref{eq: figure of merit}, we display the angle and displacement field where that occurs in columns four and five, the gaps above and below the first conduction band in columns six and seven, the bandwidth of that band in column eight, and the corresponding trace condition in column nine.  }
    \label{tab:summary noninteracting}
\end{table}

Referring to Table \ref{tab:summary noninteracting} for guidance, we see the following trends for the antiparallel configuration:
\begin{enumerate}
    \item The $\mathcal{C} = 2$ phase that exists in the negative $\Delta$ region around $\theta \approx 1.2^\circ$ is robust as it can be seen with and without relaxation and with a larger or smaller $\gamma_0$ value. 
    \item Disregarding lattice relaxation increases the bandwidths of the first conduction band, decrease the gap above that band while increases the gap below it. Not including lattice relaxation also decreases the phase space area of this $\mathcal{C} = 2$ phase. 
    \item The absence of lattice relaxation increases the trace condition violation. 
    \item Reducing $\gamma_0$ decreases $\delta_\mathrm{above}$ and increases $\delta_\mathrm{below}.$ It does not seem to affect the $W$ much. $\lambda$ is slightly decreased with reduced $\gamma_0.$
\end{enumerate}
For the parallel configuration, the $\mathcal{C} = 2$ phase has the following trends, many of which are opposite to the trends of the $\mathcal{C} = 2$ phase in the antiparallel configuration:
\begin{enumerate}
    \item The $\mathcal{C} = 2$ phase that exists in the negative $\Delta$ region around $\theta \approx 1.0^\circ$ is robust as it can be seen with and without relaxation and with a larger or smaller $\gamma_0$ value. 
    \item Removing lattice relaxation slightly increases the phase space area of this phase, increases $\delta_\mathrm{above},$ decreases $\delta_\mathrm{below},$ and increases $W.$
    \item The absence of lattice relaxation increases the trace condition violation. 
\end{enumerate}
For the parallel configuration, the $\mathcal{C} = 3$ phase has the following trends
\begin{enumerate}
    \item The $\mathcal{C} = 3$ phase is robust at $\gamma_0 = 3100$ meV with and without lattice relaxation, but is fragile at $\gamma_0 = 2700$ meV since it appears prominently with relaxation and essentially disappears without relaxation.  
    \item Turing off lattice relaxation decreases both $\delta_\mathrm{above}$ and $\delta_\mathrm{below}$ and increases $W.$
    \item The trace condition violation more than doubles when lattice relaxation is removed. 
\end{enumerate}

All in all, the $\mathcal{C} = 2$ is quite stable in the non-interacting limit. The $\mathcal{C} = 3$ is more sensitive to changing $\gamma_0$ and including or not including lattice relaxation. We end this section with a cautionary note that the preceding analysis using non-interacting band structures only provides some guidance to the likelihood of stabilizing non-trivial Chern phases. However, this analysis is not definitive, and because of that, in the succeeding sections, we carefully examine the effects of Coulomb repulsion on stabilizing non-trivial Chern phases using the self-consistent mean-field.

\section{Layer-Dependent Coulomb Potential Energy}
\label{sec: Layer-Dependent Coulomb Potential Energy}

\subsection{Green's Function Derivation}
\label{sec: Green's Function Derivation}

\begin{figure}
    \centering
    \includegraphics[width=0.7\linewidth]{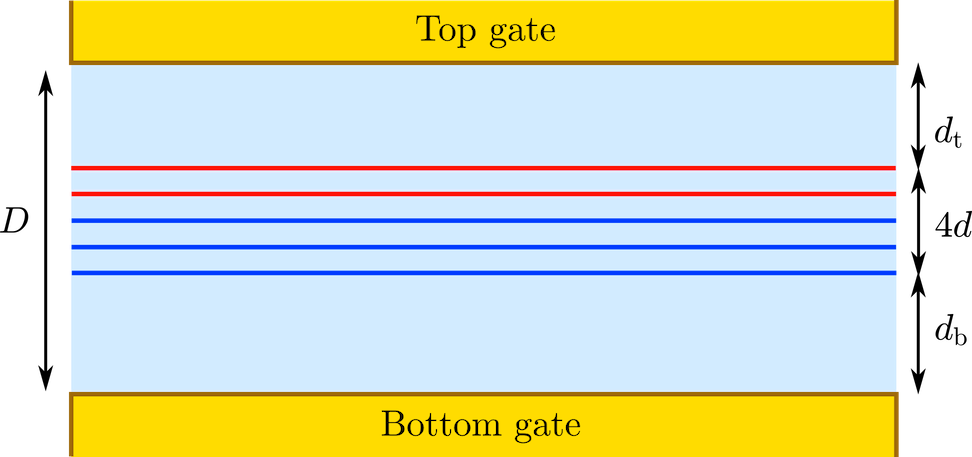}
    \caption{\textbf{Schematic representation of a dual-gate experiment.} The blue (red) lines represent layers of the trilayer stack (bilayer stack). }
    \label{fig:gate_schematic}
\end{figure}

\begin{figure}[t]
    \centering
    \includegraphics[width=0.8\linewidth]{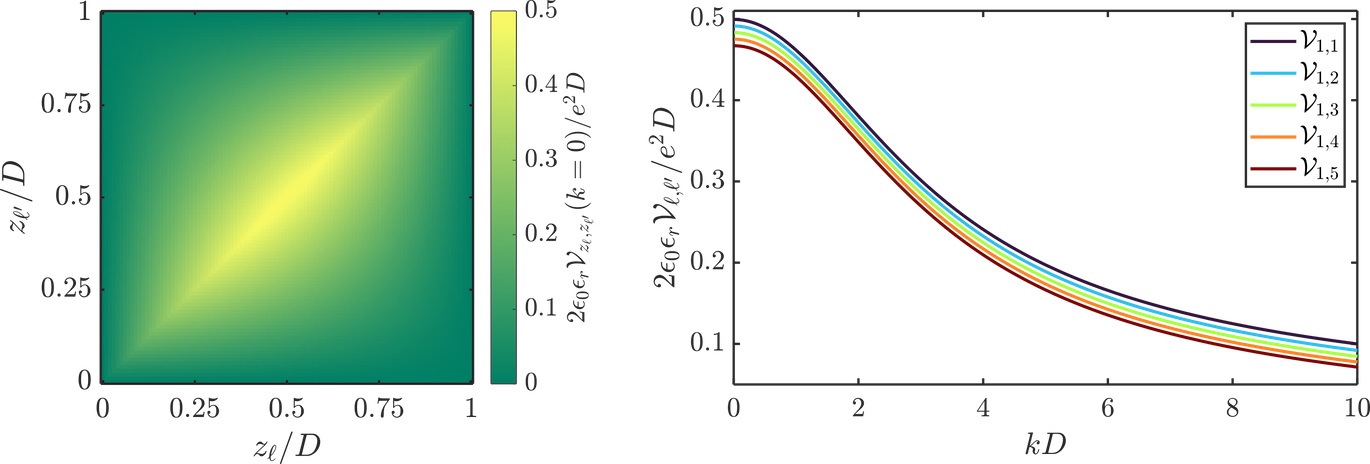}
    \caption{\textbf{Layer-dependent Coulomb potential energy.} (Left) Coulomb potential energy at $k=0$ for different locations of the two layers, where $0<z_\ell<D$ and $0<z_{\ell'}<D$. (Right) Momentum dependence of the Coulomb potential energy. Here, we use $d/D = 3.35/400.$  When $\ell = \ell',$ the Coulomb potential is essentially $e^2\tanh(kD/2)/2\epsilon_0 \epsilon_rk$ since the gates are much farther away from the active material than the thickness of said material. However, for $\ell \neq \ell',$ the deviation from the layer-independent potential is not negligible. In the present case, that deviation is about $7\%$. Hence, it is important to consider layer dependence to even capture qualitative physics.  }
    \label{fig:coulomb_dependent}
\end{figure}
A distinguishing feature of our work is the employment of a layer-dependent electrostatic potential. In this section, we derive that Coulomb potential energy between two charges trapped inside a dual-gate setup. To that end, we solve a related problem of a $\delta$-function charge in between two infinite metallic plates located at the $z = 0$ and $z = D$ planes. The electrostatic potential $\varphi(\mathbf{r},z,z')$ can be integrated from Poisson's equation 
\begin{equation}
    \nabla^2 \varphi(\mathbf{r},z,z') = \frac{e}{\epsilon_0\epsilon_r} \delta^{(2)}(\mathbf{r}-\mathbf{r}')\delta(z-z'),
\end{equation}
where $\epsilon_0\epsilon_r$ is the relative permittivity, $\mathbf{r}=(x,y)$ is the two-dimensional vector, and $(\mathbf{r}',z')$ is the location of the charge ($0<z'<D$). For simplicity, let us absorb $e/\epsilon_0\epsilon_r$ into $\varphi$ for now and resurrect it only when necessary. The metallic boundary conditions are $ \varphi(\mathbf{r},z=0,z') = \varphi(\mathbf{r},z = D,z') = 0.$ We note that the equation $\nabla^2 \varphi(\mathbf{r},z,z') = \delta^{(2)}(\mathbf{r}-\mathbf{r}')\delta(z-z')$ is just the familiar Green's function equation for the Laplace operator. We assume that solutions take the form 
\begin{equation}
    \varphi(\mathbf{r},z,z') = \sum_{n=1}^\infty \varphi_n(\mathbf{r},z') \sin \left( \frac{\pi n}{D} z \right).
\end{equation}
This ansatz clearly satisfies the boundary conditions. It is simply an expansion using a Fourier sine series with coefficients $ \varphi_n(\mathbf{r},z') $ to be determined. Substituting this ansatz into Poisson's equation, we obtain 
\begin{equation}
    \frac{\partial^2\varphi_n(\mathbf{r},z')}{\partial x^2}+\frac{\partial^2\varphi_n(\mathbf{r},z')}{\partial y^2}- \frac{\pi^2n^2}{D^2}\varphi_n(\mathbf{r},z') = \frac{2}{D}\sin \left( \frac{\pi n}{D} z' \right)\delta^{(2)}(\mathbf{r}-\mathbf{r}').
\end{equation}
We now expand $\varphi_n(\mathbf{r},z')$ as a Fourier transform:
$\varphi_n(\mathbf{r},z') = \int \frac{d^2 \mathbf{k}}{(2\pi)^2} e^{i \mathbf{k}\cdot \mathbf{r}} \varphi_n(\mathbf{k},z')$ and $
    \varphi_n(\mathbf{k},z') = \int d^2 \mathbf{r} e^{-i \mathbf{k}\cdot \mathbf{r}} \varphi_n(\mathbf{r},z').$ Substituting this ansatz into Poisson's equation, we obtain an \textit{algebraic} equation
\begin{equation}
    \left[-k^2- \frac{\pi^2n^2}{D^2} \right]\varphi_n(\mathbf{k},z') = \frac{2}{D}\sin \left( \frac{\pi n}{D} z' \right) e^{-i\mathbf{k} \cdot \mathbf{r}'},
\end{equation}
where $k = |\mathbf{k}|.$ Thus, the coefficients are 
\begin{equation}
    \varphi_n(\mathbf{k},z') = -\left[k^2+ \frac{\pi^2n^2}{D^2} \right]^{-1}\frac{2}{D}\sin \left( \frac{\pi n}{D} z' \right) e^{-i\mathbf{k} \cdot \mathbf{r}'}.
\end{equation}
The full Green's function is therefore
\begin{equation}
\begin{split}
    \varphi(\mathbf{r},z,z') &=  \int \frac{d^2 \mathbf{k}}{(2\pi)^2} e^{i \mathbf{k}\cdot (\mathbf{r}-\mathbf{r}')} \varphi(\mathbf{k},z,z'),\\
    \varphi(\mathbf{k},z,z') &=- \frac{1}{D} \sum_{n=-\infty}^\infty \left[k^2+ \frac{\pi^2n^2}{D^2} \right]^{-1}\sin \left( \frac{\pi n}{D} z' \right)  \sin \left( \frac{\pi n}{D} z \right).    
\end{split}
\end{equation}
To do the sum, we can use Poisson's summation formula: $\sum_{n \in \mathbb{Z}}f(n) = \sum_{n \in \mathbb{Z}}\hat{f}(n), $ where $\hat{f}(\omega)$ (the hat here denotes Fourier transform and not operator) is the Fourier transform of $f(t)$ \footnote{Here is a quick proof of the formula, which can be found in any introductory text on Fourier analysis. Let $f(t)$ be a function whose Fourier transform $\hat{f}(\omega) = \int_\mathbb{R}e^{-2\pi i \omega t} f(t) dt$ exists. Define a periodic extension of $f(t)$ with period $1:$ $P(t) = \sum_{n\in \mathbb{Z}} f(t+n).$ Because this function is periodic, it has a Fourier series whose coefficients are given by $a_n = \int_0^1 P(t) e^{-2\pi i n t} dt = \int_0^1\sum_{m\in \mathbb{Z}} f(t+m) e^{-2\pi i n t} dt = \int_0^1\sum_{m\in \mathbb{Z}} f(t+m) e^{-2\pi i n (t+m)} dt = \int_\mathbb{R} f(t) e^{-2\pi i n t} dt = \hat{f}(n).$ Consequently, we have $P(t) = \sum_{n\in \mathbb{Z}} f(t+n) = \sum_{n \in \mathbb{Z}} \hat{f}(n) e^{2\pi i nt}.$ Evaluating this at $t = 0$ gives the desired result.
}. In our case, we have 
\begin{equation}
\begin{split}
    f(t) &= \left[k^2+ \frac{\pi^2t^2}{D^2} \right]^{-1}\sin \left( \frac{\pi t}{D} z' \right)  \sin \left( \frac{\pi t}{D} z \right) = \frac{1}{2}\left[k^2+ \frac{\pi^2t^2}{D^2} \right]^{-1} \left[\cos \left( \frac{\pi (z-z')}{D}t \right)  -  \cos \left( \frac{\pi (z+z')}{D}t \right)\right]    \\
    &= \frac{1}{4}\left[k^2+ \frac{\pi^2t^2}{D^2} \right]^{-1} \left[ e^{\pi i z_-t/D}  + e^{-\pi i z_-t/D} -  e^{\pi i z_+t/D} - e^{-\pi i z_+t/D}\right],
\end{split}
\end{equation}
where $z_\pm = z \pm z'.$ The Fourier transform of this function is
\begin{equation}
    \hat{f}(\omega) = \frac{D}{4k} \left( \exp \left[-2kD \left| \omega -  \frac{z_-}{2 D} \right|  \right] + \exp \left[-2kD \left| \omega +  \frac{z_-}{2 D} \right|  \right] - \exp \left[-2kD \left| \omega -  \frac{z_+}{2 D} \right|  \right] - \exp \left[-2kD \left| \omega +  \frac{z_+}{2 D} \right|  \right]  \right).
\end{equation}
This is calculated as follows. We consider a function of the following form, with various constants  to be determined later, 
\begin{equation}
\begin{split}
    f(t) &= \frac{e^{i\beta t}}{k^2 + \alpha^2 t^2} = b\int_\mathbb{R} e^{2 \pi i \omega t} e^{-a\left|\omega-\omega'\right|} d \omega = b \left[\int_{\omega'}^\infty e^{2 \pi i \omega t} e^{-a\left(\omega-\omega'\right)} d \omega + \int^{\omega'}_{-\infty} e^{2 \pi i \omega t} e^{+a\left(\omega-\omega'\right)} d \omega\right]    \\
    &=  b e^{2\pi i \omega' t}\left[\int_{0}^\infty e^{2 \pi i \omega t} e^{-a\omega} d \omega + \int^{0}_{-\infty} e^{2 \pi i \omega t} e^{+a\omega} d \omega\right]    =  b e^{2\pi i \omega' t}\left[\frac{2a}{a^2+(2\pi t)^2}\right].    
\end{split}
\end{equation}
From this, we find $\omega' = \beta/2\pi, a = 2 \pi  k/\alpha,$ and $b= \pi /\alpha  k.$ Consequently, the Fourier transform of $f(t)$ is 
\begin{equation}
    \hat{f}(\omega) = \frac{\pi }{\alpha  k} \exp \left[-\frac{2 \pi  k}{\alpha } \left| \omega -  \frac{\beta}{2 \pi} \right|  \right].
\end{equation}
Now, matching $\alpha$ and $\beta$ with $D$ and $z_\pm,$ we obtain the desired result. It is clear that $\hat{f}(\omega)$ is even in both $z_\pm.$ Therefore, we can write $z_\pm$ as $|z_\pm|$ (actually, $z_+ >0$ so the absolute value on $z_+$ is not necessary). Furthermore, we also note that $|z_\pm|/2D < 1.$ This allows us to compute the sums exactly. For example,
\begin{equation}
    \begin{split}
        \sum_{n=-\infty}^\infty\exp \left[-2kD \left| n -  \frac{|z_-|}{2 D} \right|  \right] &= \sum_{n=1}^\infty\exp \left[-2kD \left( n -  \frac{|z_-|}{2 D} \right)  \right] + \sum_{n=-\infty}^0\exp \left[2kD \left( n -  \frac{|z_-|}{2 D} \right)  \right] \\
        &= \csch(k D) \cosh (k (D-|z_-|)) .
    \end{split}
\end{equation}
Putting the pieces together, we find
\begin{equation}
    \sum_{n=-\infty}^\infty \hat{f}(n) = \frac{D\csch(k D)}{2k} \left[ \cosh (k \left[D-|z_-|\right]) -  \cosh (k \left[D-|z_+|\right]) \right].
\end{equation}
Rewriting the full Green's function in terms of this resumed result, we find a very simple formula, with electric charge $e$ and permittivity $\epsilon_0\epsilon_r$ restored:
\begin{equation}
\begin{split}
    \varphi(\mathbf{k},z,z') &= - \frac{e\csch(k D)}{2\epsilon_0\epsilon_r k} \left[ \cosh (k \left[D-|z_-|\right]) -  \cosh (k \left[D-|z_+|\right]) \right].    
\end{split}
\end{equation}
Clearly, $\varphi(\mathbf{k}, z=0,z') = \varphi(\mathbf{k}, z=D,z') = 0$ as desired. The Coulomb potential energy is just $\mathcal{V}(\mathbf{k},z,z') = -e \varphi(\mathbf{k},z,z')$ \cite{Huang2023Spin,Kol2023Electrostatic}:
\begin{equation}
\label{eq: Coulomb potential energy}
\mathcal{V}(\mathbf{k},z,z') = \frac{e^2\csch(k D)}{2\epsilon_0\epsilon_r k} \left[ \cosh (k \left[D-|z-z'|\right]) -  \cosh (k \left[D-|z+z'|\right]) \right].    
\end{equation}
This is the Coulomb potential energy of two charges, one located at $z$ and one located at $z'$, inside two metallic plates a distance $D$ apart. As a consistency check, the symmetric-gate setup where $z=z'=D/2$ has the potential energy $\mathcal{V}(\mathbf{k},z=z'=D/2) =  \frac{e^2}{2\epsilon_0\epsilon_r k} \tanh(k D/2) .$ As a final check, we confirm that the Green's function we obtained actually solves Poisson's equation with a $\delta$-function source:
$\nabla^2\varphi(\mathbf{r}-\mathbf{r}',z,z') =  0$ if $z \neq z'$ since $\partial_z^2\varphi(\mathbf{k},z,z') = +k^2\varphi(\mathbf{k},z,z').$ Exactly at $z = z',$ we have a jump discontinuity when $\mathbf{r} = \mathbf{r}':$ $\lim_{z \rightarrow (z')^+} \partial_z \varphi(\mathbf{r}-\mathbf{r}',z,z') - \lim_{z \rightarrow (z')^-} \partial_z\varphi(\mathbf{r}-\mathbf{r}',z,z') = \frac{e}{\epsilon_0\epsilon_r} \delta^{(2)}(\mathbf{r}-\mathbf{r}').$ We end this section by considering the $k \rightarrow 0$ limit. We do this two different ways and show that they agree. In the first, we simply take the limit of $\mathcal{V}(\mathbf{k},z,z')$
\begin{equation}
\label{eq: k=0 limit 1}
    \lim_{k\rightarrow 0} \mathcal{V}(\mathbf{k},z,z') = \frac{e^2}{2\epsilon_0\epsilon_r}\frac{  2 D| z+z'| -2 D | z-z'|+| z-z'| ^2 -| z+z'|^2 }{2 D} = \frac{e^2}{\epsilon_0\epsilon_r} \left[ \min \left(z,z'\right) - \frac{zz'}{D} \right].
\end{equation}
Apparently, this must equal to the following for $0<z'< D$ and $0 \leq z \leq D:$
\begin{equation}
\label{eq: k=0 limit 2}
     \frac{2D}{\pi^2} \sum_{n=1}^\infty \frac{1}{n^2}\sin \left( \frac{\pi n}{D} z' \right)  \sin \left( \frac{\pi n}{D} z \right) =  \frac{D}{2\pi^2} \left[\mathrm{Li}_2\left(e^{\frac{i \pi  }{D}(z-z')}\right)-\mathrm{Li}_2\left(e^{\frac{i \pi  }{D}(z+z')}\right)+\mathrm{Li}_2\left(e^{-\frac{i \pi  }{D}(z-z')}\right)-\mathrm{Li}_2\left(e^{-\frac{i \pi  }{D}(z+z')}\right)\right],
\end{equation}
where $\mathrm{Li}_2(z)$ is the dilogarithm. It is not obvious that Eqs. \eqref{eq: k=0 limit 1} and \eqref{eq: k=0 limit 2} are equal to each other, but one can use the identity $\mathrm{Li}_2(z) + \mathrm{Li}_2(z^{-1}) = - \pi^2/6 - [\log(-z)]^2/2$ to show that they do indeed agree \cite{zagier2007dilogarithm}. We have also checked numerically that they agree. It is worth pointing out that if the metallic gates are held at non-zero potentials $\varphi_0 \neq 0,$ as long as both gates have the same potential, then all we  need to do is add this constant to the electrostatic potential $\varphi(\mathbf{r},z,z') \mapsto \varphi(\mathbf{r},z,z') + \varphi_0.$ This has no effect on the Coulomb potential energy since such an overall shift in the electrostatic potential is accounted for in the electron occupation. For metallic gates held at unequal potentials, without loss of generality, let us assume $\varphi(\mathbf{r},z=0,z') = 0$ and $ \varphi(\mathbf{r},z=D,z') = \varphi_0.$ This changes the potential to $\varphi(\mathbf{r}-\mathbf{r}',z,z') \mapsto \varphi(\mathbf{r}-\mathbf{r}',z,z') + \varphi_0z/D.$ The term $\varphi_0z/D$ is independent of $\mathbf{r}$ and $\mathbf{r}'$. It is most conveniently represented in the Hamiltonian as a layer-dependent electrostatic potential energy. Therefore, the Coulomb potential energy in Eq. \eqref{eq: Coulomb potential energy} is the most general form for our purpose. For our system of interest, we use the setup shown in Fig. \ref{fig:gate_schematic}. Accordingly, Eq. \eqref{eq: Coulomb potential energy} should be modified such that 
\begin{equation}
    \begin{split}
        D &= d_\mathrm{b}+d_\mathrm{t} +4d,\\
        z_\ell &= d_\mathrm{b} + (\ell-1)d,
    \end{split}
\end{equation}
where $d_\mathrm{b}$ ($d_\mathrm{t}$) is the distance from the bottom (top) to the nearest graphene layer and $d = 3.35$ \AA $ $ is the separation distance between adjacent graphene layers. Using this notation, we write $\mathcal{V}_{\ell, \ell'}(\mathbf{k}) = \mathcal{V}\left(\mathbf{k}, z_\ell, z_{\ell'}\right).$ To simplify, we assume that $d_\mathrm{t} = d_\mathrm{b}$ so that we only need to specify the total gate-to-gate distance $D.$ In this work, we primarily use $D = 400$ \AA \hspace{1pt} (the distance from the middle of the five-layer stack to one of the gates is $200$ \AA). For these values, the Coulomb potential energy is shown in Fig. \ref{fig:coulomb_dependent}. We see that when $\ell \neq \ell',$ the Coulomb potential is slightly suppressed (about $7\%$ at maximum) for all momenta. When $\ell = \ell',$ we find that the Coulomb potential is essentially equal to the simple $(e^2/2\epsilon_0 \epsilon_r)\tanh(kD/2)/k$ potential.

\subsection{Screening in the Classical Limit}

\begin{figure}
    \centering
    \includegraphics[width=0.9\linewidth]{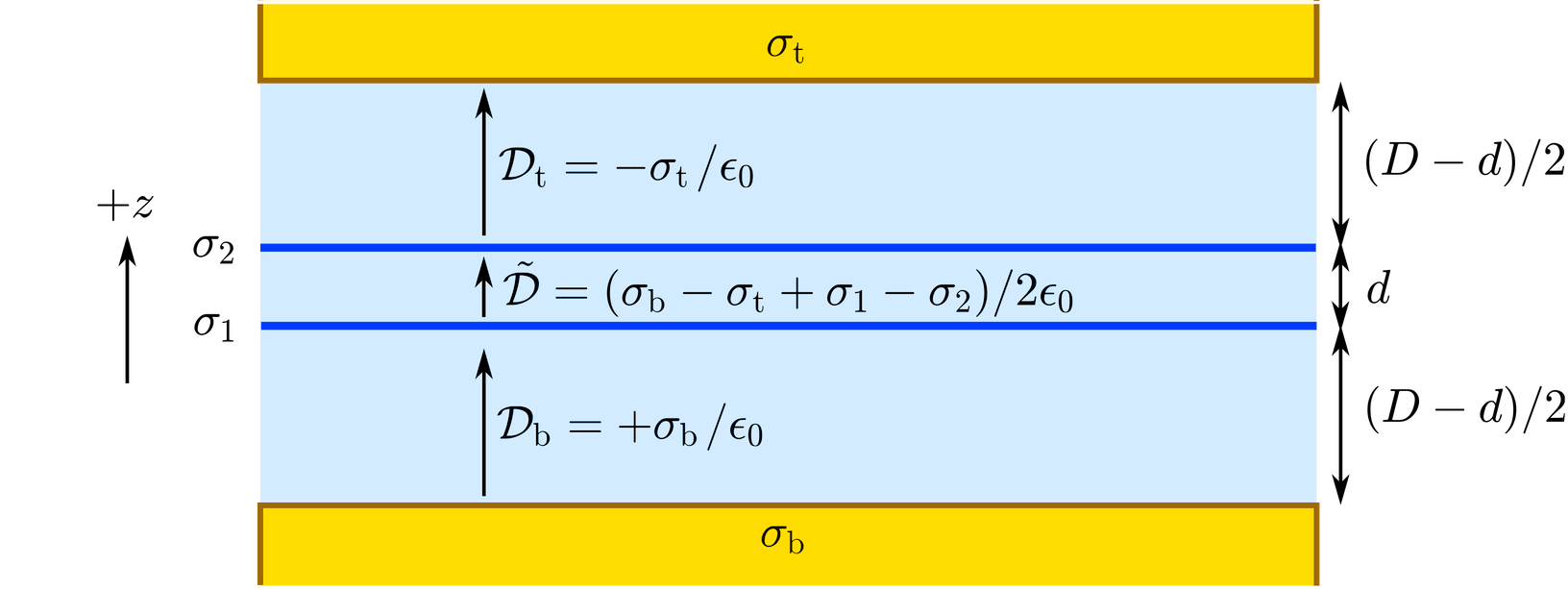}
    \caption{\textbf{Parallel-plate capacitor model.} Two active layers are encapsulated between a top and bottom metallic gate. The charge densities $\sigma$ and associated displacement fields $\mathcal{D}$ are indicated.}
    \label{fig:capacitors}
\end{figure}

In this section, we consider the $\mathbf{k} = \mathbf{0}$ component of the layer-dependent Coulomb potential, which corresponds to uniform charge densities. We show that the layer-dependent Coulomb potential agrees exactly with classical considerations, boosting confidence in the correctness of our formulas. For this purpose, we consider a parallel-plate capacitor model with two gates and two layers of the active material (the generalization to many layers is already included in the formulas of the previous section and is unnecessary for the classical consideration here). We consider the geometry shown in Fig. \ref{fig:capacitors}, which is the same as the geometry in Fig. \ref{fig:gate_schematic} in the limit with only two active layers. The charge densities on the bottom and top gates are $\sigma_\mathrm{b}$ and $\sigma_\mathrm{t}$ and on the first and second layers are $\sigma_1$ and $\sigma_2.$ The associated displacement fields in the regions below and above the active material are
\begin{subequations}
    \begin{align}
        \mathcal{D}_\mathrm{b} &= +\frac{\sigma_\mathrm{b}}{\epsilon_0},\\
        \mathcal{D}_\mathrm{t} &= -\frac{\sigma_\mathrm{t}}{\epsilon_0}.
    \end{align}
\end{subequations}
The field in between the two layers of the active material can be found using Gauss's law:
\begin{subequations}
    \begin{align}
        \tilde{\mathcal{D}} - \mathcal{D}_\mathrm{b} = \frac{\sigma_1}{\epsilon_0}, \\
        \mathcal{D}_\mathrm{t} - \tilde{\mathcal{D}} = \frac{\sigma_2}{\epsilon_0}.
    \end{align}
\end{subequations}
Solving for $\tilde{\mathcal{D}},$ we find
\begin{equation}
    \tilde{\mathcal{D}} = \frac{\mathcal{D}_\mathrm{b}+\mathcal{D}_\mathrm{t}}{2} + \frac{\sigma_1-\sigma_2}{2\epsilon_0} = \mathcal{D}_\mathrm{ave} + \frac{\sigma_1-\sigma_2}{2\epsilon_0},
\end{equation}
where $\mathcal{D}_\mathrm{ave} = (\mathcal{D}_\mathrm{b}+\mathcal{D}_\mathrm{t})/2  = (\sigma_\mathrm{b}-\sigma_\mathrm{t})/2\epsilon_0$ is the average displacement field often quoted in experimental works as \textit{the} displacement field experimentally controlled by varying gate biases. Using classical electrostatics, we find that the areal energy density is 
\begin{equation}
\label{eq: potential energy electric}
    \mathcal{U} = \frac{\epsilon_0}{2\epsilon_r} \int |\mathcal{D}(z)|^2dz = \frac{1}{2\epsilon_0\epsilon_r} \frac{D-d}{2} \left(\sigma_\mathrm{b}^2 + \sigma_\mathrm{t}^2  \right) + \frac{d}{2\epsilon_0\epsilon_r} \frac{(\sigma_\mathrm{b}-\sigma_\mathrm{t}+\sigma_1-\sigma_2)^2}{4}.
\end{equation}
It is convenient to group terms based on the total density $\sigma = \sigma_\mathrm{t}+\sigma_\mathrm{b} = -\sigma_1-\sigma_2$ and the density fluctuations $\sigma_\mathrm{t}-\sigma_\mathrm{b} = \delta\sigma_\mathrm{gate}$ and $\sigma_2-\sigma_1 = \delta\sigma_\mathrm{sample}.$ We have enforced charge neutrality in the system. We can then rewrite the potential energy density as
\begin{equation}
\label{eq: potential in terms of delta sigma}
    \mathcal{U} = \frac{1}{8\epsilon_0\epsilon_r} \left(D-d\right) \left(\sigma^2 + \delta \sigma_\mathrm{gate}^2  \right) + \frac{d}{8\epsilon_0\epsilon_r} (\delta \sigma_\mathrm{gate}+\delta \sigma_\mathrm{sample})^2.
\end{equation}
We emphasize that this is \textit{not} just the energy density of interaction between the charges on the active material but is the \textit{total} energy density of putting together this charge configuration, including the energy it costs to deposit charges on the gates and their interaction with the charges on the active material. For now, we assume that the gates are held at the same potential. We show that the screened Coulomb potential allows us to calculate this total energy density in the Hartree approximation. To do that, we need to determine the potential difference between the top and bottom gates in the presence of charges as indicated and then set that potential difference equal to zero to determine the relationship between the charges. We find difference in the electrostatic potential between the top and bottom gates to be
\begin{equation}
\label{eq: potential difference}
    \begin{split}
        \Delta \varphi = \varphi_\mathrm{t} - \varphi_\mathrm{b}= - \frac{\tilde{\mathcal{D}}}{\epsilon_r}d- \frac{\mathcal{D}_\mathrm{t}+\mathcal{D}_\mathrm{b}}{\epsilon_r}\frac{D-d}{2} = \frac{\delta \sigma_\mathrm{sample}}{2\epsilon_0\epsilon_r}d + \frac{\delta \sigma_\mathrm{gate}}{2\epsilon_0\epsilon_r}D = 0 \rightarrow d\delta \sigma_\mathrm{sample} = -D \delta \sigma_\mathrm{gate}.
    \end{split}
\end{equation}
Using this restriction, Eq. \eqref{eq: potential energy electric} can be rewritten as 
\begin{equation}
\label{eq: potential energy electric 2 }
    \mathcal{U} = \frac{1}{8\epsilon_0\epsilon_r} \left(D-d\right) \sigma^2 + \frac{1}{8\epsilon_0\epsilon_r} \frac{d(D-d)}{D} \delta\sigma_\mathrm{sample}^2.
\end{equation}
In terms of the screened Coulomb potential, the energy density of system is given by 
\begin{equation}
    \mathcal{U} = \frac{1}{2e^2} \sum_{\ell,\ell'} \mathcal{V}_{\ell,\ell'}(\mathbf{k}=\mathbf{0}) \sigma_\ell \sigma_{\ell'},
\end{equation}
where we have chosen zero potential at the middle of the stack. Recalling that 
\begin{equation}
    \begin{split}
        \mathcal{V}_{1,1}(\mathbf{0}) = \mathcal{V}_{2,2}(\mathbf{0}) = \frac{e^2}{4\epsilon_0\epsilon_rD} \left[ D^2-d^2 \right], \quad \mathcal{V}_{1,2}(\mathbf{0}) = \mathcal{V}_{2,1}(\mathbf{0}) = \frac{e^2}{4\epsilon_0\epsilon_rD} \left[ D-d \right]^2,       
    \end{split}
\end{equation}
we find 
\begin{equation}
\begin{split}
    \mathcal{U} &= \frac{1}{8\epsilon_0\epsilon_rD} \left[ D^2-d^2 \right] \left(\sigma_1^2+\sigma_2^2\right) + \frac{1}{4\epsilon_0\epsilon_rD} \left[ D-d \right]^2\sigma_1\sigma_2, \\
    &= \frac{1}{8\epsilon_0\epsilon_r} (D-d) \sigma^2 + \frac{1}{8\epsilon_0\epsilon_r} \frac{d(D-d)}{D} \delta\sigma_\mathrm{sample}^2,  
\end{split}
\end{equation}
which agrees exactly with the other method, showing consistency between the two.

Next, we consider the situation of unequal potentials on the metallic gates. This case is slightly more complicated due to inhomogeneous boundary conditions.  The electrostatic potential energy in Eq. \eqref{eq: potential energy electric} is correct irrespective of whether the two metallic gates have the same potential. This is because that formula calculates the energy directly from the energy density of electric fields. Consequently, it does not suffer from gauge ambiguity hidden in formulas involving potentials that require fixing some reference point. For us, potentials are measured relative to ground. To check consistency with the potential approach, we calculate the energy density using the following expression
\begin{equation}
\label{eq: potential energy from potential}
    \mathcal{U} = \frac{1}{2V_\mathrm{sys}}\int \rho(\mathbf{r},z) \varphi(\mathbf{r},z) d^2\mathbf{r}dz + \frac{1}{2}\varphi_\mathrm{t}\sigma_\mathrm{t} +  \frac{1}{2}\varphi_\mathrm{b} \sigma_\mathrm{b},
\end{equation}
where $V_\mathrm{sys}$ is the area of the system, the charge density distribution inside the gates is given by 
\begin{equation}
\label{eq: charge density}
    \rho(\mathbf{r},z)  = \rho(z) = \sigma_1 \delta(z-[D-d]/2) + \sigma_2\delta(z-[D+d]/2)
\end{equation}
and the electrostatic potential can be found by convolving the Green's function in Sec. \ref{sec: Green's Function Derivation} with the charge density in Eq. \eqref{eq: charge density}
\begin{equation}
    \varphi(\mathbf{r},z) = -\frac{1}{e}\int \rho(z') \varphi(\mathbf{r}-\mathbf{r}',z,z') d^2 \mathbf{r}' dz' + \frac{\epsilon_0\epsilon_r}{e} \int \left[ \varphi_\mathrm{t} \frac{\partial \varphi(\mathbf{r}-\mathbf{r}',z,z') }{\partial z'}\bigg\vert_{z'=D} -\varphi_\mathrm{b} \frac{\partial \varphi(\mathbf{r}-\mathbf{r}',z,z') }{\partial z'}\bigg\vert_{z'=0}  \right]d^2 \mathbf{r}'
\end{equation}
subject to Dirichlet boundary conditions $\varphi(\mathbf{r},z=D) = \varphi_\mathrm{t}$ and $\varphi(\mathbf{r},z=0) = \varphi_\mathrm{b}$. In Eq. \eqref{eq: potential energy electric}, we have separated two contributions to the energy: (1) the integral contribution contains effects coming from the charges in between the gates, and (2) the latter two contributions contain effects coming from the charges at the gates. From this division, it is simple to observe that even if there were absolutely no charges in between the gates, i.e. $\rho(\mathbf{r},z) = 0,$ the electrostatic potential energy would still be non-zero due purely to charges on the gates setting up an electric field. It is worth emphasizing that the electrostatic potential $\varphi(\mathbf{r},z)$ contains information about the entire setup, i.e. all charges everywhere. So the distinction drawn here is only a rough division for bookkeeping purposes. It is absolutely \textit{not} correct to say that the integral term in Eq. \eqref{eq: potential energy electric} contains only contributions coming from interactions among charges on the active material.

Because the charge density is uniform laterally, translational invariance dictates that the potential does not depend on the in-plane coordinates. Therefore, integration over $\mathbf{r}$ and $\mathbf{r}'$ allows us to replace the real-space potential with the $\mathbf{k}=\mathbf{0}$ component of the Fourier-transformed potential. Henceforth, all potentials are in reciprocal space at $\mathbf{k} = 0$. Performing this integration, Eq. \eqref{eq: potential energy from potential} can be simplified to 
\begin{equation}
    \mathcal{U} = \frac{1}{2} \sigma_1 \varphi(z=D/2-d/2) + \frac{1}{2} \sigma_2 \varphi(z=D/2+d/2) + \frac{1}{2}\varphi_\mathrm{t}\sigma_\mathrm{t} + \frac{1}{2}\varphi_\mathrm{b}\sigma_\mathrm{b},
\end{equation}
where the potential at layer $1$ is 
\begin{equation}
\label{eq: layer 1 potential}
    \begin{split}
        \varphi(z=D/2-d/2) = \frac{1}{\epsilon_0\epsilon_r} \left[  \frac{\sigma_1}{D} \left(\frac{D^2}{4}-\frac{d^2}{4} \right) + \frac{\sigma_2}{D} \left( \frac{D}{2}-\frac{d}{2} \right)^2        \right]+ \frac{\varphi_\mathrm{b} + \varphi_\mathrm{t}}{2}  - \frac{d}{D}\frac{\varphi_\mathrm{t} - \varphi_\mathrm{b}}{2} ,
    \end{split}
\end{equation}
and potential at layer $2$ is 
\begin{equation}
\label{eq: layer 2 potential}
    \begin{split}
        \varphi(z = D/2+d/2) = \frac{1}{\epsilon_0\epsilon_r} \left[ \frac{\sigma_1}{D} \left( \frac{D}{2}-\frac{d}{2} \right)^2 + \frac{\sigma_2}{D} \left(\frac{D^2}{4}-\frac{d^2}{4} \right) \right] + \frac{\varphi_\mathrm{b} + \varphi_\mathrm{t}}{2}  + \frac{d}{D}\frac{\varphi_\mathrm{t} - \varphi_\mathrm{b}}{2}.
    \end{split}
\end{equation}
We note that these functions have the correct limits as $d\rightarrow D.$ The terms in square brackets in Eqs. \eqref{eq: layer 1 potential} and \eqref{eq: layer 2 potential} are the same terms originating from the homogeneous Green's function as they are independent of the gate potentials; the terms involving $\varphi_\mathrm{t}+\varphi_\mathrm{b}$ are constant shifts in potential needed to fix the boundary conditions; the terms involving $\varphi_\mathrm{t} - \varphi_\mathrm{b}$ are of the form $\Delta\varphi z/D$, linear functions of position. Indeed, Eqs. \eqref{eq: layer 1 potential} and \eqref{eq: layer 2 potential} are quite simple that one could have easily guessed them without needing to compute the Green's function at all, but the purpose of the exercise here is to check that the Green's function  produces the correct result.  Using the fact the that $\varphi_\mathrm{b}+ \varphi_\mathrm{t} = \sigma(D-d)/2\epsilon_0\epsilon_r,$ we find
\begin{equation}
    \begin{split}
        \mathcal{U} &= \frac{1}{8\epsilon_0\epsilon_rD} \left[ D^2-d^2 \right] \left(\sigma_1^2+\sigma_2^2\right) + \frac{1}{4\epsilon_0\epsilon_rD} \left[ D-d \right]^2\sigma_1\sigma_2 - \frac{1}{8\epsilon_0\epsilon_r}\sigma^2(D-d) + \frac{d}{4D} \delta \sigma_\mathrm{sample} \Delta\varphi + \frac{1}{2}\varphi_\mathrm{t}\sigma_\mathrm{t} + \frac{1}{2}\varphi_\mathrm{b}\sigma_\mathrm{b}\\
        &= \frac{1}{8\epsilon_0\epsilon_rD} \left[ D^2-d^2 \right] \left(\sigma_1^2+\sigma_2^2\right) + \frac{1}{4\epsilon_0\epsilon_rD} \left[ D-d \right]^2\sigma_1\sigma_2  + \frac{1}{4D}\left[d \delta \sigma_\mathrm{sample}  + D\delta \sigma_\mathrm{gate}\right]\Delta \varphi \\
        &= \frac{1}{8\epsilon_0\epsilon_rD} \left[ D^2-d^2 \right] \left(\sigma_1^2+\sigma_2^2\right) + \frac{1}{4\epsilon_0\epsilon_rD} \left[ D-d \right]^2\sigma_1\sigma_2  + \frac{\epsilon_0\epsilon_r}{2D}(\Delta \varphi)^2\\
        &= \frac{1}{8\epsilon_0\epsilon_r} \left(D-d\right)\sigma^2 + \frac{1}{8\epsilon_0\epsilon_rD} \left[dD-d^2 \right] \delta \sigma_\mathrm{sample}^2 + \frac{\epsilon_0\epsilon_r}{2D}(\Delta \varphi)^2.
    \end{split}
\end{equation}
Now, using Eq. \eqref{eq: potential difference}, we can eliminate $\Delta \varphi$ to find
\begin{equation}
\begin{split}
     \mathcal{U} &= \frac{1}{8\epsilon_0\epsilon_r} \left(D-d\right)\sigma^2 + \frac{1}{8\epsilon_0\epsilon_r} \left[2d\sigma_\mathrm{sample}\sigma_\mathrm{gate}  + D \sigma_\mathrm{gate}^2+d\sigma_\mathrm{sample}^2 \right],\\
    &=\frac{1}{8\epsilon_0\epsilon_r} \left(D-d\right) \left(\sigma^2 + \delta \sigma_\mathrm{gate}^2  \right) + \frac{d}{8\epsilon_0\epsilon_r} (\delta \sigma_\mathrm{gate}+\delta \sigma_\mathrm{sample})^2,   
\end{split}
\end{equation}
which agrees exactly with Eq. \eqref{eq: potential in terms of delta sigma}, again showing that our approach is correct even at unequal potential values at the gates.

One might worry that in order to minimize the total electrostatic energy that one needs to not only rearrange the charges on the active material but also redistribute the charges at the gates as well. We end this section by remarking that this is not necessary as long as the gate potentials are fixed. This can be seen by noting that the potential energy at fixed potential difference between the gates can be written as 
\begin{equation}
\label{eq: energy without gate charges}
    \mathcal{U} = \frac{1}{2e^2} \sum_{\ell,\ell'} \mathcal{V}_{\ell,\ell'}(\mathbf{k}=\mathbf{0}) \sigma_\ell \sigma_{\ell'}  + \frac{\epsilon_0\epsilon_r}{2D}(\Delta \varphi)^2,
\end{equation}
where the last term is a fixed constant once the gate potentials are specified. This difference in the gate potentials is related to the $\Delta$ parameter in the single-particle Hamiltonian. In Eq. \eqref{eq: energy without gate charges}, importantly, there is no information about the gate charge densities. Therefore, minimization of the Coulomb potential energy can be done by varying only the electron densities in the active material. This justifies the implementation of the gate-screened Coulomb potential in the HF algorithm, where no information about the gate charge densities appears.

\section{Self-Consistent Mean-Field Theory}

To study electron-electron interactions, it is most convenient to adopt second-quantized notation. First, we rewrite the kinetic energy in second-quantized notation
\begin{equation}
    \hat{\mathcal{K}} = \sum_{\mathbf{k}\in \mathrm{mBZ}, \mathbf{G}, \mathbf{G}'} \sum_{\mu,\mu'} \hat{c}_{\mu, \mathbf{k}+\mathbf{G}}^\dagger \mathbb{K}^{\mu,\mu'}_{\mathbf{G},\mathbf{G}'}(\mathbf{k}) \hat{c}_{\mu',\mathbf{k} + \mathbf{G}'},
\end{equation}
where $\mathbb{K}^{\mu,\mu'}_{\mathbf{G},\mathbf{G}'}(\mathbf{k})$ is same matrix defined in Sec. \ref{sec: Continuum Hamiltonian} with a slight change in notation to emphasize the dependence on reciprocal lattice vectors, $\hat{c}_{\mu, \mathbf{k}+\mathbf{G}}^\dagger $ creates a plane-wave state $\ket{\psi_{\mu,\mathbf{k} + \mathbf{G}}}$ with $\mu$ denoting all the internal indices, which include $\lbrace s (\text{spin}), \nu (\text{valley}), \sigma (\text{sublattice}), \ell (\text{layer}) \rbrace.$ Throughout, we use lower-case momenta to denote momenta confined within a moir\'{e}-Brillouin zone (mBZ) and capitalized momenta to denote moir\'{e} reciprocal lattice vectors. We can diagonalize the kinetic energy by finding eigenstates $\ket{\Psi_{n,\xi,\mathbf{k}}} = e^{i\mathbf{k} \cdot \hat{\mathbf{r}}} \ket{u_{n,\xi,\mathbf{k}}} = \sum_{\mathbf{G},  \sigma, \ell} \phi_{n,\xi,\sigma, \ell, \mathbf{k}+\mathbf{G}} \ket{\psi_{\xi,\sigma,\ell,\mathbf{k}+ \mathbf{G}}}$ with energy $E_{n,\xi}(\mathbf{k}),$ where $n$ is band index, $\xi \in \lbrace s,\nu \rbrace$ denotes only the non-spatial indices, and $\phi_{n,\xi,\sigma, \ell, \mathbf{k}+\mathbf{G}}$ are numerical coefficients obtained from diagonalizing $\mathbb{K}.$ The operators that create $\ket{\Psi_{n,\xi,\mathbf{k}}}$ are denoted $\hat{\Psi}^\dagger_{n,\xi,\mathbf{k}}.$ Normalization is enforced by $\sum_{\mathbf{G},\sigma,\ell}\left|\phi_{n,\xi,\sigma,\ell,\mathbf{k} + \mathbf{G}}\right|^2 = 1.$ In this band basis, the kinetic energy operator is diagonal
\begin{equation}
    \hat{\mathcal{K}} = \sum_{\mathbf{k} \in \mathrm{mBZ}}\sum_{n,\xi} E_{n,\xi}(\mathbf{k}) \hat{\Psi}_{n,\xi,\mathbf{k}}^\dagger \hat{\Psi}_{n,\xi,\mathbf{k}}.
\end{equation}
Moving onto the four-fermion Coulomb interaction, we can also write it in the band basis as 
\begin{equation}
    \begin{split}
      \frac{1}{2} & \sum_{\ell,\ell',\alpha, \alpha'}\int d^2 \mathbf{r} d^2\mathbf{r}' \mathcal{V}_{\ell, \ell'}(\mathbf{r}-\mathbf{r}') :  \hat{\rho}_{\alpha,\ell}(\mathbf{r})  \hat{\rho}_{\alpha',\ell'}(\mathbf{r}') :  \\
       = \frac{1}{2}& \sum_{\substack{\mathbf{k},\mathbf{p},\mathbf{q} \in \mathrm{mBZ}\\n_1,n_2,n_3,n_4 \\ \xi_1, \xi_2}} \mathbb{V}^{n_1,n_2,n_3,n_4}_{\xi_1,\xi_2}(\mathbf{k},\mathbf{p},\mathbf{q}) \hat{\Psi}_{n_1, \xi_1, \mathbf{k}+\mathbf{q}}^\dagger \hat{\Psi}_{n_2, \xi_2, \mathbf{p}-\mathbf{q}}^\dagger \hat{\Psi}_{n_4, \xi_2, \mathbf{p}}\hat{\Psi}_{n_3, \xi_1, \mathbf{k}} ,
    \end{split}
\end{equation}
where $\alpha = \lbrace s,\nu,\sigma \rbrace$ to single out the layer index $\ell$ and  the matrix elements are computed using the one-particle wave functions:
\begin{equation}
    \begin{split}
        \mathbb{V}^{n_1,n_2,n_3,n_4}_{\xi_1,\xi_2}(\mathbf{k},\mathbf{p},\mathbf{q}) = \frac{1}{V_\mathrm{sys}} \sum_{\mathbf{Q},\ell_1,\ell_2}  &\left[\sum_{\mathbf{G},\sigma_1} \phi^\dagger_{n_1,\xi_1,\sigma_1,\ell_1,\mathbf{k}+\mathbf{G}+\mathbf{q}+\mathbf{Q}} \phi_{n_3,\xi_1,\sigma_1,\ell_1,\mathbf{k}+\mathbf{G}} \right] \mathcal{V}_{\ell_1,\ell_2}(\mathbf{q}+\mathbf{Q}) \times \\
        \times &\left[\sum_{\mathbf{P},\sigma_2} \phi^\dagger_{n_2,\xi_2,\sigma_2,\ell_2,\mathbf{p}+\mathbf{P}-\mathbf{q}-\mathbf{Q}} \phi_{n_4,\xi_2,\sigma_2,\ell_2,\mathbf{p}+\mathbf{P}} \right].
    \end{split}
\end{equation}
The layer-dependent screened Coulomb potential energy $\mathcal{V}_{\ell_1,\ell_2}(\mathbf{k})$ is given in Eq. \eqref{eq: Coulomb potential energy} of Sec. \ref{sec: Layer-Dependent Coulomb Potential Energy}. We can simplify by defining layer-resolved form factors
\begin{equation}
\label{eq: form factors}
    \begin{split}
        \mathbb{\Lambda}^{n,n'}_{\xi,\ell}(\mathbf{k},\mathbf{q}+\mathbf{Q}) = \sum_{\mathbf{G},\sigma} \phi^\dagger_{n,\xi,\sigma,\ell,\mathbf{k}+\mathbf{G}+\mathbf{q}+\mathbf{Q}} \phi_{n',\xi,\sigma,\ell,\mathbf{k}+\mathbf{G}} = \bra{u_{n,\xi,\mathbf{k}+\mathbf{q}}} e^{i \mathbf{Q} \cdot \hat{\mathbf{r}}} \mathbb{P}_\ell \ket{u_{n',\xi,\mathbf{k}}},
    \end{split}
\end{equation}
where $\mathbb{P}_\ell$ is the projector onto the $\ell$ layer. These form factors obey $\left[ \mathbb{\Lambda}^{n,n'}_{\xi,\ell}(\mathbf{k},\mathbf{q}+\mathbf{Q})\right]^\dagger = \mathbb{\Lambda}^{n',n}_{\xi,\ell}(\mathbf{k}+\mathbf{q},-\mathbf{q}-\mathbf{Q}).$ With these form factors, we can write the matrix elements as 
\begin{equation}
\label{eq: Coulomb matrix elements}
    \begin{split}
        \mathbb{V}^{n_1,n_2,n_3,n_4}_{\xi_1,\xi_2}(\mathbf{k},\mathbf{p},\mathbf{q}) &= \frac{1}{V_\mathrm{sys}} \sum_{\mathbf{Q},\ell_1,\ell_2} \mathbb{\Lambda}^{n_1,n_3}_{\xi_1,\ell_1}(\mathbf{k},\mathbf{q}+\mathbf{Q}) \mathcal{V}_{\ell_1,\ell_2}(\mathbf{q}+\mathbf{Q}) \mathbb{\Lambda}^{n_2,n_4}_{\xi_2,\ell_2}(\mathbf{p},-\mathbf{q}-\mathbf{Q}) \\
        &= \frac{1}{V_\mathrm{sys}} \sum_{\mathbf{Q},\ell_1,\ell_2} \mathbb{\Lambda}^{n_1,n_3}_{\xi_1,\ell_1}(\mathbf{k},\mathbf{q}+\mathbf{Q}) \mathcal{V}_{\ell_1,\ell_2}(\mathbf{q}+\mathbf{Q}) \left[\mathbb{\Lambda}^{n_4,n_2}_{\xi_2,\ell_2}(\mathbf{p}-\mathbf{q},\mathbf{q}+\mathbf{Q}) \right]^\dagger.
    \end{split}
\end{equation}
We note that the form factors $\mathbb{\Lambda}^{n,n'}_{\xi,\ell}(\mathbf{k},\mathbf{q})$ are periodic in the first momentum $\mathbf{k}$ but \textit{not} in the second momentum $\mathbf{q}$ assuming that the energy eigenfunctions are calculated with enough $\mathbf{G}$ vectors to ensure convergence; meanwhile, the band-projected Coulomb matrix elements $\mathcal{V}^{n_1,n_2,n_3,n_4}_{\xi_1,\xi_2}(\mathbf{k},\mathbf{p},\mathbf{q})$ are periodic in all three momenta, again assuming that enough $\mathbf{Q}$ vectors are included in the sum. Let us now implement the HF decoupling  assuming that moir\'{e} translational symmetry is \textit{not} broken:
\begin{equation}
\label{eq: Hartree Fock decoupling}
    \begin{split}
        \frac{1}{2}&\sum_{\substack{\mathbf{k},\mathbf{p},\mathbf{q} \in \mathrm{mBZ}\\n_1,n_2,n_3,n_4 \\ \xi_1, \xi_2}} \mathbb{V}^{n_1,n_2,n_3,n_4}_{\xi_1,\xi_2}(\mathbf{k},\mathbf{p},\mathbf{q}) \hat{\Psi}_{n_1, \xi_1, \mathbf{k}+\mathbf{q}}^\dagger \hat{\Psi}_{n_2, \xi_2, \mathbf{p}-\mathbf{q}}^\dagger \hat{\Psi}_{n_4, \xi_2, \mathbf{p}}\hat{\Psi}_{n_3, \xi_1, \mathbf{k}} \\
        \approx &\sum_{\substack{\mathbf{k},\mathbf{p} \in \mathrm{mBZ}\\n_1,n_2,n_3,n_4 \\ \xi_1, \xi_2}} \mathbb{V}^{n_1,n_2,n_3,n_4}_{\xi_1,\xi_2}(\mathbf{k},\mathbf{p},\mathbf{0})\langle  \hat{\Psi}_{n_1, \xi_1, \mathbf{k}}^\dagger \hat{\Psi}_{n_3, \xi_1, \mathbf{k}} \rangle\hat{\Psi}_{n_2, \xi_2, \mathbf{p}}^\dagger \hat{\Psi}_{n_4, \xi_2, \mathbf{p}} \\
        - &\sum_{\substack{\mathbf{k},\mathbf{p} \in \mathrm{mBZ}\\n_1,n_2,n_3,n_4 \\ \xi_1, \xi_2}} \mathbb{V}^{n_1,n_2,n_3,n_4}_{\xi_1,\xi_2}(\mathbf{k},\mathbf{p},\mathbf{p}-\mathbf{k}) \langle \hat{\Psi}_{n_1, \xi_1, \mathbf{p}}^\dagger \hat{\Psi}_{n_4, \xi_2, \mathbf{p}} \rangle \hat{\Psi}_{n_2, \xi_2, \mathbf{k}}^\dagger \hat{\Psi}_{n_3, \xi_1, \mathbf{k}} + ...
    \end{split}
\end{equation}
Defining the density matrix $\mathbb{D}^{n,n'}_{\xi,\xi'}(\mathbf{p}) = \langle \hat{\Psi}_{n, \xi, \mathbf{p}}^\dagger \hat{\Psi}_{n', \xi', \mathbf{p}} \rangle,$ we can now write the full Coulomb potential energy in the HF approximation as 
\begin{equation}
\begin{split}
    \hat{\mathcal{V}} &\approx  \sum_{\substack{\mathbf{k},\mathbf{p} \in \mathrm{mBZ}\\n_1,n,n_3,n' \\ \xi_1, \xi}} \mathbb{V}^{n_1,n,n_3,n'}_{\xi_1,\xi}(\mathbf{p},\mathbf{k},\mathbf{0}) \mathbb{D}^{n_1,n_3}_{\xi_1,\xi_1}(\mathbf{p}) \hat{\Psi}_{n, \xi, \mathbf{k}}^\dagger \hat{\Psi}_{n', \xi, \mathbf{k}} - \sum_{\substack{\mathbf{k},\mathbf{p} \in \mathrm{mBZ}\\n_1,n,n',n_4 \\ \xi', \xi}} \mathbb{V}^{n_1,n,n',n_4}_{\xi',\xi}(\mathbf{k},\mathbf{p},\mathbf{p}-\mathbf{k}) \mathbb{D}_{\xi',\xi}^{n_1,n_4}(\mathbf{p})\hat{\Psi}_{n, \xi, \mathbf{k}}^\dagger \hat{\Psi}_{n', \xi', \mathbf{k}}   + ...,
\end{split}
\end{equation}
where the ellipsis stands for (ignored) constants that can only shift the overall spectrum. There is much redundancy in the Coulomb matrix elements. Such redundancy can be exploited to speed up numerical implementation. To see this, we write the matrix elements explicitly:
\begin{equation}
\label{eq: explicit formulas}
\begin{split}
     \mathbb{V}^{n_1,n_2,n_3,n_4}_{\xi_1,\xi_2}(\mathbf{k},\mathbf{p},\mathbf{p}-\mathbf{k}) &= \frac{1}{V_\mathrm{sys}} \sum_{\mathbf{Q},\ell_1,\ell_2} \bra{u_{n_1,\xi_1,\mathbf{p}}} e^{i \mathbf{Q} \cdot \hat{\mathbf{r}}} \mathbb{P}_{\ell_1} \ket{u_{n_3,\xi_1,\mathbf{k}}} \mathcal{V}_{\ell_1,\ell_2}(\mathbf{p}-\mathbf{k}+\mathbf{Q}) \left[\bra{u_{n_4,\xi_2,\mathbf{p}}} e^{i \mathbf{Q} \cdot \hat{\mathbf{r}}} \mathbb{P}_{\ell_2} \ket{u_{n_2,\xi_2,\mathbf{k}}}  \right]^\dagger,   \\
     \mathbb{V}^{n_1,n_2,n_3,n_4}_{\xi_1,\xi_2}(\mathbf{k},\mathbf{p},\mathbf{0}) &= \frac{1}{V_\mathrm{sys}} \sum_{\mathbf{Q},\ell_1,\ell_2} \bra{u_{n_1,\xi_1,\mathbf{k}}} e^{i \mathbf{Q} \cdot \hat{\mathbf{r}}} \mathbb{P}_{\ell_1} \ket{u_{n_3,\xi_1,\mathbf{k}}} \mathcal{V}_{\ell_1,\ell_2}(\mathbf{Q}) \left[\bra{u_{n_4,\xi_2,\mathbf{p}}} e^{i \mathbf{Q} \cdot \hat{\mathbf{r}}} \mathbb{P}_{\ell_2} \ket{u_{n_2,\xi_2,\mathbf{p}}}  \right]^\dagger.   
\end{split}
\end{equation} 
From the first line of Eq. \eqref{eq: explicit formulas}, we find $\mathbb{V}^{n_1,n_2,n_3,n_4}_{\xi_1,\xi_2}(\mathbf{k},\mathbf{p},\mathbf{p}-\mathbf{k}) = \left[ \mathbb{V}^{n_3,n_4,n_1,n_2}_{\xi_1,\xi_2}(\mathbf{p},\mathbf{k},\mathbf{k}-\mathbf{p}) \right]^\dagger = \left[ \mathbb{V}^{n_4,n_3,n_2,n_1}_{\xi_2,\xi_1}(\mathbf{k},\mathbf{p},\mathbf{p}-\mathbf{k}) \right]^\dagger$. In the first equality, band indices $(n_1 \leftrightarrow n_3, n_2 \leftrightarrow n_4)$ and momenta $(\mathbf{k} \leftrightarrow \mathbf{p})$ are exchanged, while  valley-spin indices are fixed. In the second equality, band indices $(n_1 \leftrightarrow n_4, n_2 \leftrightarrow n_3)$ and valley-spin indices $(\xi_1 \leftrightarrow \xi_2)$ are exchanged, while momenta are fixed. From the second line of Eq. \eqref{eq: explicit formulas}, we find $\mathbb{V}^{n_1,n_2,n_3,n_4}_{\xi_1,\xi_2}(\mathbf{k},\mathbf{p},\mathbf{0}) = \left[ \mathbb{V}^{n_3,n_4,n_1,n_2}_{\xi_1,\xi_2}(\mathbf{k},\mathbf{p},\mathbf{0}) \right]^\dagger = \left[ \mathbb{V}^{n_4,n_3,n_2,n_1}_{\xi_2,\xi_1}(\mathbf{p},\mathbf{k},\mathbf{0}) \right]^\dagger$. In the first equality, band indices $(n_1 \leftrightarrow n_3, n_2 \leftrightarrow n_4)$ are exchanged, while  valley-spin indices and momenta are fixed. In the second equality, band indices $(n_1 \leftrightarrow n_4, n_2 \leftrightarrow n_3)$, valley-spin indices $(\xi_1 \leftrightarrow \xi_2)$, and momenta $(\mathbf{k} \leftrightarrow \mathbf{p})$ are all exchanged. These properties are used in the derivation of Eq. \eqref{eq: Hartree Fock decoupling} to eliminate factors of $\frac{1}{2}$. Collecting terms, we can write the mean-field Hamiltonian as 
\begin{equation}
\begin{split}
    \hat{\mathcal{H}} &= \sum_{\substack{\mathbf{k} \in \mathrm{mBZ} \\ n,n', \xi, \xi'}} \hat{\Psi}^\dagger_{n,\xi,\mathbf{k}} \left[  \mathbb{K}^{n,n'}_{\xi,\xi'}(\mathbf{k}) + \mathbb{H}^{n,n'}_{\xi,\xi'}(\mathbf{k})+ \mathbb{F}^{n,n'}_{\xi,\xi'}(\mathbf{k}) \right]\hat{\Psi}_{n',\xi',\mathbf{k}},
\end{split}
\end{equation}
where the kinetic energy, Hartree potential, and Fock potential matrices are 
\begin{subequations}
\label{eq: HF equations}
    \begin{align}
            \mathbb{K}^{n,n'}_{\xi,\xi'}(\mathbf{k}) &= \delta_{\xi,\xi'} \delta_{n,n'} E_{n,\xi}(\mathbf{k}) , \\
    \mathbb{H}^{n,n'}_{\xi,\xi'}(\mathbf{k}) &= +\delta_{\xi,\xi'} \sum_{\substack{\mathbf{p} \in \mathrm{mBZ}\\ n_1, n_3, \xi_1}} \mathbb{V}^{n_1,n,n_3,n'}_{\xi_1,\xi}(\mathbf{p},\mathbf{k},\mathbf{0}) \mathbb{D}^{n_1,n_3}_{\xi_1,\xi_1}(\mathbf{p}), \\
     \mathbb{F}^{n,n'}_{\xi,\xi'}(\mathbf{k}) &= -\sum_{\substack{\mathbf{p} \in \mathrm{mBZ}\\ n_1, n_4}} \mathbb{V}^{n_1,n,n',n_4}_{\xi',\xi}(\mathbf{k},\mathbf{p},\mathbf{p}-\mathbf{k}) \mathbb{D}_{\xi',\xi}^{n_1,n_4}(\mathbf{p}).
    \end{align}
\end{subequations}
The Hartree and Fock terms depend on the density matrix; so they must be solved self-consistently. On the other hand, the kinetic piece is a one-particle term that does not depend on density; it does not need to be solved self-consistently.

\subsection{Background Subtraction}

Some amount of electron-electron interactions should have already been accounted for in \textit{ab-initio} calculations to obtain the non-interacting band structure at charge neutrality in zero displacement field. To avoid double counting the effects of electron-electron interactions, we have to implement a background subtraction in HF renormalization. There is not yet a unique way to perform such a background subtraction; there are several possible candidates assumed in the literature. In this work, we consider a simple background subtraction that assumes the background charge density is that of a uniform half-filled lattice of isolated atoms. The reference density for such a background is $\mathbb{D}_\mathrm{ref}(\mathbf{p}) = \frac{1}{2} \delta_{n,n'} \delta_{\xi, \xi'}.$ Using this, we replace the density matrix in Eq. \eqref{eq: HF equations} by $\delta \mathbb{D} = \mathbb{D} - \mathbb{D}_\mathrm{ref},$ that is, only charge fluctuations away from the neutral background contribute to the HF Hamiltonian. 

\subsection{Convergence Criterion}

In our iterative algorithm, no special updating method of the density matrix is used to accelerate convergence. Instead, at each iterative step, we simply recalculate the density matrix by filling the appropriate number of lowest energy eigenstates of the HF Hamiltonian obtained from the previous iteration. We find that even without using a special updating method, every calculation converged in a reasonable amount of time. In our simulations, convergence at step $n$ of the calculation is defined as satisfying a criterion whereby the density matrix in the next step is not changing by much. We take this criterion to be $\norm{\mathbb{D}_n-\mathbb{D}_{n+1}} < 10^{-8},$ where $\mathbb{D}_n$ is the density matrix at step $n$ and the matrix norm is taken to be the absolute value of the maximal matrix element.

\subsection{Initial Ansatz for Density Matrix}

In this work, we only consider filling factor of one, leaving other possibilities to future studies. Said differently, we electron dope the system with one particle per moir\'{e} unit cell. We use two different ansatzes for the density matrix to search for different ground states using the HF algorithm. In the first, we are after an isospin-polarized ground state, and therefore, the initial ansatz is a density matrix that is half-filled equally in all four flavors plus an extra $N_\mathbf{k}^2$ electrons placed into the remaining available lowest-energy states in only one of the four flavors. In the second, we seek the symmetric ground state, and therefore, the initial ansatz is a density matrix that is filled equally in all four flavors. To determine which ground state is favored, we compare their total energy and choose the one with the lower energy. To ensure consistency, we confirm that $C_{3z}$ is numerically preserved for all converged solutions. Ideally, to determine the ``true" ground state, one has to, the very at least, perform many repeated searches with randomized ansatz for the initial density matrix and then compare their energies to choose the best solution. However, because we prefer to work with relatively large grids and to include a large number of bands to ensure convergence, doing randomized searches is currently computationally prohibitive for us. With that said, since our work is focused on the possibility of inducing isolated narrow Chern bands at filling factor of one, our choice of initial density matrices is physically motivated. Because the energy differences between the symmetric ground states and the $\mathcal{T}$-broken ground states for wide areas of parameter space are relatively large (on the order of millielectronvolts), we expect these $\mathcal{T}$-broken ground phases to be robust. Thus, the qualitative conclusion of the \textit{existence} of isolated narrow Chern bands should survive in more refined mean-field searches even if the phase boundaries may shift.

\section{Self-Consistent Mean-Field Phase Diagrams}

All of our simulations are done with a grid of $N_\mathbf{k} \times N_\mathbf{k} = 18\times18$. All results are simulated with at least 6 bands per isospin. However, many of the data points presented in the main text are obtained with 8 bands per isospin. We have checked that even a $N_\mathbf{k} \times N_\mathbf{k} = 12\times 12$ grid gives qualitatively similar results to those with a $N_\mathbf{k} \times N_\mathbf{k} = 18\times18$ grid. Therefore, we believe that our calculations have converged. Of course, definitive conclusion about convergence is difficult to make since it would require, in addition to increasing mesh sizes, some sort of analytic argument about the thermodynamic limit \cite{Huang2024Self}. Increasing beyond $N_\mathbf{k} \times N_\mathbf{k} = 18\times18$ is possible, but is costly and is beyond our available computational resources. There is a wide range of values for the dielectric constant used in theoretical works  for twisted systems. There does not seem to be any consensus on what that value should be, even when given the experimental setup. Here, for the majority of the simulations, we use $\epsilon_r=6$ and $\epsilon_r=12$ for the dielectric constant. While the smaller value of $\epsilon_r=6$ is closer to the experimental value of the dielectric constant of hexagonal boron nitride, the larger value of $\epsilon_r=12$ is probably more realistic because it may account for some effects of internal screening. All in all, dielectric screening does indeed alter the phase boundaries of our system, but does not affect to a significant degree the existence of various phases of interest. For consistency, all results have been checked to respect $C_{3z}$ symmetry and for cases of the symmetric state, have also been checked to additionally respect $\mathcal{T}$ symmetry.

The phase diagrams show a variety of phases. We characterize them based on the following criteria:
\begin{enumerate}
    \item \textbf{$\mathcal{T}$-preserving metal}: In our HF calculations, we compare the energies of the converged solutions obtained from an initial ansatzes which populate equally all four isopsins and those of the converged solutions calculated from initial ansatzes which populates only one isospin. Whenever the energy of the symmetric solution is lower (or very close to the symmetry-broken solution), we assign that state a $\mathcal{T}$-preserving metal. This state retains all symmetries of the non-interacting model, and  is necessarily metallic because filling one electron per moir\'{e} unit cell with a fourfold degeneracy means that each band is only one-fourth filled. 
    \item \textbf{$\mathcal{T}$-breaking metal}: This state is obtained from the symmetry-broken ansatz. Therefore, it breaks time-reversal symmetry by unequally filling the four otherwise degenerate bands. We diagnose it as metallic by calculating the gap to excited states. When that gap is zero, the state is metallic. 
    \item \textbf{Correlated trivial insulator}: This state emerges from the symmetry-broken ansatz. Coulomb interactions stabilize the separation of the first conduction band in a single isopsin flavor from the higher-energy bands. When the gap to excited states finite, i.e. $\delta_\mathrm{above} > 0$ in the notation of Sec. \ref{sec:Non-Interacting Phase Diagrams}, we have an insulating state. If $\sum_{n\in \mathrm{occupied}}\mathcal{C}_n = 0,$ where $n$ is band index, we call that state a trivial insulator. 
     \item \textbf{Anomalous Chern insulator}: This state differs from the correlated trivial insulator only in that $\sum_{n\in \mathrm{occupied}}\mathcal{C}_n \neq 0$. However, the band immediately below the correlated gap is \textit{not} isolated. So we cannot assign that band a Chern number. In that regard, we do not anticipate this state to be a favorable precursor to a fractional Chern state upon hole doping. Nonetheless, when the chemical potential is parked in the gap above that band, a non-zero Hall conductivity can be measured.
     \item \textbf{Anomalous Chern insulator with isolated Chern band}: This state is a refined characterization of the anomalous Chern insulator state with $\delta_\mathrm{below} >0.$ In other words, this state not only has quantized Hall conductivity but it also features an isolated narrow band with finite Chern number upon doping of which a fractional Chern insulator (FCI) state may emerge. As a consequence, this state can be labeled by two quantum numbers: the Chern number of the band which is isolated, $\mathcal{C},$ and the Chern number associated with the Hall conductivity of the gap right above that band, $\sum_{n\in \mathrm{occupied}}\mathcal{C}_n$. These two quantum numbers usually agree for the majority of phase space. However, they do not necessarily have to the same, and indeed, there are points in phase space where they do not agree. Since our interest is in the isolated Chern band, for brevity and clarity, we show only $\mathcal{C}$ in our phase diagrams.
\end{enumerate}

\subsection{Antiparallel Stacking Order}

In this section, we study the phase diagrams of the antiparallel stacking order using the four parameter sets listed in Table \ref{tab:hopping parameters}.

\subsubsection{With Some Lattice Relaxation}

\begin{figure}[t!]
\includegraphics[width=0.90\linewidth]{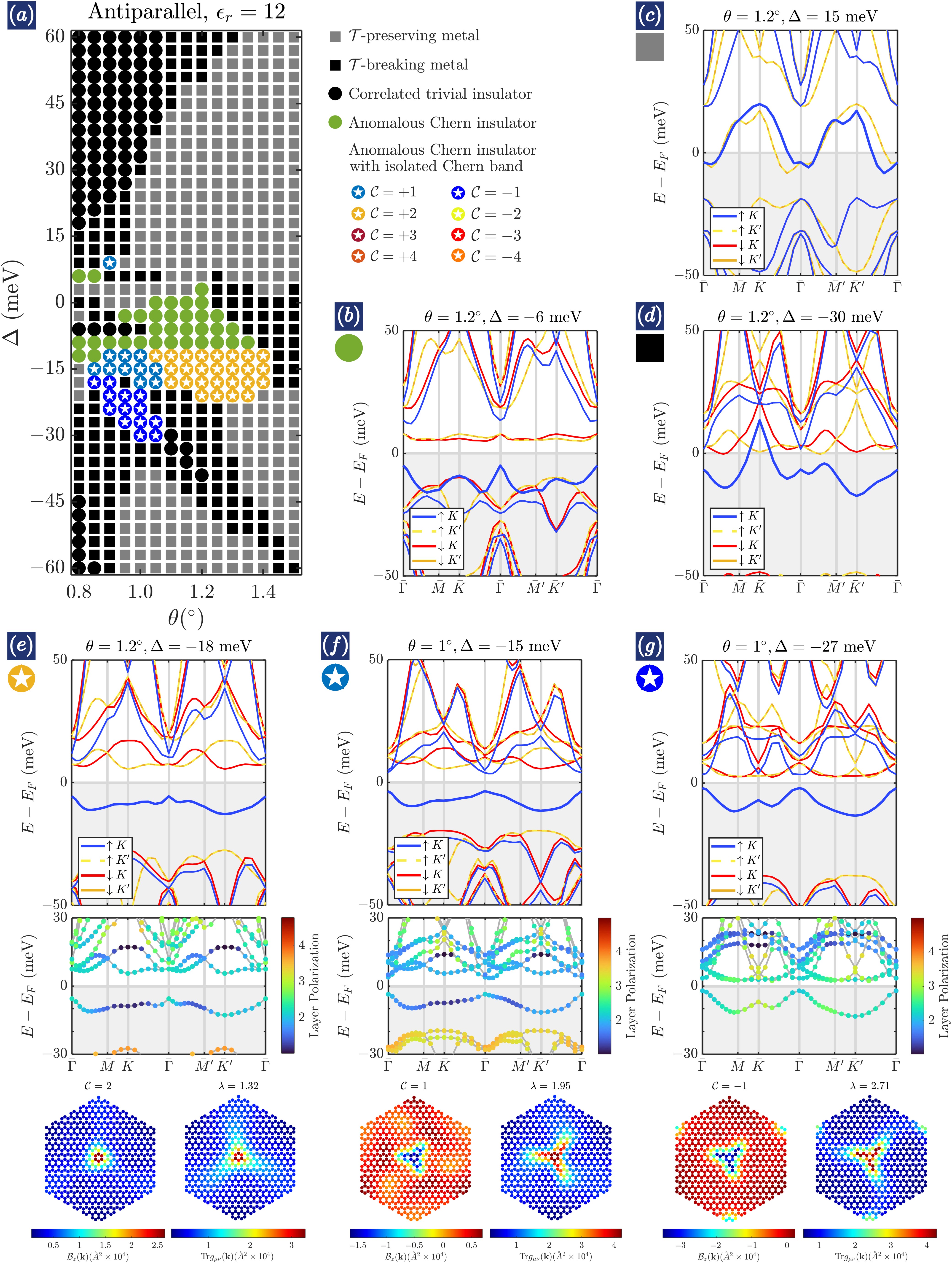}
\caption{\textbf{Mean-field phase diagram for the antiparallel configuration with lattice relaxation and $\epsilon_r = 12.$} (a) Phase diagram as a function of $\theta$ and $\Delta$. The results here are simulated using parameters from {\fontfamily{cmtt}\selectfont Param\_one}. Representative band structures of the different phases are shown in (b)-(g): an anomalous Chern insulator with no gap below the band closest to chemical potential  (b), a $\mathcal{T}$-preserving metal (c), a $\mathcal{T}$-breaking metal (d), and an anomalous Chern insulator with isolated Chern band with $\mathcal{C} = +2$ (e), $\mathcal{C} = +1$ (f), or $\mathcal{C} = -1$ (g). For (b)-(d), we only show the isospin polarization of the energy bands. For (e)-(g), we additionally color code the layer polarization of the Bloch states; furthermore, we show the corresponding Berry curvature and quantum metric of the band closest to chemical potential. }
\label{fig:HF_antiparallel_relaxation_e12_largevF_phase_diagram}
\end{figure}

\begin{figure}[t!]
    \centering
    \includegraphics[width=1\linewidth]{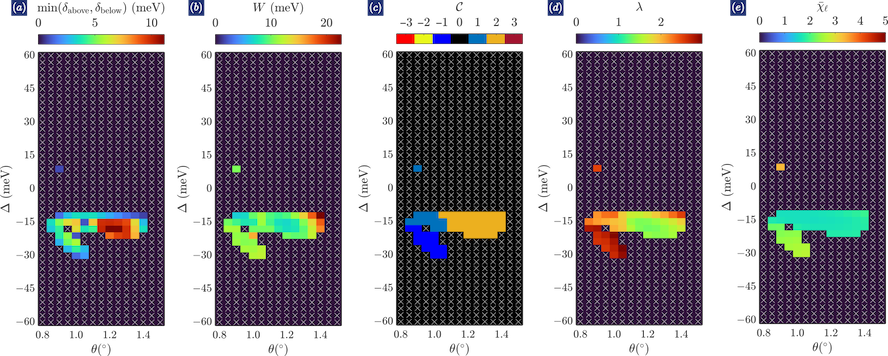}
    \caption{\textbf{Further spectral and topological characterization of anomalous Chern phases in Fig. \ref{fig:HF_antiparallel_relaxation_e12_largevF_phase_diagram}.} For anomalous Chern phases with an isolated Chern band, we show following properties of that isolated band: spectral gap (a), bandwidth (b), Chern number (c), trace condition violation (d), and average layer polarization (e).}
    \label{fig:HF_antiparallel_relaxation_e12_largevF_phase_diagram_further_characterization}
\end{figure}

\begin{figure}[t!]
    \centering
    \includegraphics[width=0.8\linewidth]{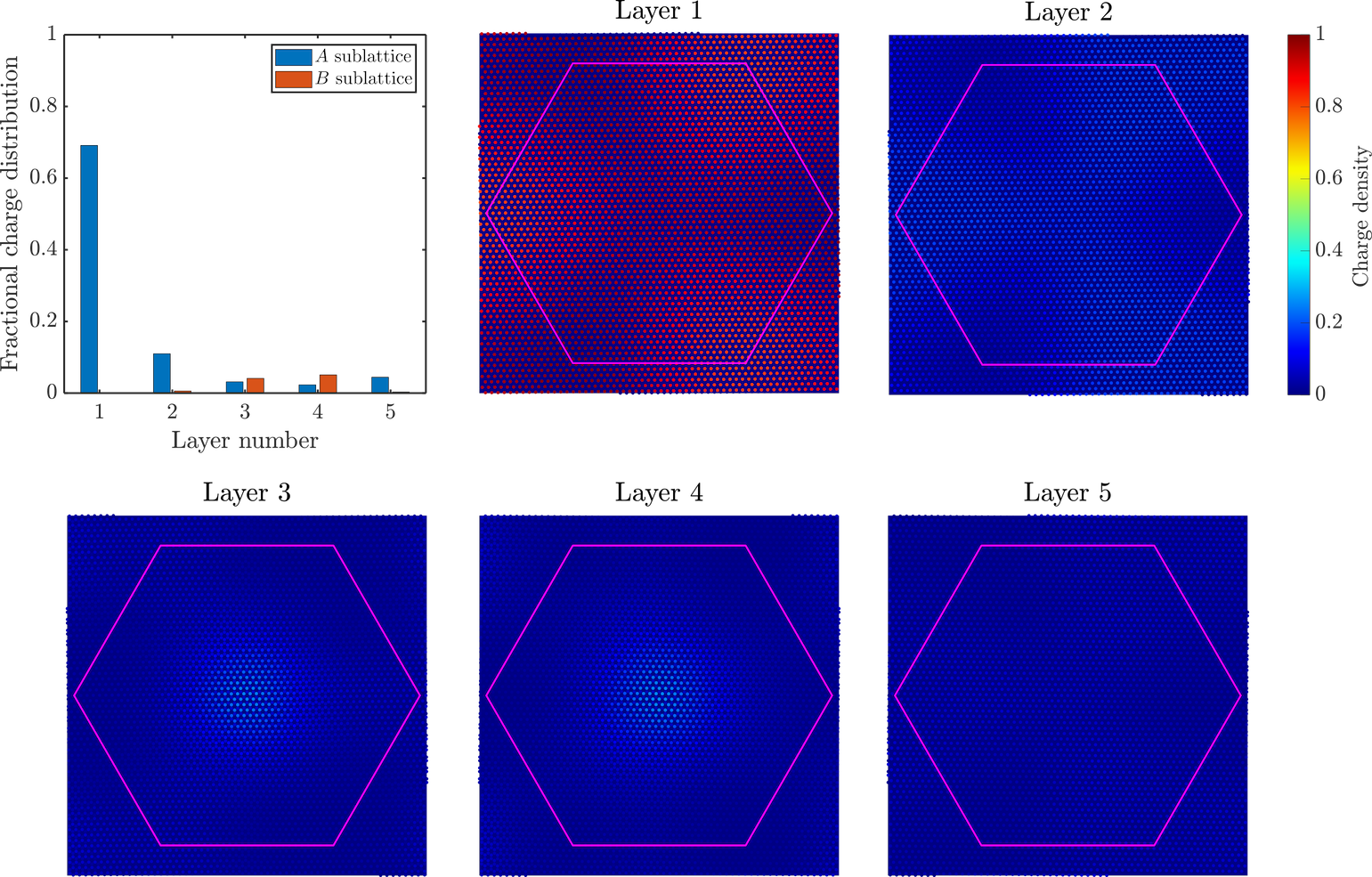}
    \caption{\textbf{Charge density at $\theta = 1.20^\circ$ and $\Delta = -18$ meV in Fig. \ref{fig:HF_antiparallel_relaxation_e12_largevF_phase_diagram}.} In the bar graph, we plot the fractional charge distribution on each layer and sublattice summed over all $\mathbf{k}$ in the narrow band in Fig. \ref{fig:HF_antiparallel_relaxation_e12_largevF_phase_diagram}(e), demonstrating a majority population on layer 1. In the density plots, we show the real-space distribution of the summed charge density for that same band. Most of the charge density is concentrated on layer 1. There is slightly more charge concentrated at the $\alpha$ and $\beta$ regions compared to at other regions. If one carefully inspects these density plots, one can discern the sublattice polarization on each layer. }
    \label{fig:charge_density_batch1}
\end{figure}

We first study the phase diagrams simulated using {\fontfamily{cmtt}\selectfont Param\_one}, which uses a large Fermi velocity $\gamma_0 = -3100$ meV and unequal interlayer hoppings $\lbrace \gamma_\mathrm{AA},\gamma_\mathrm{AB}\rbrace = \lbrace 100,120\rbrace$ meV to include some effects of lattice relaxation. The phase diagram for $\epsilon_r = 12$ is shown in Fig. \ref{fig:HF_antiparallel_relaxation_e12_largevF_phase_diagram}(a). The phase diagram is unsurprisingly complex, featuring all of the phases aforementioned, including metallic and insulating phases, some of which preserve $\mathcal{T}$-symmetry and some of which break $\mathcal{T}$-symmetry spontaneously. In \ref{fig:HF_antiparallel_relaxation_e12_largevF_phase_diagram}(b,c,d), we show a representative band structure, polarized by isospin, for an anomalous Chern insulator without a gap below, a $\mathcal{T}$-preserving metal, and a $\mathcal{T}$-breaking metal respectively. These three phases are not the focus of our study; so we do not characterize these phases further. Moving onto the Chern insulating phases where there is an isolated Chern band right below Fermi level $E_F,$ we find that for the antiparallel configuration reported here, there are three different phases with $\mathcal{C} = \pm 1$ and $\mathcal{C}=2.$ For these phases, we show the band structures both polarized by isospin and by layer index in \ref{fig:HF_antiparallel_relaxation_e12_largevF_phase_diagram}(e,f,g). It is clear that all of these phases have average electron densities primarily localized on the trilayer substack. We shall characterize the lateral distribution of the electron density later. For each of the Chern phases, we also show the associated Berry curvature and quantum metric distribution in $\mathbf{k}$-space for the representative values of $(\theta,\Delta).$ Compared to the $|\mathcal{C}|=1$ phase, the $\mathcal{C}=2$ phase has smaller $\lambda,$ which is favorable for the emergence of fractionalized phases. For this phase, we observe in Fig. \ref{fig:HF_antiparallel_relaxation_e12_largevF_phase_diagram}(e) that the Chern band, chosen for the optimal $\mathfrak{f.o.m.},$ is relatively narrow with Berry curvature localized around the $\bar{\Gamma}$ point in momentum space.  For this state, we have $\lbrace\theta,\Delta,\delta_\mathrm{above},\delta_\mathrm{below},W,\lambda \rbrace = \lbrace 1.20^\circ,  -18 \text{ meV} ,  11.13 \text{ meV},  14.62 \text{ meV},  7.25\text{ meV},    1.32 \rbrace.$

Next, we focus on just the Chern phases with an isolated band. We further characterize these phases in Fig. \ref{fig:HF_antiparallel_relaxation_e12_largevF_phase_diagram_further_characterization}. Here, $\times$ decates $(\theta,\Delta)$ where other less interesting phases stabilize. For values of $(\theta,\Delta)$ where we have a spectrally-isolated Chern band, we show the band gap of that band in Fig. \ref{fig:HF_antiparallel_relaxation_e12_largevF_phase_diagram_further_characterization}(a), the bandwidth in Fig. \ref{fig:HF_antiparallel_relaxation_e12_largevF_phase_diagram_further_characterization}(b), the Chern number in Fig. \ref{fig:HF_antiparallel_relaxation_e12_largevF_phase_diagram_further_characterization}(c), the trace condition violation in Fig. \ref{fig:HF_antiparallel_relaxation_e12_largevF_phase_diagram_further_characterization}(d), and the band-average layer polarization in Fig. \ref{fig:HF_antiparallel_relaxation_e12_largevF_phase_diagram_further_characterization}(e). The band average layer polarization is defined as $\bar{\chi}_\ell=  \frac{1}{N_\mathbf{k}^2}\sum_{\mathbf{k}} \chi_\ell(\mathbf{k}),$ where $\chi_\ell(\mathbf{k})$ is the layer polarization for state $\mathbf{k}$ in the band of interest.    This is an average measure for the charge localization of an entire band, not just a single Bloch state, and thus provides a further characterization of the charge distribution. From Fig. \ref{fig:HF_antiparallel_relaxation_e12_largevF_phase_diagram_further_characterization}, we see that as a general trend, the $\mathcal{C}=2$ phase features smaller bandwidths, larger band gaps, and smaller trace condition violations compared to the $\mathcal{C}=\pm 1$ phases. Therefore, this phase is not only interesting because it contains a higher Chern number, but it is also interesting because it contains favorable spectral and quantum geometric properties for further strong electron-electron effects at partial filling.

To characterize the lateral charge distribution, we choose to focus only on the $\mathcal{C}=2$ phase. Using the value of $(\theta,\Delta)$ that yields optimal $\mathfrak{f.o.m.},$ we plot the charge density in real space summed over all states in the band of interest for all five layers, resolved down to the sublattices, in Fig. \ref{fig:charge_density_batch1}. First, we show the fractional share of the charge for each sublattice on each layer in the bar graph. As is clear, most of the charges are localized on the $A$ sublattice on layer 1. However, we see that there are non-zero, albeit much smaller, charge densities on all the other layers. This is reflected in the real-space charge distribution. On layer 1, we see a weak modulation of the charge density on the moir\'{e} scale showing slightly higher concentrations at the $\alpha$ and $\beta$ regions. The pattern on layer 2 is similar to the pattern on layer 1, though layer 2 has significantly less charge. On layers 3 and 4, the charge is concentrated at the $\alpha$ regions. On the basis of the charge density summed over states in the active band, we can conclude that this is a Hall crystal because it features a charge density that not only occupies layers unequally but also fluctuates laterally on the moir\'{e} scale. However, we should note that the fluctuation is only mild. Also, we have not summed all states in the other valence bands, which if done, may change the charge density profile.

\begin{figure}[t!]
\includegraphics[width=0.90\linewidth]{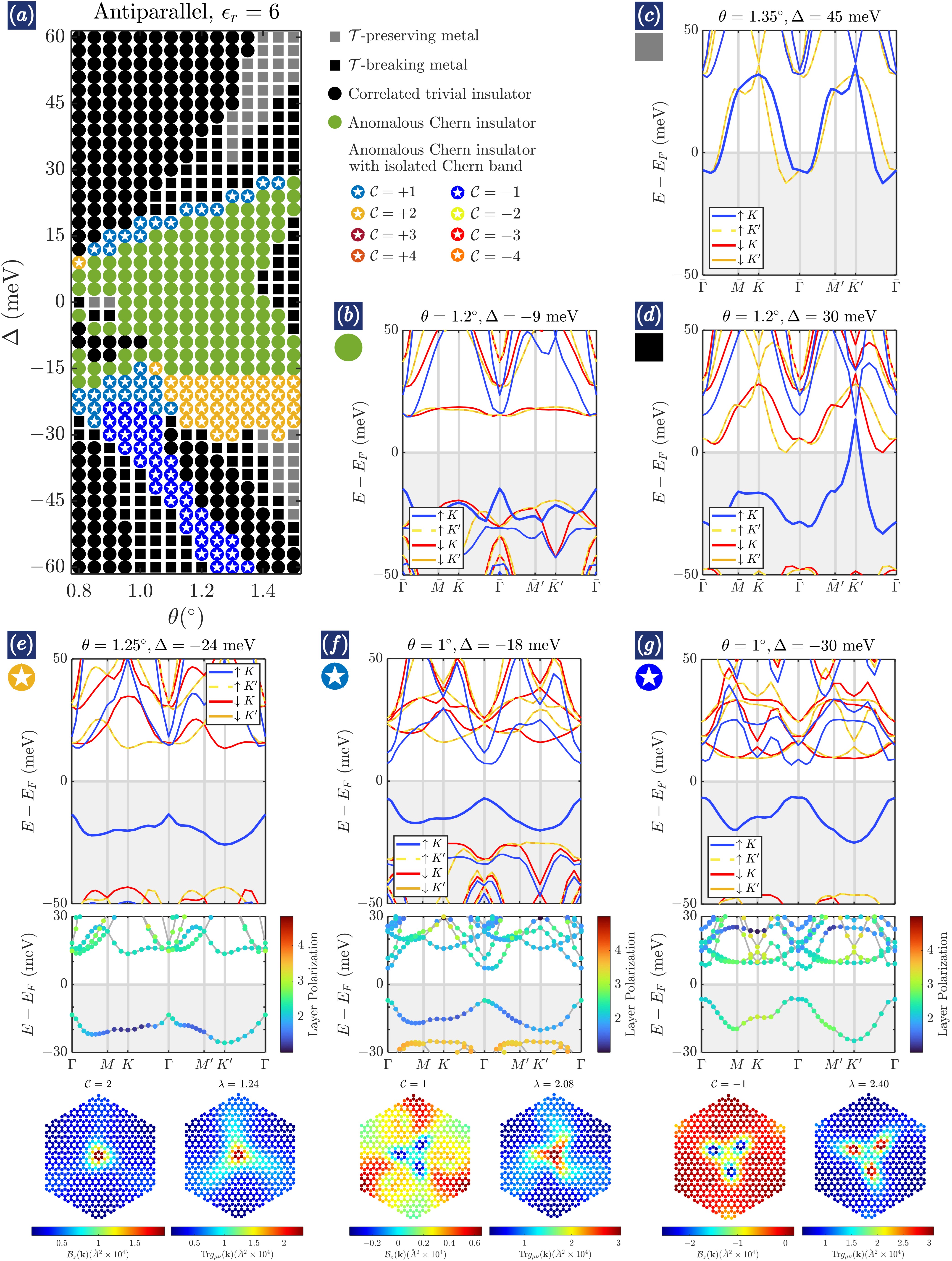}
\caption{\textbf{Mean-field phase diagram for the antiparallel configuration with lattice relaxation and $\epsilon_r = 6.$} (a) Phase diagram as a function of $\theta$ and $\Delta$. The results here are simulated using parameters from {\fontfamily{cmtt}\selectfont Param\_one}. Representative band structures of the different phases are shown in (b)-(g): an anomalous Chern insulator with no gap below the band closest to chemical potential  (b), a $\mathcal{T}$-preserving metal (c), a $\mathcal{T}$-breaking metal (d), and an anomalous Chern insulator with isolated Chern band with $\mathcal{C} = +2$ (e), $\mathcal{C} = +1$ (f), or $\mathcal{C} = -1$ (g). For (b)-(d), we only show the isospin polarization of the energy bands. For (e)-(g), we additionally color code the layer polarization of the Bloch states; furthermore, we show the corresponding Berry curvature and quantum metric of the band closest to chemical potential. }
\label{fig:HF_antiparallel_relaxation_e6_largevF_phase_diagram}
\end{figure}

\begin{figure}[t!]
    \centering
    \includegraphics[width=1\linewidth]{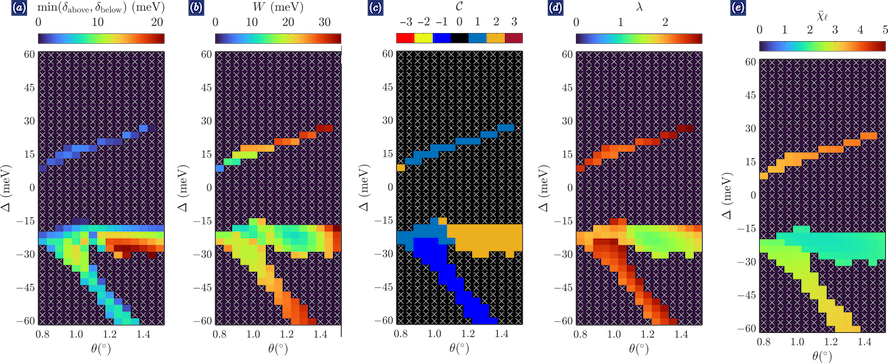}
    \caption{\textbf{Further spectral and topological characterization of anomalous Chern phases in Fig. \ref{fig:HF_antiparallel_relaxation_e6_largevF_phase_diagram}.} For anomalous Chern phases with an isolated Chern band, we show following properties of that isolated band: spectral gap (a), bandwidth (b), Chern number (c), trace condition violation (d), and average layer polarization (e).}
    \label{fig:HF_antiparallel_relaxation_e6_largevF_phase_diagram_further_characterization}
\end{figure}

\begin{figure}[t!]
    \centering
    \includegraphics[width=0.8\linewidth]{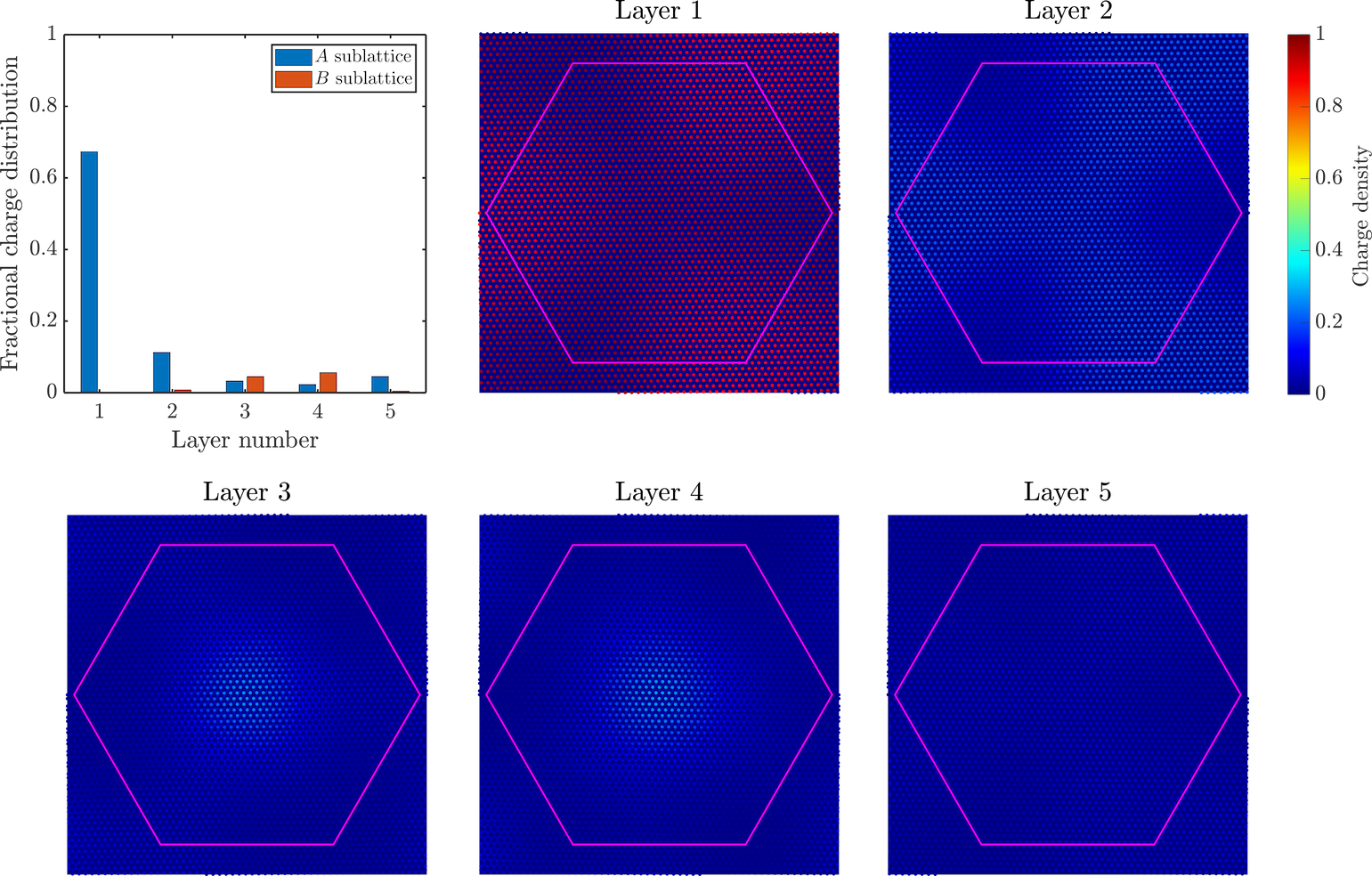}
    \caption{\textbf{Charge density at $\theta = 1.25^\circ$ and $\Delta = -24$ meV in Fig. \ref{fig:HF_antiparallel_relaxation_e6_largevF_phase_diagram}.} In the bar graph, we plot the fractional charge distribution on each layer and sublattice summed over all $\mathbf{k}$ in the narrow band in Fig. \ref{fig:HF_antiparallel_relaxation_e6_largevF_phase_diagram}(e), demonstrating a majority population on layer 1. In the density plots, we show the real-space distribution of the summed charge density for that same band. Most of the charge density is on layer 1. $\alpha$ and $\beta$ regions host slightly more charge density than other regions. We note that this figure is nearly identical to Fig. \ref{fig:charge_density_batch1}, demonstrating that at least for the $\mathcal{C} = 2$ phase of interest, results simulated with $\epsilon_r=6$ and $\epsilon_r=12$ share the same charge density profile. }
    \label{fig:charge_density_batch2}
\end{figure}

We now move onto the phase diagram simulated using {\fontfamily{cmtt}\selectfont Param\_one} with $\epsilon_r=6.$ The results are shown in Figs. \ref{fig:HF_antiparallel_relaxation_e6_largevF_phase_diagram},\ref{fig:HF_antiparallel_relaxation_e6_largevF_phase_diagram_further_characterization}, and \ref{fig:charge_density_batch2} analogous to Figs. \ref{fig:HF_antiparallel_relaxation_e12_largevF_phase_diagram},\ref{fig:HF_antiparallel_relaxation_e12_largevF_phase_diagram_further_characterization}, and \ref{fig:charge_density_batch1} for $\epsilon_r=12.$ In the present case, because $\epsilon_r=6$ is smaller, Coulomb interactions are twice as strong. Not surprisingly then, we observe in the phase diagram of Fig. \ref{fig:HF_antiparallel_relaxation_e6_largevF_phase_diagram} that symmetry-broken phases dominate phase space. The regions of Chern phases are enlarged compared to the simulations with $\epsilon_r = 12.$ However, all qualitative features remain the same as before, suggesting that our conclusions are robust in a range of electron-electron interactions. Here, the optimal $\mathfrak{f.o.m.}$ occurs $\lbrace\theta,\Delta,\delta_\mathrm{above},\delta_\mathrm{below},W,\lambda \rbrace = \lbrace 1.25^\circ,  -24 \text{ meV},   26.94 \text{ meV},   17.37 \text{ meV},   12.36 \text{ meV},    1.24 \rbrace.$ The optimal trace condition violation here is slightly better than for $\epsilon_r = 12.$ The charge density distribution shown in Fig. \ref{fig:charge_density_batch2} is nearly indistinguishable from the charge distribution shown in Fig. \ref{fig:charge_density_batch1} (though they are not identical as the reader can attest upon careful inspection).

\subsubsection{Without Lattice Relaxation}

\begin{figure}[t!]
\includegraphics[width=1\linewidth]{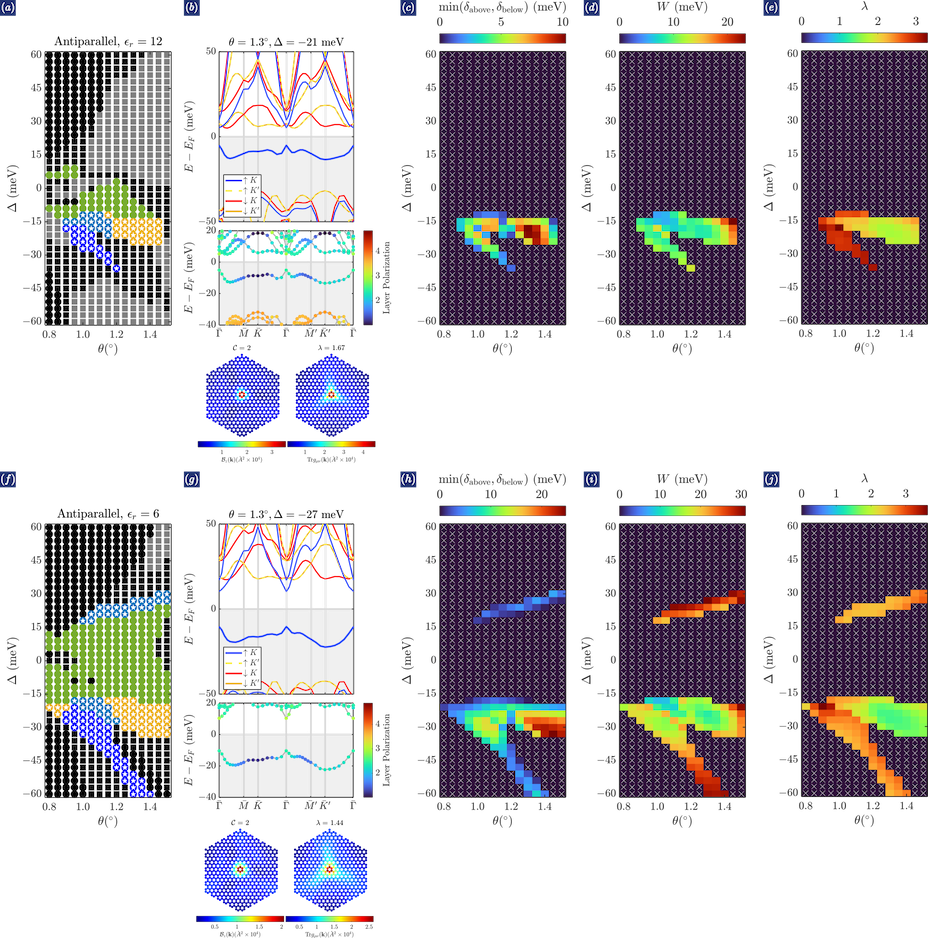}
\caption{\textbf{Mean-field phase diagram for the antiparallel configuration without lattice relaxation.} We show the phase diagram and its further characterization for $\epsilon_r=12$ in the top panel and for $\epsilon_r=6$ in the bottom panel.  The results here are simulated using parameters from {\fontfamily{cmtt}\selectfont Param\_two}. In each panel, we show the phase diagram in $(\theta,\Delta)$ space, the band structures and associated Berry curvature and quantum metric distribution for an optimal $\mathfrak{f.o.m.}$ phase point, and further characterization of the $\mathcal{C}=2$ phase including the band gap, bandwidth, and trace condition violation. }
\label{fig:antiparallel_no_relaxation}
\end{figure}

In this section, we analyze the phase diagrams simulated with {\fontfamily{cmtt}\selectfont Param\_two} where $\lbrace \gamma_\mathrm{AA},\gamma_\mathrm{AB}\rbrace = \lbrace 120,120 \rbrace$ meV. This batch of results, shown in Fig. \ref{fig:antiparallel_no_relaxation}, represents a twisted structure without any lattice relaxation. It is an idealization that we do \textit{not} expect an experimental platform to respect. It is nonetheless good to analyze this case to assess the stability of our results. In brief, all relevant trends observed with lattice relaxation are also observed without, again demonstrating robustness of our results. In more details, the top panel of Fig. \ref{fig:antiparallel_no_relaxation} shows results for $\epsilon_r=12$ while the bottom panel shows results for $\epsilon_r = 6.$ In both cases, we observe a prominent $\mathcal{C}=2$ phase at negative $\Delta$ over a range of angles. Increasing Coulomb interactions enlarges the phase-space area of this phase. For $\epsilon_r=12,$ optimal $\mathfrak{f.o.m.}$ occurs at $\lbrace\theta,\Delta,\delta_\mathrm{above},\delta_\mathrm{below},W,\lambda \rbrace = \lbrace 1.30^\circ,  -21 \text{ meV},   10.14 \text{ meV},   18.33 \text{ meV},    8.37 \text{ meV},    1.67 \rbrace.$ For $\epsilon_r = 6,$ optimal $\mathfrak{f.o.m.}$ occurs at $\lbrace\theta,\Delta,\delta_\mathrm{above},\delta_\mathrm{below},W,\lambda \rbrace = \lbrace 1.30^\circ, -27 \text{ meV},  20.68 \text{ meV},   19.40 \text{ meV},   12.02 \text{ meV},    1.44 \rbrace.$ The band structures for these optimal states are shown in Fig. \ref{fig:antiparallel_no_relaxation}(b,g). We notice that the trace condition violations here are generally larger than the results with lattice relaxation. Here, we also observe that the $\mathcal{C}=2$ phase exhibits enhanced Berry curvature at the $\bar{\Gamma}$ point in momentum space.  Also, optimal angles are pushed higher without lattice relaxation compared to with lattice relaxation. Therefore, it appears that lattice relaxation is favorable to the stabilization of the $\mathcal{C}=2$ phase we seek.

\subsubsection{With Reduced Fermi Velocity $\gamma_0$}

\begin{figure}[t!]
\includegraphics[width=1\linewidth]{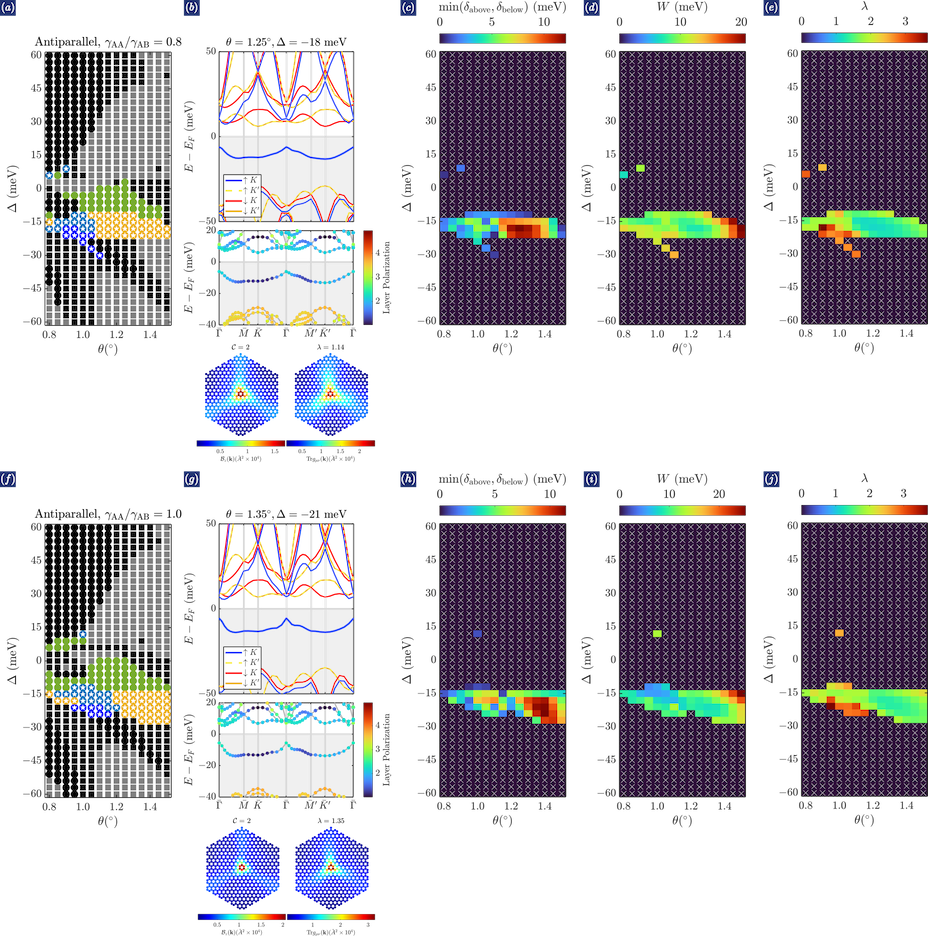}
\caption{\textbf{Mean-field phase diagram for the antiparallel configuration with reduced $\gamma_0$.} We show the phase diagram and its further characterization with relaxation ({\fontfamily{cmtt}\selectfont Param\_three}) in the top panel and without relaxation ({\fontfamily{cmtt}\selectfont Param\_four}) in the bottom panel. In each panel, we show the phase diagram in $(\theta,\Delta)$ space, the band structures and associated Berry curvature and quantum metric distribution for an optimal $\mathfrak{f.o.m.}$ phase point, and further characterization of the $\mathcal{C}=2$ phase including the band gap, bandwidth, and trace condition violation. Here, $\epsilon_r = 12.$ }
\label{fig:antiparallel_reduced_vF}
\end{figure}

Here, we study the phase diagrams generated with parameters from {\fontfamily{cmtt}\selectfont Param\_three} and {\fontfamily{cmtt}\selectfont Param\_four}. The Fermi velocity has been reduced to $\gamma_0=-2700$ meV. The results with lattice relaxation $\lbrace \gamma_\mathrm{AA},\gamma_\mathrm{AB}\rbrace = \lbrace 80,100 \rbrace$ meV are shown in the top panel of Fig. \ref{fig:antiparallel_reduced_vF}, while the results without lattice relaxation $\lbrace \gamma_\mathrm{AA},\gamma_\mathrm{AB}\rbrace = \lbrace 100,100 \rbrace$ meV are shown in the bottom panel of Fig. \ref{fig:antiparallel_reduced_vF}. For those sets of simulations, we set $\epsilon_r = 12.$ We observe $\mathcal{C} = 2$ phases in both scenarios. We see that lattice relaxation pushes the range of angles where this phase is observed to higher values. With lattice relaxation, optimal $\mathfrak{f.o.m.}$ occurs at $\lbrace\theta,\Delta,\delta_\mathrm{above},\delta_\mathrm{below},W,\lambda \rbrace = \lbrace 1.25^\circ,  -18 \text{ meV},   12.09 \text{ meV},   15.76 \text{ meV},    7.13 \text{ meV} ,   1.14 \rbrace.$ Without lattice relaxation, optimal $\mathfrak{f.o.m.}$ occurs at $\lbrace\theta,\Delta,\delta_\mathrm{above},\delta_\mathrm{below},W,\lambda \rbrace = \lbrace 1.35^\circ,  -21 \text{ meV},   11.10 \text{ meV},   20.64 \text{ meV},    8.43 \text{ meV},    1.35 \rbrace.$ We again observe concentration of the Berry curvature at the zone center for the $\mathcal{C} = 2$ states, consistent with previous results using different sets of parameters.

\subsection{Parallel Stacking Order}

In this section, we study the phase diagrams of the parallel stacking order using the four parameter sets listed in Table \ref{tab:hopping parameters}. This stacking order is richer than the antiparallel stacking order because it features $\mathcal{C}=2$ and $\mathcal{C}=3$ phases. As we shall show, it is also more sensitive to parameter choice than the previous analysis.

\subsubsection{With Some Lattice Relaxation}

\begin{figure}[t!]
\includegraphics[width=0.85\linewidth]{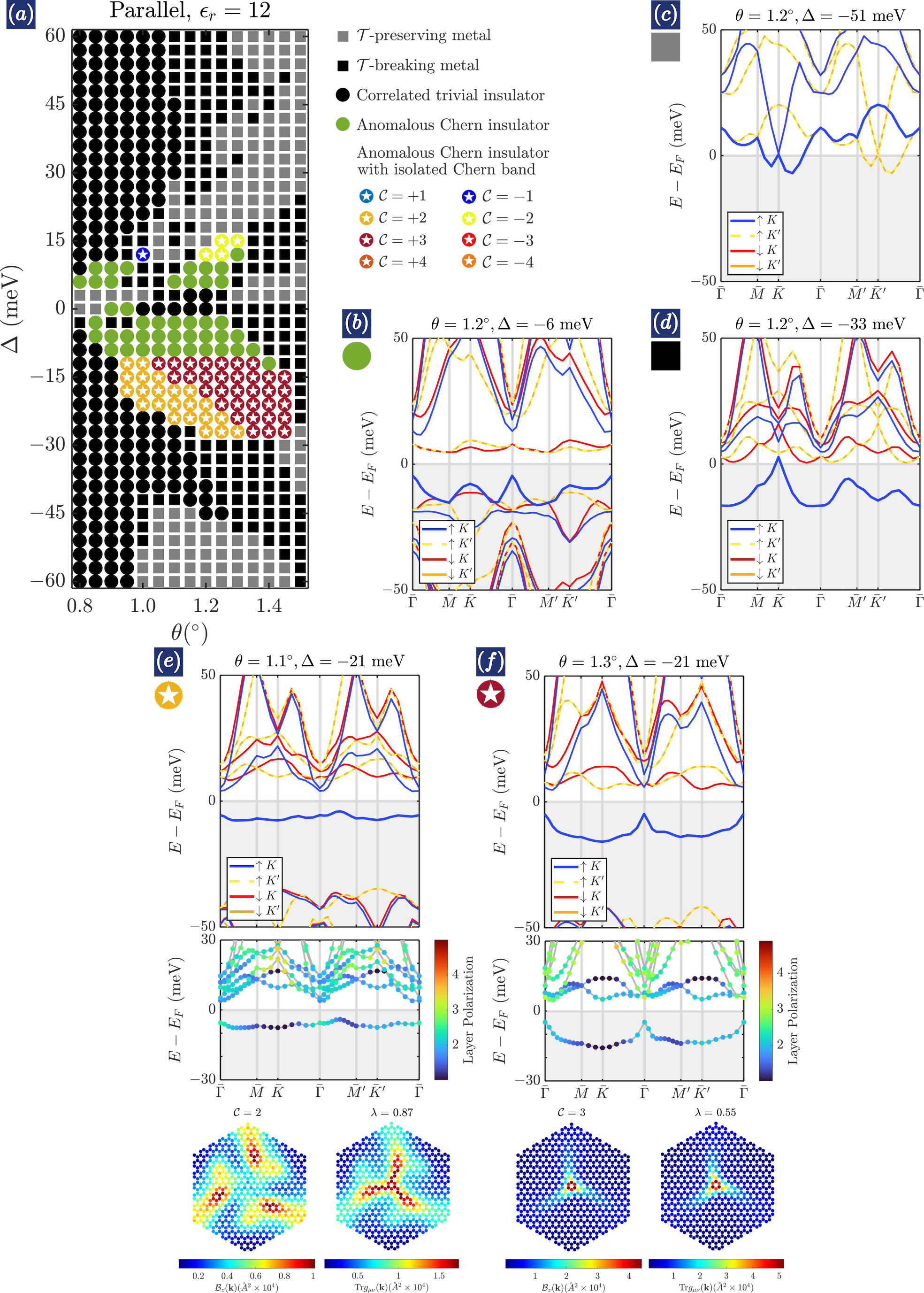}
\caption{\textbf{Mean-field phase diagram for the parallel configuration with lattice relaxation and $\epsilon_r = 12.$} (a) Phase diagram as a function of $\theta$ and $\Delta$. The results here are simulated using parameters from {\fontfamily{cmtt}\selectfont Param\_one}. Representative band structures of the different phases are shown in (b)-(f): an anomalous Chern insulator with no gap below the band closest to chemical potential  (b), a $\mathcal{T}$-preserving metal (c), a $\mathcal{T}$-breaking metal (d), and an anomalous Chern insulator with isolated Chern band with $\mathcal{C} = +2$ (e) or $\mathcal{C} = +3$ (f). For (b)-(d), we only show the isospin polarization of the energy bands. For (e)-(f), we additionally color code the layer polarization of the Bloch states; furthermore, we show the corresponding Berry curvature and quantum metric of the band closest to chemical potential. }
\label{fig:HF_parallel_relaxation_e12_largevF_phase_diagram}
\end{figure}

Using parameters from {\fontfamily{cmtt}\selectfont Param\_one}, we simulate the results shown in Fig. \ref{fig:HF_parallel_relaxation_e12_largevF_phase_diagram} with $\epsilon_r = 12.$ Similar to the antiparallel case, we observe incredible complexity in the evolution of different phases in Fig. \ref{fig:HF_parallel_relaxation_e12_largevF_phase_diagram}(a). This is because the band structure for this system is quite involved. In Fig. \ref{fig:HF_parallel_relaxation_e12_largevF_phase_diagram}(b,c,d), we show an anomalous Chern insulator without an isolator band, a $\mathcal{T}$-preserving metal, and a $\mathcal{T}$-breaking metal respectively.  Unlike in the antiparallel configuration, we observe two prominent higher-Chern phases: $\mathcal{C}=2$ and $\mathcal{C} = 3$. The $\mathcal{C}=2$ phase occurs at lower angles than the $\mathcal{C}=3$ phase. Both phases roughly occupy the same areas in phase space. The optimal $\mathcal{C}=2$ state occurs at $\lbrace\theta,\Delta,\delta_\mathrm{above},\delta_\mathrm{below},W,\lambda \rbrace = \lbrace 1.10^\circ,  -21 \text{ meV},    7.85 \text{ meV},   27.04 \text{ meV},    3.75 \text{ meV},    0.87 \rbrace.$ The band structure for this state is shown in Fig. \ref{fig:HF_parallel_relaxation_e12_largevF_phase_diagram}(e). We draw particular attention to the very narrow band right below Fermi level. The Berry curvature for this state is more delocalized throughout the mBZ compared to the other nontrivial states that we have shown. The trace condition violation here is $\lambda = 0.87.$ We also see that the Bloch states for the active band are localized mostly in the trilayer substack. For the $\mathcal{C}=3$ phase, the optimal state occurs at $\lbrace\theta,\Delta,\delta_\mathrm{above},\delta_\mathrm{below},W,\lambda \rbrace = \lbrace 1.30^\circ,  -21 \text{ meV},    9.54 \text{ meV},   25.55 \text{ meV},   11.04 \text{ meV},    0.55 \rbrace.$ The band structure for this state is shown in Fig. \ref{fig:HF_parallel_relaxation_e12_largevF_phase_diagram}(f). The band of interest has some dispersion after HF renormalization.  The Berry curvature is centered around the $\bar{\Gamma}$ point. The trace condition violation is especially small here, only around $\lambda = 0.55.$

\begin{figure}[t!]
    \centering
    \includegraphics[width=1\linewidth]{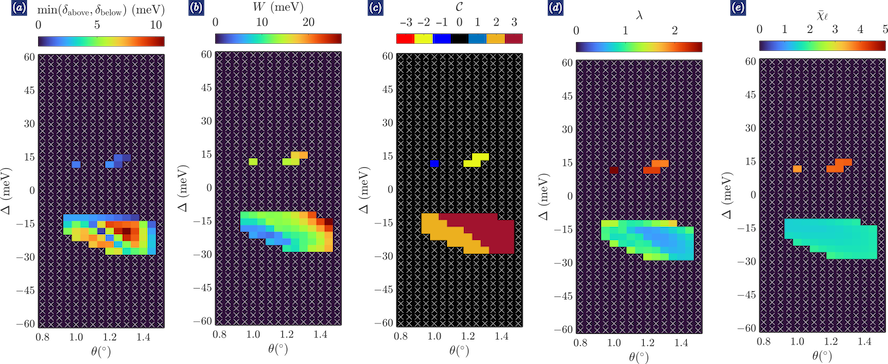}
    \caption{\textbf{Further spectral and topological characterization of anomalous Chern phases in Fig. \ref{fig:HF_parallel_relaxation_e12_largevF_phase_diagram}.} For anomalous Chern phases with an isolated Chern band, we show following properties of that isolated band: spectral gap (a), bandwidth (b), Chern number (c), trace condition violation (d), and average layer polarization (e).}
    \label{fig:HF_parallel_relaxation_e12_largevF_phase_diagram_further_characterization}
\end{figure}

Focusing on the higher-Chern-nontrivial phases, we show further spectral and topological characterization in Fig. \ref{fig:HF_parallel_relaxation_e12_largevF_phase_diagram_further_characterization}. Here, $\times$ decates $(\theta,\Delta)$ where other less interesting phases stabilize. For values of $(\theta,\Delta)$ where we have a spectrally-isolated Chern band, we show the band gap of that band in Fig. \ref{fig:HF_parallel_relaxation_e12_largevF_phase_diagram_further_characterization}(a), the bandwidth in Fig. \ref{fig:HF_parallel_relaxation_e12_largevF_phase_diagram_further_characterization}(b), the Chern number in Fig. \ref{fig:HF_parallel_relaxation_e12_largevF_phase_diagram_further_characterization}(c), the trace condition violation in Fig. \ref{fig:HF_parallel_relaxation_e12_largevF_phase_diagram_further_characterization}(d), and the band-average layer polarization in Fig. \ref{fig:HF_parallel_relaxation_e12_largevF_phase_diagram_further_characterization}(e). Some trends are evident, the $\mathcal{C}=2$ phase has smaller band gaps and bandwidths than the $\mathcal{C}=3$ phase. However, the $\mathcal{C}=2$ phase has large $\lambda$ than the the $\mathcal{C}=3$ phase. Both phases have average $\bar{\chi}_\ell \lesssim2,$ including that they both harbor majority charge density in the trilayer substack as opposed to the bilayer substack.

\begin{figure}[t!]
    \centering
    \includegraphics[width=0.8\linewidth]{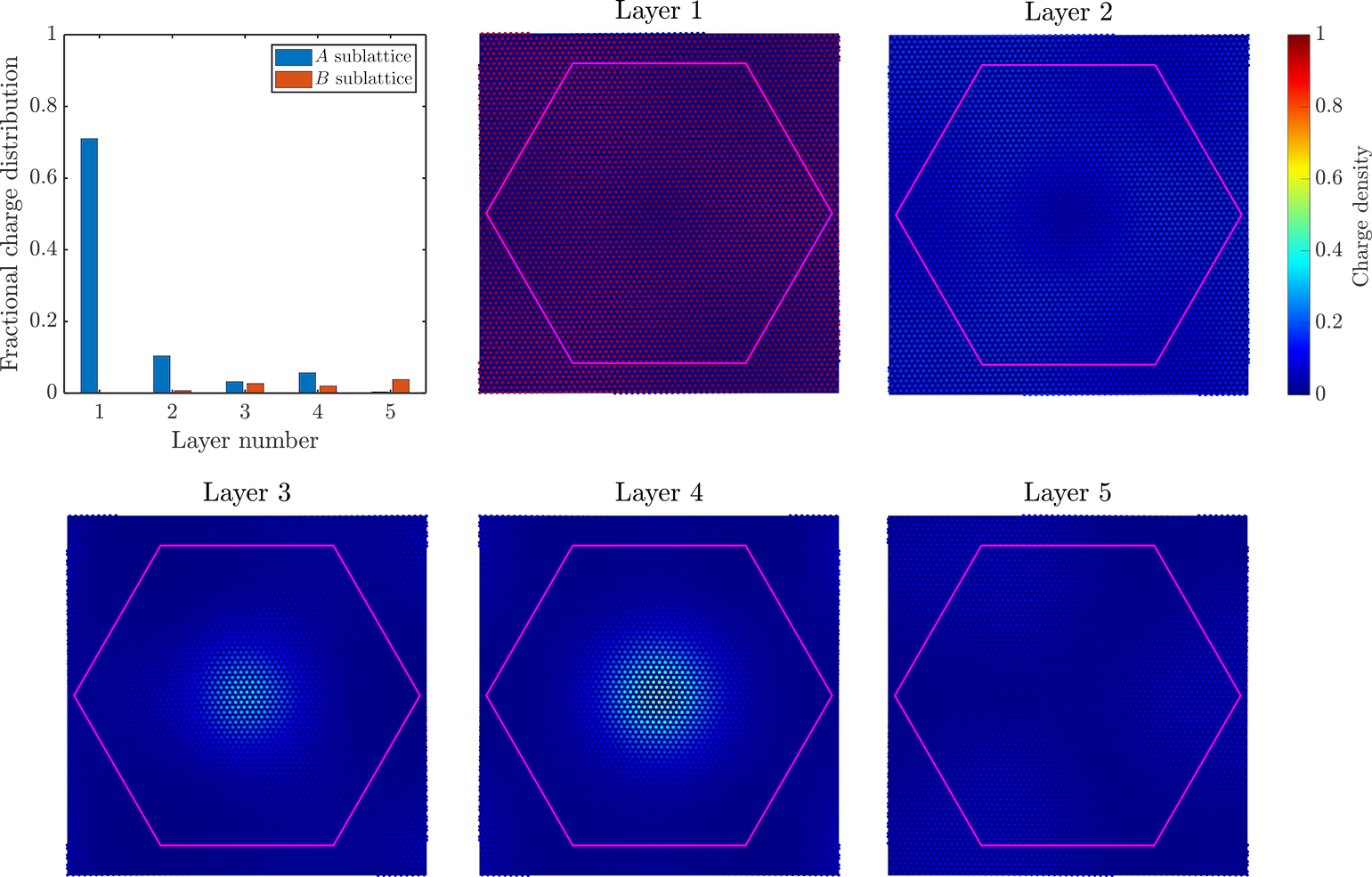}
    \caption{\textbf{Charge density at $\theta = 1.10^\circ$ and $\Delta = -21$ meV in Fig. \ref{fig:HF_parallel_relaxation_e12_largevF_phase_diagram}.} In the bar graph, we plot the fractional charge distribution on each layer and sublattice summed over all $\mathbf{k}$ in the narrow band in Fig. \ref{fig:HF_parallel_relaxation_e12_largevF_phase_diagram}(e), demonstrating a majority population on layer 1. In the density plots, we show the real-space distribution of the summed charge density for that same band. Most of the charge density is on layer 1, concentrated more at $\alpha$ and $\gamma$ regions than other regions. }
    \label{fig:charge_density_batch4}
\end{figure}

\begin{figure}[t!]
    \centering
    \includegraphics[width=0.8\linewidth]{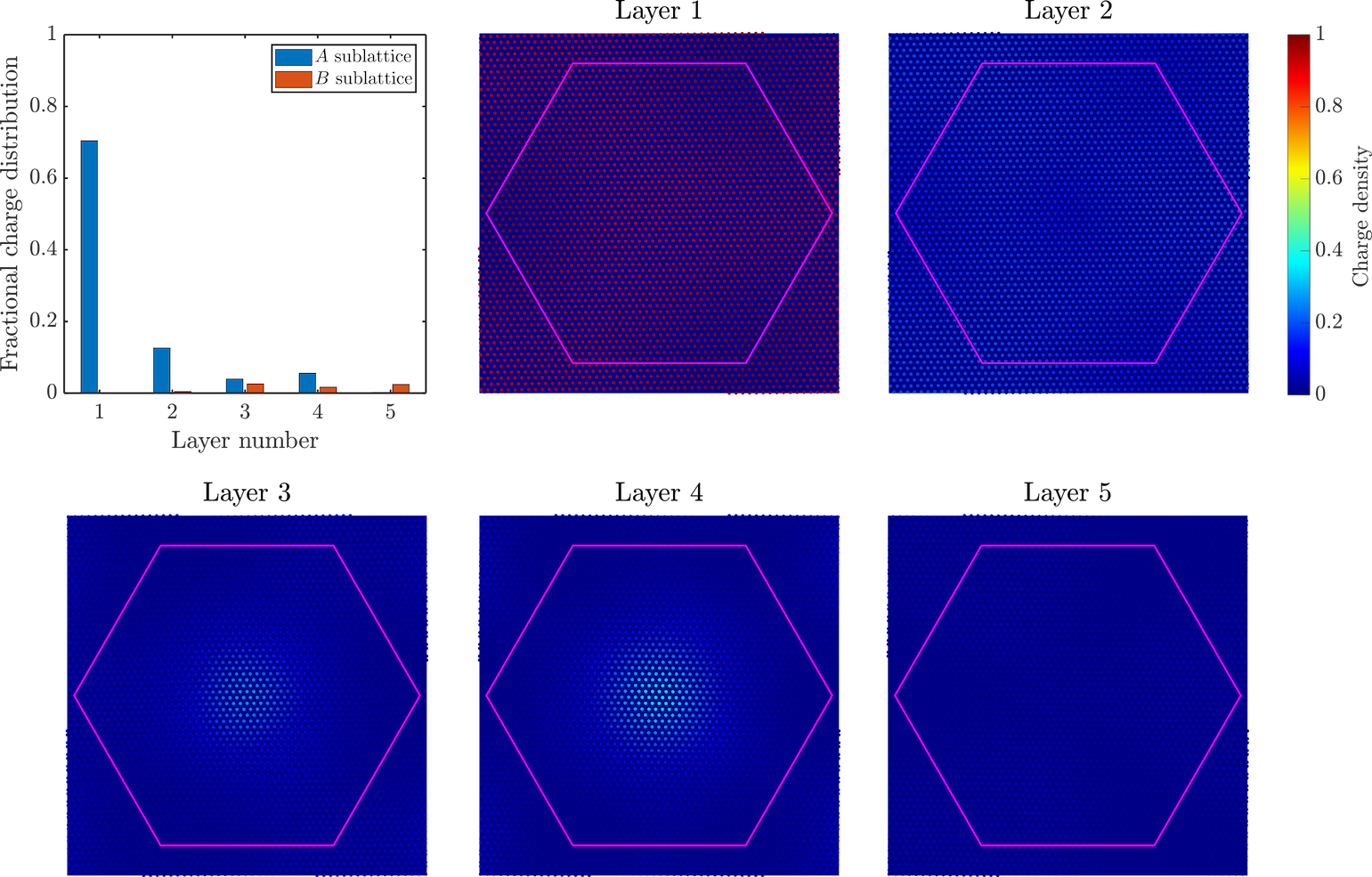}
    \caption{\textbf{Charge density at $\theta = 1.30^\circ$ and $\Delta = -21$ meV in Fig. \ref{fig:HF_parallel_relaxation_e12_largevF_phase_diagram}.} In the bar graph, we plot the fractional charge distribution on each layer and sublattice summed over all $\mathbf{k}$ in the narrow band in Fig. \ref{fig:HF_parallel_relaxation_e12_largevF_phase_diagram}(f), demonstrating a majority population on layer 1. In the density plots, we show the real-space distribution of the summed charge density for that same band. Most of the charge density is on layer 1, concentrated more at $\gamma$ regions than other regions. }
    \label{fig:charge_density_batch4_2}
\end{figure}

Moving onto the lateral charge distribution, we map the charge density on each of the five layers for the two optimal states aforementioned. In Fig. \ref{fig:charge_density_batch4}, we show the charge density for the $\mathcal{C}=2$ state at $\theta = 1.10^\circ$ and $\Delta = -21$ meV. From the bar graph, we see that most of the charge resides on layer 1 on the $A$ sublattice. There are smaller distributions on the other layers. Of note, on layer 5, the charge density is localized also exclusively on the $B$ sublattice. Laterally, the charge distribution on layer 1 is fairly uniform, with some inhomogeneity due to slightly higher concentrations at the $\alpha$ and $\gamma$ regions. On layer 1, the charge density is depleted at the $\alpha$ regions, while it is peaked at the $\alpha$ regions on layers 3 and 4. On layer 5, there is small charge density is localized at the $\beta$ regions. In Fig. \ref{fig:charge_density_batch4_2}, we show the charge density for the $\mathcal{C}=3$ state at $\theta = 1.30^\circ$ and $\Delta = -21$ meV. The charge density here is quite similar to that shown in Fig. \ref{fig:charge_density_batch4}. On layer 1, we see that charges coalesce slightly more at the $\gamma$ regions than other regions. On layer 2, the charge is fairly uniform. On layers 3 and 4, the charges localize at the $\alpha$ regions. On layer 5, there is very little charge, with inhomogeneity due to slight localization at the $\gamma$ regions. Indeed, in both phases, the charge is \textit{not} uniform across the layers and within each layer, featuring clear inhomogeneities. 

\begin{figure}[t!]
\includegraphics[width=0.85\linewidth]{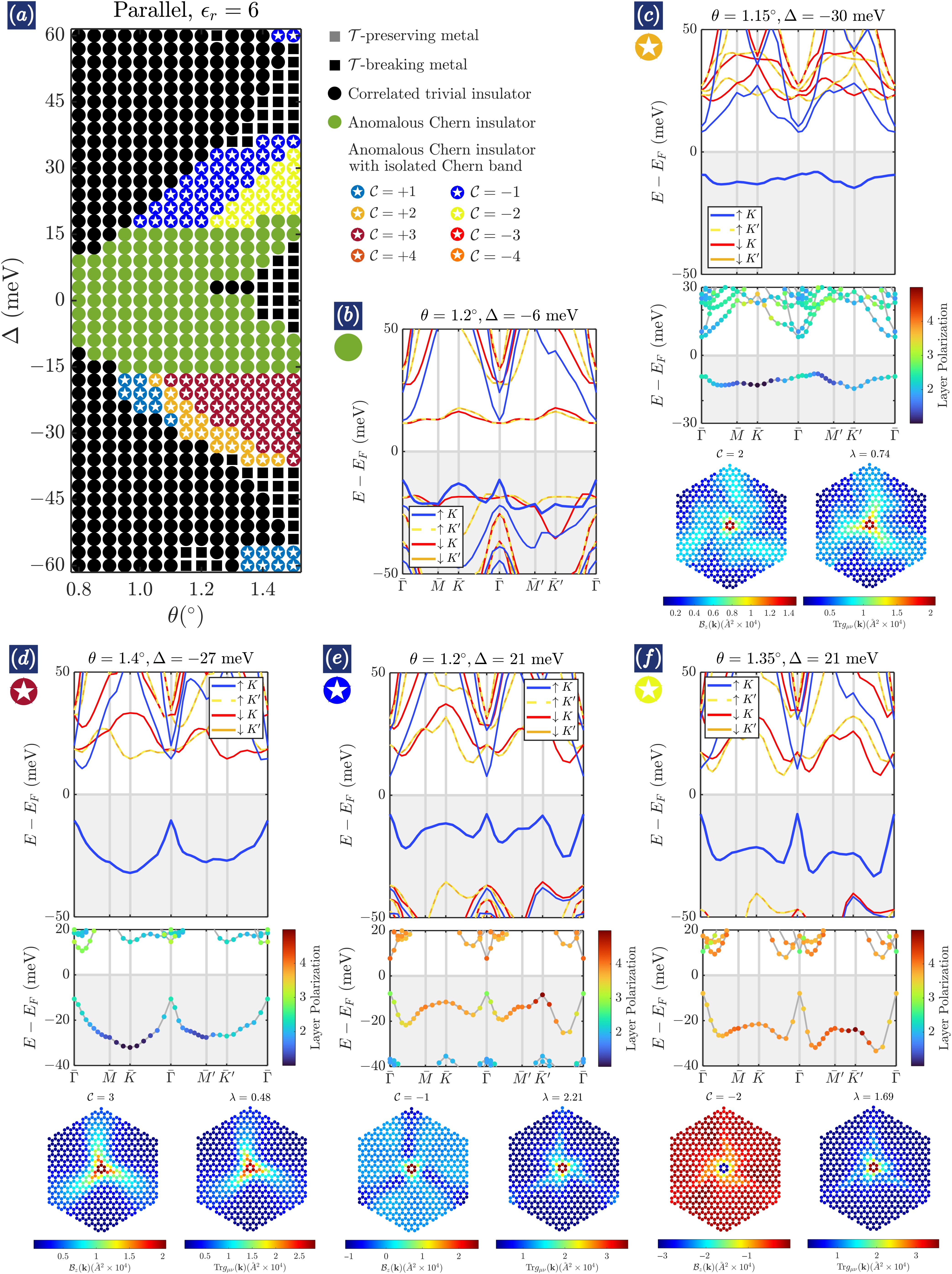}
\caption{\textbf{Mean-field phase diagram for the parallel configuration with lattice relaxation and $\epsilon_r = 6.$} (a) Phase diagram as a function of $\theta$ and $\Delta$. The results here are simulated using parameters from {\fontfamily{cmtt}\selectfont Param\_one}. Representative band structures of the different phases are shown in (b)-(f): an anomalous Chern insulator with no gap below the band closest to chemical potential  (b) and  an anomalous Chern insulator with isolated Chern band with $\mathcal{C} = +2$ (c), $\mathcal{C} = +3$ (d), $\mathcal{C} = -1$ (e), and $\mathcal{C} = -2$ (f). For (c)-(f), we color code the layer polarization of the Bloch states; furthermore, we show the corresponding Berry curvature and quantum metric of the band closest to chemical potential. }
\label{fig:HF_parallel_relaxation_e6_largevF_phase_diagram}
\end{figure}

\begin{figure}[t!]
    \centering
    \includegraphics[width=1\linewidth]{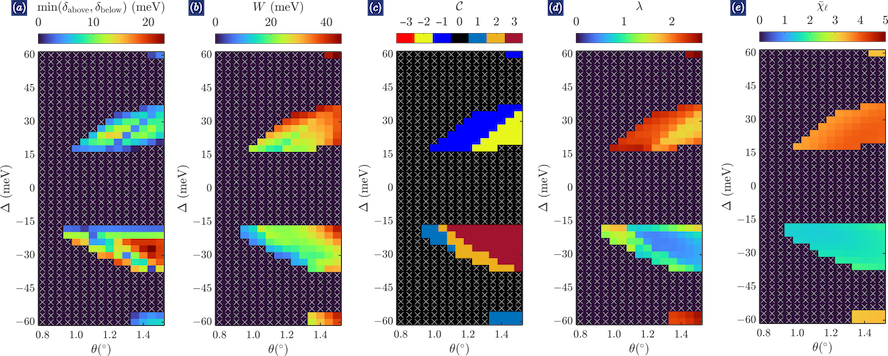}
    \caption{\textbf{Further spectral and topological characterization of anomalous Chern phases in Fig. \ref{fig:HF_parallel_relaxation_e6_largevF_phase_diagram}.} For anomalous Chern phases with an isolated Chern band, we show following properties of that isolated band: spectral gap (a), bandwidth (b), Chern number (c), trace condition violation (d), and average layer polarization (e).}
    \label{fig:HF_parallel_relaxation_e6_largevF_phase_diagram_further_characterization}
\end{figure}

\begin{figure}[t!]
    \centering
    \includegraphics[width=0.8\linewidth]{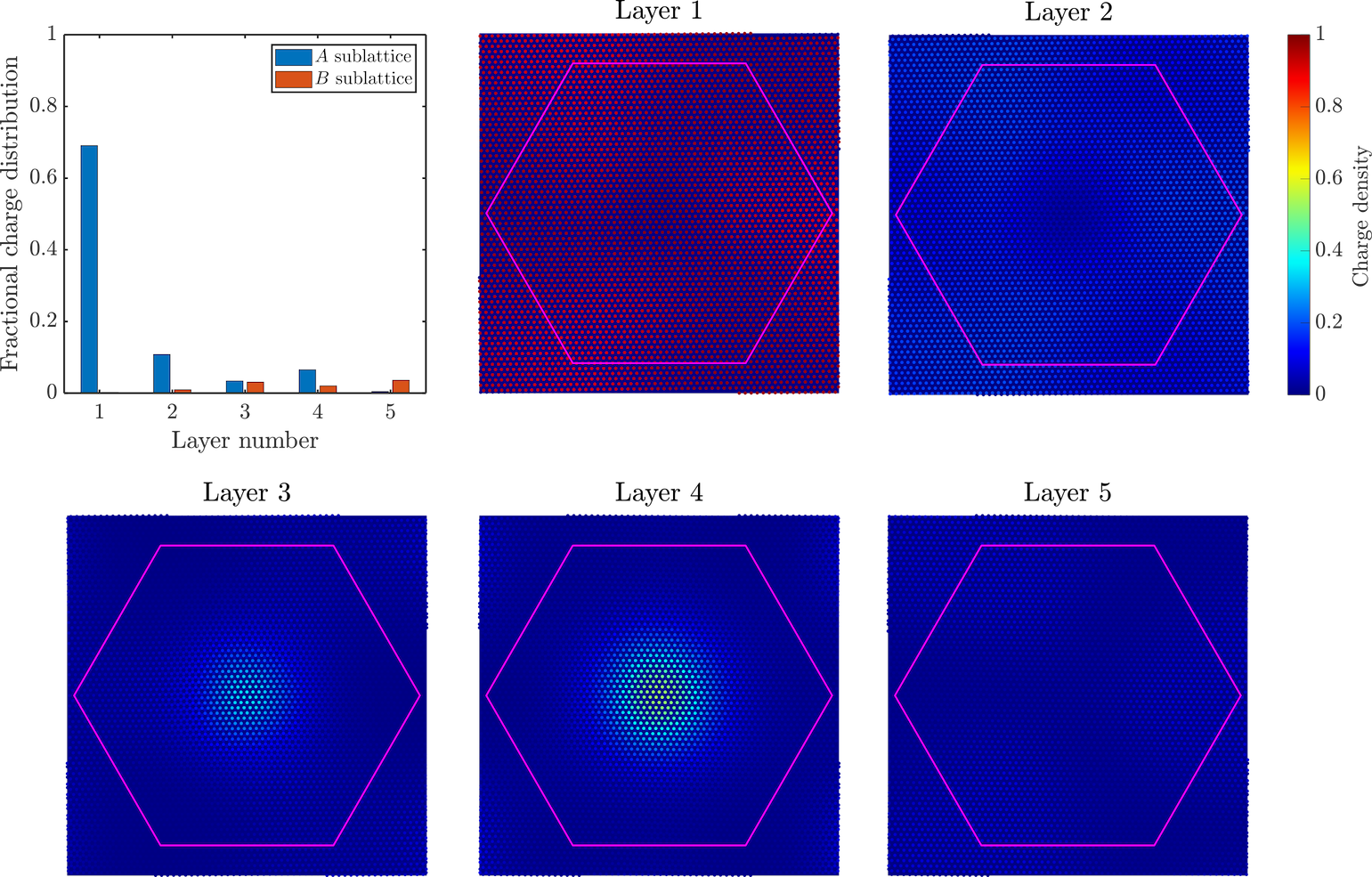}
    \caption{\textbf{Charge density at $\theta = 1.15^\circ$ and $\Delta = -30$ meV in Fig. \ref{fig:HF_parallel_relaxation_e6_largevF_phase_diagram}.} In the bar graph, we plot the fractional charge distribution on each layer and sublattice summed over all $\mathbf{k}$ in the narrow band in Fig. \ref{fig:HF_parallel_relaxation_e6_largevF_phase_diagram}(c), demonstrating a majority population on layer 1. In the density plots, we show the real-space distribution of the summed charge density for that same band. Most of the charge density is on layer 1, concentrated more at $\alpha$ and $\gamma$ regions than other regions. }
    \label{fig:charge_density_batch3}
\end{figure}

\begin{figure}[t!]
    \centering
    \includegraphics[width=0.8\linewidth]{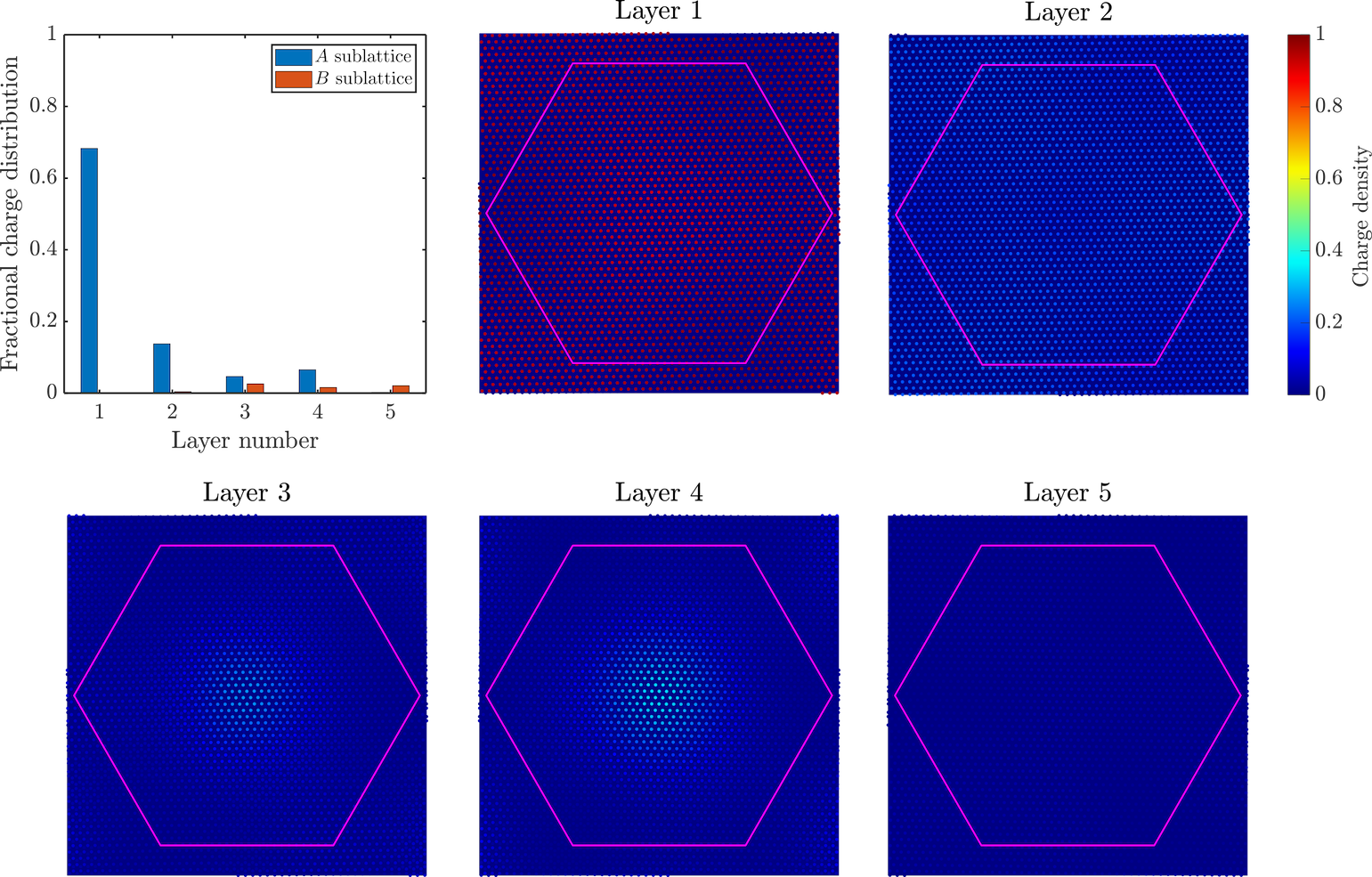}
    \caption{\textbf{Charge density at $\theta = 1.40^\circ$ and $\Delta = -27$ meV in Fig. \ref{fig:HF_parallel_relaxation_e6_largevF_phase_diagram}.} In the bar graph, we plot the fractional charge distribution on each layer and sublattice summed over all $\mathbf{k}$ in the narrow band in Fig. \ref{fig:HF_parallel_relaxation_e6_largevF_phase_diagram}(d), demonstrating a majority population on layer 1. In the density plots, we show the real-space distribution of the summed charge density for that same band. Most of the charge density is on layer 1, concentrated more at $\gamma$ regions than other regions. }
    \label{fig:charge_density_batch3_2}
\end{figure}

Next, we consider the phase diagram simulated with $\epsilon_r = 6$ for the parallel configuration using {\fontfamily{cmtt}\selectfont Param\_one}. The results are shown in Fig. \ref{fig:HF_parallel_relaxation_e6_largevF_phase_diagram}. In Fig. \ref{fig:HF_parallel_relaxation_e6_largevF_phase_diagram}(a), we notice that because Coulomb interactions are stronger, the spontaneously-formed phases outcompete the symmetric phase. Remarkably, we find four prominent Chern phases with $\mathcal{C}=-1,$$\mathcal{C}=-2,$ $\mathcal{C}=2,$ and $\mathcal{C}=3.$ The first two occur at positive $\Delta$ while the last two occur at negative $\Delta.$  The band structures for the $\mathcal{C}=-1$ and $\mathcal{C}=-2$ phases are shown in  Fig. \ref{fig:HF_parallel_relaxation_e6_largevF_phase_diagram}(e,f) respectively. For the $\mathcal{C}=-1$ phase, the trace condition violation is higher than the other phases; it is $\lambda=2.21$ for the present case. For the $\mathcal{C}=-2$ phase, the active band is quite dispersive with gaps to other bands that are not as large. These two phases are localized primarily, but not exclusively, on the bilayer substack. It is also worth noting that these two phases do not exist when $\epsilon_r = 12.$ So they are not stable to changing Coulomb interactions in the range of parameters of interest. For the $\mathcal{C}=2$ and $\mathcal{C}=3$ phases, we show the band structures with optimal $\mathfrak{f.o.m.}$ in Fig. \ref{fig:HF_parallel_relaxation_e6_largevF_phase_diagram}(c,d). For the $\mathcal{C}=2,$ we observe a  narrow isolated band with large gaps both above and below said band. The Berry curvature distribution and quantum metric do peak around the $\bar{\Gamma}$ point but also have non-negligible spread throughout the mBZ. The optimal $\mathfrak{f.o.m.}$ here is $\lbrace\theta,\Delta,\delta_\mathrm{above},\delta_\mathrm{below},W,\lambda \rbrace = \lbrace 1.15^\circ,  -30 \text{ meV},   16.33 \text{ meV},   40.88 \text{ meV},    6.51 \text{ meV},    0.74 \rbrace.$ For the $\mathcal{C}=3$ phase, the band is, again,  dispersive, but is is well isolated from the rest of the band structure. The optimal $\mathfrak{f.o.m.}$ here is $\lbrace\theta,\Delta,\delta_\mathrm{above},\delta_\mathrm{below},W,\lambda \rbrace = \lbrace  1.40^\circ, -27 \text{ meV},   21.22 \text{ meV},   30.13 \text{ meV},   21.44 \text{ meV},    0.48
 \rbrace.$ The Berry curvature here is also peaked at the $\bar{\Gamma}$ point. Further characterization of the Chern-nontrivial phases is shown in Fig. \ref{fig:HF_parallel_relaxation_e6_largevF_phase_diagram_further_characterization}. From there, we observe clearly at the Chern phases at positive $\Delta$ have comparatively larger trace condition violations and are localized primarily on the bilayer substack as diagonosed by $\bar{\chi}_\ell.$ For the Chern phases at negative $\Delta,$ we observe that the trace condition violation is generally smaller for the $\mathcal{C}=3$ phase compared to the $\mathcal{C}=2$ phase. The ratio of the bandwidth to band gap is smaller for the $\mathcal{C}=2$ phase than the $\mathcal{C}=3$ phase. The real-space charge densities for these two phases are shown in Figs. \ref{fig:charge_density_batch3} and \ref{fig:charge_density_batch3_2}. They are very similar to Figs. \ref{fig:charge_density_batch4} and \ref{fig:charge_density_batch4_2}. Therefore, the discussion about the latter figures apply to the former as well.

\subsubsection{Without Lattice Relaxation}

\begin{figure}[t!]
\includegraphics[width=1\linewidth]{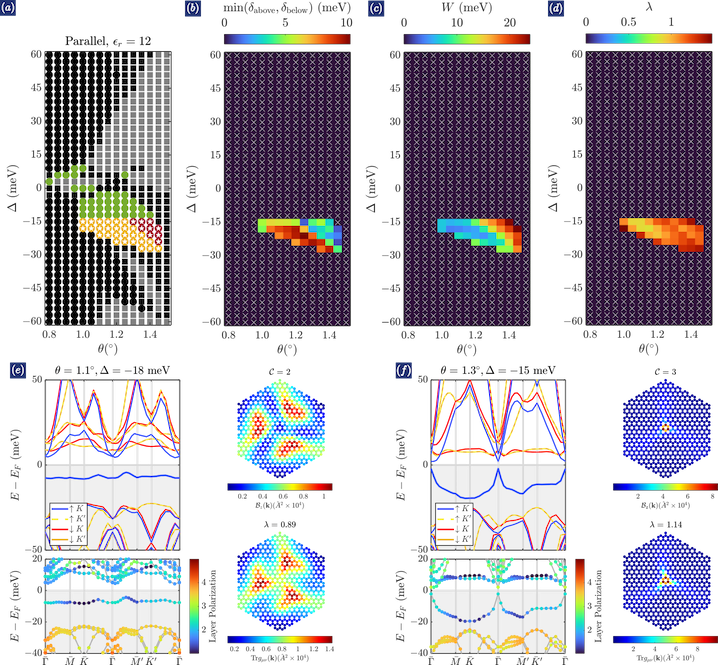}
\caption{\textbf{Mean-field phase diagram for the parallel configuration without lattice relaxation amd $\epsilon_r=12$.} The results here are simulated using {\fontfamily{cmtt}\selectfont Param\_two}. (a) Phase diagram and (b,c,d) its further characterization: band gap, bandwidth, and trace condition violation. (e) Band structure and associated Berry curvature and quantum metric distribution of an $\mathfrak{f.o.m.}$ $\mathcal{C}=2$ state. (f) Band structure and associated Berry curvature and quantum metric distribution of an $\mathfrak{f.o.m.}$ $\mathcal{C}=3$ state. }
\label{fig:parallel_no_relaxation}
\end{figure}

\begin{figure}[t!]
\includegraphics[width=1\linewidth]{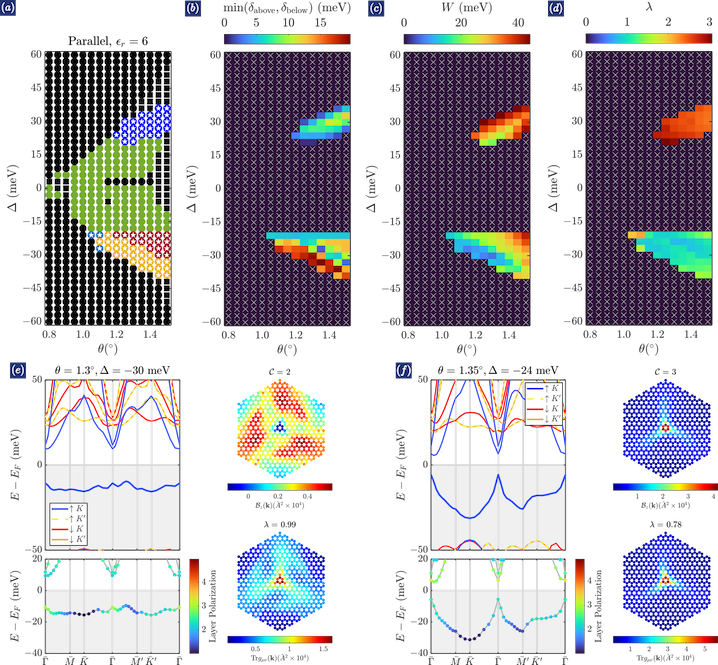}
\caption{\textbf{Mean-field phase diagram for the parallel configuration without lattice relaxation amd $\epsilon_r=6$.} The results here are simulated using {\fontfamily{cmtt}\selectfont Param\_two}. (a) Phase diagram and (b,c,d) its further characterization: band gap, bandwidth, and trace condition violation. (e) Band structure and associated Berry curvature and quantum metric distribution of an $\mathfrak{f.o.m.}$ $\mathcal{C}=2$ state. (f) Band structure and associated Berry curvature and quantum metric distribution of an $\mathfrak{f.o.m.}$ $\mathcal{C}=3$ state. }
\label{fig:parallel_no_relaxation_2}
\end{figure}

Here, we analyze the phase diagrams for the parallel configuration simulated using {\fontfamily{cmtt}\selectfont Param\_two} which sets $\lbrace \gamma_\mathrm{AA},\gamma_\mathrm{AB}\rbrace=\lbrace 120,120 \rbrace$ meV. This parameter set represents the absence of lattice relaxation. We first focus on $\epsilon_r = 12.$ In Fig. \ref{fig:parallel_no_relaxation}, we show the results for $\epsilon_r=12$ and, in Fig. \ref{fig:parallel_no_relaxation_2}, we show the results for $\epsilon_r=6.$ From Fig. \ref{fig:parallel_no_relaxation}(a), we observe that the $\mathcal{C}=3$ phase has been reduced significantly in size compared to the results with lattice relaxation. However, the $\mathcal{C}=2$ phase remains prominent without lattice relaxation. As before, the $\mathcal{C}=2$ phase features a relatively narrow and isolated band with favorable quantum geometry. In  Fig. \ref{fig:parallel_no_relaxation}(e), we show a representative band structure from the optimal state with $\mathfrak{f.o.m.}$ here is $\lbrace\theta,\Delta,\delta_\mathrm{above},\delta_\mathrm{below},W,\lambda \rbrace = \lbrace  1.10^\circ,  -18 \text{ meV},    8.59 \text{ meV},   14.32\text{ meV},    3.92\text{ meV},    0.89
 \rbrace.$ The $\mathcal{C}=3$ phase here does not look appealing here because it does not have as small a trace condition violation compared to the results with lattice relaxation. The band of interest is dispersive and is not well isolated from the rest of the band structure. Even the optimal state for this phase, shown in \ref{fig:parallel_no_relaxation}(f), does not appear optimal from the point of view of being conducive to further electron-electron interaction effects at partial doping. The optimal state here is $\lbrace\theta,\Delta,\delta_\mathrm{above},\delta_\mathrm{below},W,\lambda \rbrace = \lbrace   1.30^\circ,  -15 \text{ meV},   4.29  \text{ meV},    5.37  \text{ meV},   17.40 \text{ meV},    1.14
 \rbrace.$ Therefore, we observe that the lattice relaxation is important for the existence of the $\mathcal{C}=3$ phase at $\epsilon_r = 12.$ The $\mathcal{C}=2$ phase remains robust without lattice relaxation at $\epsilon_r=12.$

 We now turn our attention to the results in Fig. \ref{fig:parallel_no_relaxation_2} simulated with $\epsilon_r=6.$ Here, Coulomb interactions are stronger so symmetry-breaking phases dominate. We again observe Chern phases at both positive and negative $\Delta$ but for reasons described in the previous section with lattice relaxation, we only focus on the phases at negative $\Delta. $ At this value of $\epsilon_r,$ we recover both the $\mathcal{C}=2$ and $\mathcal{C}=3$ to their former prominence. The $\mathcal{C}=2$ phase again has quite narrow bands with favorable quantum geometry, as shown in \ref{fig:parallel_no_relaxation_2}(e). The optimal state here occurs at  $\lbrace\theta,\Delta,\delta_\mathrm{above},\delta_\mathrm{below},W,\lambda \rbrace = \lbrace   1.30^\circ,  -30 \text{ meV},   19.04 \text{ meV},   33.48 \text{ meV},    6.13 \text{ meV},    0.99
 \rbrace.$ The $\mathcal{C}=3$ phase is still dispersive but it is isolated from the rest of the energy spectrum, as shown in \ref{fig:parallel_no_relaxation_2}(f). The Berry curvature and quantum metric are also centered at $\bar{\Gamma}.$ The optimal state here occurs at  $\lbrace\theta,\Delta,\delta_\mathrm{above},\delta_\mathrm{below},W,\lambda \rbrace = \lbrace 1.35^\circ, -24 \text{ meV}, 11.25 \text{ meV},12.63 \text{ meV},25.64 \text{ meV},    0.78\rbrace.$ The existence of both the $\mathcal{C}=2$ and $\mathcal{C}=3$ phases at $\epsilon_r=6$ suggest that these phases are stable against the inclusion of lattice relaxation or lack thereof. However, one needs to tune the strength of Coulomb interactions in order to stabilize them. Lattice relaxation is definitely helpful to that effort.

\subsubsection{With Reduced Fermi Velocity $\gamma_0$}

\begin{figure}[t!]
\includegraphics[width=1\linewidth]{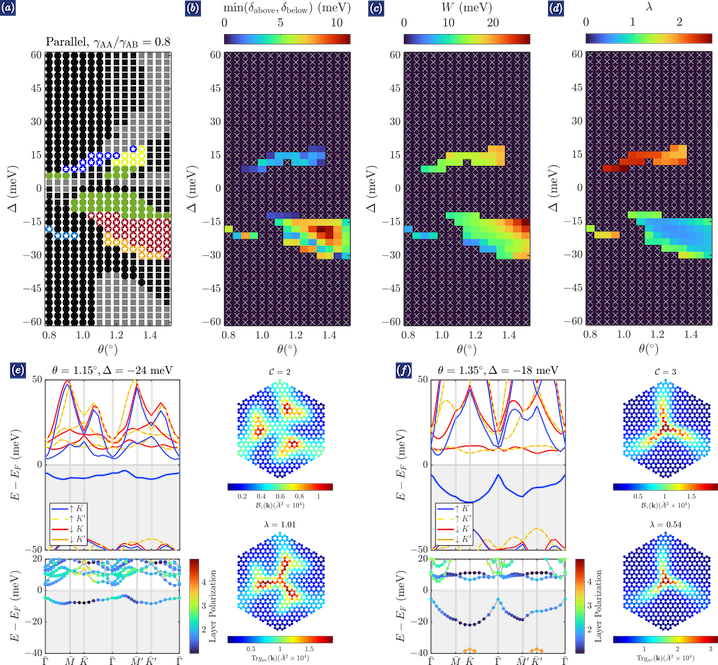}
\caption{\textbf{Mean-field phase diagram for the parallel configuration with reduced $\gamma_0$ and $\lbrace\gamma_\mathrm{AA},\gamma_\mathrm{AB}\rbrace=\lbrace80,100\rbrace$ meV.} The results here are simulated using {\fontfamily{cmtt}\selectfont Param\_three}. (a) Phase diagram and (b,c,d) its further characterization: band gap, bandwidth, and trace condition violation. (e) Band structure and associated Berry curvature and quantum metric distribution of an $\mathfrak{f.o.m.}$ $\mathcal{C}=2$ state. (f) Band structure and associated Berry curvature and quantum metric distribution of an $\mathfrak{f.o.m.}$ $\mathcal{C}=3$ state.  Here, $\epsilon_r = 12.$  }
\label{fig:parallel_reduced_vF}
\end{figure}

\begin{figure}[t!]
\includegraphics[width=1\linewidth]{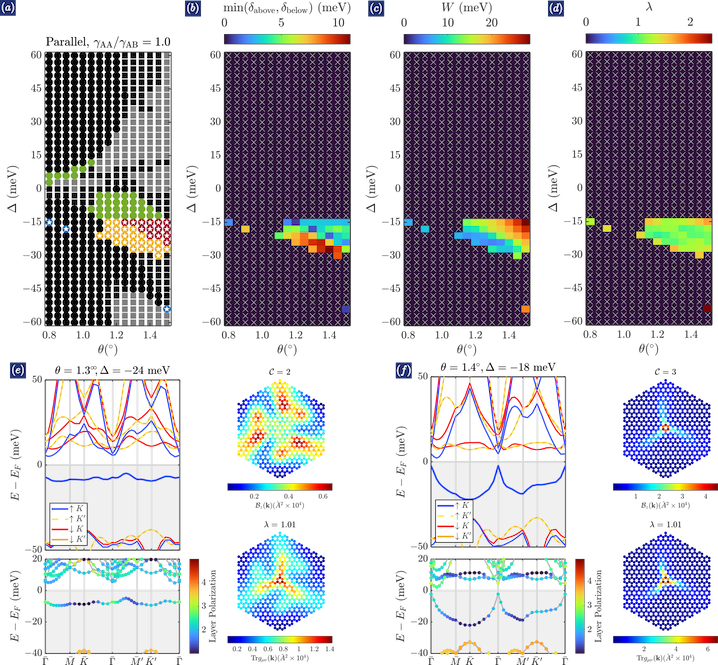}
\caption{\textbf{Mean-field phase diagram for the parallel configuration with reduced $\gamma_0$ and $\lbrace\gamma_\mathrm{AA},\gamma_\mathrm{AB}\rbrace=\lbrace100,100\rbrace$ meV.} The results here are simulated using {\fontfamily{cmtt}\selectfont Param\_four}. (a) Phase diagram and (b,c,d) its further characterization: band gap, bandwidth, and trace condition violation. (e) Band structure and associated Berry curvature and quantum metric distribution of an $\mathfrak{f.o.m.}$ $\mathcal{C}=2$ state. (f) Band structure and associated Berry curvature and quantum metric distribution of an $\mathfrak{f.o.m.}$ $\mathcal{C}=3$ state. Here, $\epsilon_r = 12.$   }
\label{fig:parallel_reduced_vF_2}
\end{figure}

Finally, we examine the phase diagrams simulated with {\fontfamily{cmtt}\selectfont Param\_three} and {\fontfamily{cmtt}\selectfont Param\_four} as shown in Figs. \ref{fig:parallel_reduced_vF} and \ref{fig:parallel_reduced_vF_2} respectively. These two parameter sets have reduced $\gamma_0=-2700$ meV. The former includes some lattice relaxation while the latter does not. For both simulations, we use $\epsilon_r = 12.$ We observe in both cases that the $\mathcal{C}=2$ and $\mathcal{C}=3$ phases exist. However, with lattice relaxation, the $\mathcal{C}=3$ is much more prominent. In both cases, the qualitative features observed before remain: the $\mathcal{C}=2$ state has narrow, well-isolated bands with favorable quantum geometry, while the $\mathcal{C}=3$ state also has favorable quantum geometry, but its bands are more dispersive. 
With lattice relaxation, the optimal $\mathcal{C}=2$ state occurs at $\lbrace\theta,\Delta,\delta_\mathrm{above},\delta_\mathrm{below},W,\lambda \rbrace = \lbrace 1.15^\circ,  -24 \text{ meV},  6.50 \text{ meV},  35.02 \text{ meV},    5.10 \text{ meV},    1.01\rbrace,$ while the optimal $\mathcal{C}=3$ state occurs at $\lbrace\theta,\Delta,\delta_\mathrm{above},\delta_\mathrm{below},W,\lambda \rbrace = \lbrace 1.35^\circ,  -18 \text{ meV},   11.22 \text{ meV},   15.27 \text{ meV},   16.32 \text{ meV},    0.54\rbrace.$ With lattice relaxation, the optimal $\mathcal{C}=2$ state occurs at $\lbrace\theta,\Delta,\delta_\mathrm{above},\delta_\mathrm{below},W,\lambda \rbrace = \lbrace 1.30^\circ,  -24 \text{ meV},    9.68 \text{ meV},   28.45 \text{ meV},   4.65 \text{ meV},    1.01\rbrace,$ while the optimal $\mathcal{C}=3$ state occurs at $\lbrace\theta,\Delta,\delta_\mathrm{above},\delta_\mathrm{below},W,\lambda \rbrace = \lbrace 1.40^\circ,  -18 \text{ meV},    4.45 \text{ meV},   10.56 \text{ meV},  19.84 \text{ meV},    1.01\rbrace.$

\subsection{Summary of Observations}

We close this section with a few remarks summarizing the general trends observed from the extensive simulation results reported. Summary information about the optimal $\mathfrak{f.o.m.}$ states is tabulated in Tables \ref{tab:C2 antiparallel}, \ref{tab:C2 parallel}, and \ref{tab:C3 parallel}. We shall use these tables for guidance. All of the relevant higher-Chern phases are observed at negative $\Delta$ where the charge density of the band of interest is localized primarily on layer 1 of the trilayer substack. For the antiparallel configuration, we observe only a $\mathcal{C}=2$ phase, while for the parallel configuration, we observe a $\mathcal{C}=2$ and a $\mathcal{C}=3$ phase. The $\mathcal{C}=2$ phase in the antiparallel configuration appears most robust to changes to parameters. The phases in the parallel configuration are more sensitive to parameter changes, as detailed below.

For the $\mathcal{C}=2$ phase of the antiparallel configuration, we find the following:
\begin{enumerate}
    \item The optimal angles are around $\theta \sim 1.2-1.3^\circ.$ The range of optimal displacement fields is $\Delta \in \left[ -27,-18 \right]$ meV.
    \item For $\gamma=-3100$ meV, increasing Coulomb interactions increases the phase-space area of this phase, decreases the optimal $\lambda,$ and increases the bandwidths and band gaps.
    \item Lattice relaxation appears to be beneficial to this phase as it decreases the optimal $\lambda$ while simultaneously lowers the bandwidths. The dependence of band gaps on lattice relaxation appears more complicated: it increases $\delta_\mathrm{above}$ but decreases $\delta_\mathrm{below}.$
    \item With reduced $\gamma_0=-2700$ meV, this phase remains robust. Here, we also observe that adding lattice relaxation decreases $\lambda$ and decreases $W,$ consistent with the other simulations.
\end{enumerate}

\begin{table}[]
    \centering
    \begin{tabular}{||c||c|c|c|c|c|c| c|c||}
        \hline\hline
         Parameter set & $\epsilon_r$  & \# of points & $\theta$ (degrees)  &  $\Delta$ (meV) &$\delta_\mathrm{above}$ (meV) & $\delta_\mathrm{below}$ (meV) & $W$ (meV) & $\lambda$\\
         \hline
         \hline
         {\fontfamily{cmtt}\selectfont Param\_one} & $12$  & $26$ & $1.20^\circ$  &  $-18$   &$11.26$   & $14.62$   & $7.25$   & $1.32$\\
         \hline
         {\fontfamily{cmtt}\selectfont Param\_one} & $6$  & $39$ & $1.25^\circ$  &  $-24$   &$26.94$   & $17.37$   & $12.36$   & $1.24$\\
         \hline
         {\fontfamily{cmtt}\selectfont Param\_two} & $12$  & $23$ & $1.30^\circ$  &  $-21$   &$10.14$   & $18.33$   & $8.38$   & $1.67$\\
         \hline
         {\fontfamily{cmtt}\selectfont Param\_two} & $6$  & $38$ & $1.30^\circ$  &  $-27$   &$20.68$   & $19.40$   & $12.02$   & $1.44$\\
         \hline
         {\fontfamily{cmtt}\selectfont Param\_three} & $12$  & $33$ & $1.25^\circ$  &  $-18$   &$12.09$   & $15.76$   & $7.13$   & $1.14$\\
         \hline
         {\fontfamily{cmtt}\selectfont Param\_four} & $12$  & $34$ & $1.35^\circ$  &  $-21$   &$11.10$   & $20.64$   & $8.43$   & $1.35$\\
         \hline
         \hline
    \end{tabular}
    \caption{\textbf{Summary of observations of the mean-field $\mathcal{C}=2$ phase in the antiparallel configuration.} The first column lists the parameter set used. The second column denotes the dielectric constant $\epsilon_r$. The third column provides the number of data points belonging to such phase and serves as a measure for the area of that phase within the phase diagram. For the data point within that area which minimizes the figure of merit in Eq. \ref{eq: figure of merit}, we display the angle and displacement field where that occurs in columns four and five, the gaps above and below the first conduction band in columns six and seven, the bandwidth of that band in column eight, and the corresponding trace condition in column nine.  }
    \label{tab:C2 antiparallel}
\end{table}

For the $\mathcal{C}=2$ phase of the parallel configuration, we observe the following:
\begin{enumerate}
    \item The optimal angles are around $\theta \sim 1.1-1.3^\circ.$ The range of optimal displacement fields is $\Delta \in \left[ -30,-18 \right]$ meV.
    \item The bandwidths are remarkably narrower in this phase compared to the other nontrivial Chern phases in both configurations. 
    \item The trace condition violations for this phase are generally smaller than the $\mathcal{C}=2$ phase of the antiparallel configuration but are generally larger than the $\mathcal{C}=3$ phase.
    \item Strengthening Coulomb interactions decreases $\lambda$ when lattice relaxation is kept but increases $\lambda$ when lattice relaxation is ignored. However, increasing Coulomb interactions increases band gaps and bandwidths whether or not lattice relaxation is accounted for. Therefore, increasing Coulomb interactions is helpful to this phase. However, the phase areas have a complicated relationship on the strength of Coulomb interactions. 
    \item For $\gamma=-3100$ meV, lattice relaxation does not appear to affect the phase much. While including lattice does decrease optimal $\lambda,$ it also decrease the phase areas.
    \item With reduced $\gamma_0=-2700$ meV, this phase area is reduced noticeably with relaxation, but the phase area remains relatively similar to the other parameter choices when relaxation is included.
\end{enumerate}

\begin{table}[]
    \centering
    \begin{tabular}{||c||c|c|c|c|c|c| c|c||}
        \hline\hline
         Parameter set & $\epsilon_r$  & \# of points & $\theta$ (degrees)  &  $\Delta$ (meV) &$\delta_\mathrm{above}$ (meV) & $\delta_\mathrm{below}$ (meV) & $W$ (meV) & $\lambda$\\
         \hline
         \hline
         {\fontfamily{cmtt}\selectfont Param\_one} & $12$  & $22$ & $1.10^\circ$  &  $-21$   &$7.85$   & $27.04$   & $3.75$   & $0.87$\\
         \hline
         {\fontfamily{cmtt}\selectfont Param\_one} & $6$  & $16$ & $1.15^\circ$  &  $-30$   &$16.33$   & $40.88$   & $6.51$   & $0.74$\\
         \hline
         {\fontfamily{cmtt}\selectfont Param\_two} & $12$  & $27$ & $1.10^\circ$  &  $-18$   &$8.60$   & $14.32$   & $3.92$   & $0.89$\\
         \hline
         {\fontfamily{cmtt}\selectfont Param\_two} & $6$  & $26$ & $1.30^\circ$  &  $-30$   &$19.04$   & $33.48$   & $6.13$   & $0.99$\\
         \hline
         {\fontfamily{cmtt}\selectfont Param\_three} & $12$  & $10$ & $1.15^\circ$  &  $-24$   &$6.50$   & $35.02$   & $5.10$   & $1.01$\\
         \hline
         {\fontfamily{cmtt}\selectfont Param\_four} & $12$  & $24$ & $1.30^\circ$  &  $-24$   &$9.68$   & $28.45$   & $4.65$   & $1.01$\\
         \hline
         \hline
    \end{tabular}
    \caption{\textbf{Summary of observations of the mean-field $\mathcal{C}=2$ phase in the parallel configuration.} The first column lists the parameter set used. The second column denotes the dielectric constant $\epsilon_r$. The third column provides the number of data points belonging to such phase and serves as a measure for the area of that phase within the phase diagram. For the data point within that area which minimizes the figure of merit in Eq. \ref{eq: figure of merit}, we display the angle and displacement field where that occurs in columns four and five, the gaps above and below the first conduction band in columns six and seven, the bandwidth of that band in column eight, and the corresponding trace condition in column nine.  }
    \label{tab:C2 parallel}
\end{table}

For the $\mathcal{C}=3$ phase of the parallel configuration, we note the following:
\begin{enumerate}
    \item The optimal angles are around $\theta \sim 1.3-1.4^\circ.$ The range of optimal displacement fields is $\Delta \in \left[ -27,-18 \right]$ meV.
    \item As a general trend, this phase has the lowest $\lambda$ compared to the other two phases. 
    \item This phase contains active bands that are appreciably more dispersive than the bands in the other nontrivial phases. 
    \item For $\gamma_0=-3100$ meV, lattice relaxation is favorable to this phase as it decreases $\lambda,$ decreases the bandwidths, and increases the band gaps. 
    \item Increasing Coulomb interactions stabilizes this phase, reducing $\lambda$ and increasing band gaps as well as bandwidths.
    \item With reduced $\gamma_0=-2700$ meV, this phase area is reduced noticeably without relaxation. This phase follows the opposite trend compared to the $\mathcal{C}=2$ phase in the parallel configuration.
\end{enumerate}

\begin{table}[]
    \centering
    \begin{tabular}{||c||c|c|c|c|c|c| c|c||}
        \hline\hline
         Parameter set & $\epsilon_r$  & \# of points & $\theta$ (degrees)  &  $\Delta$ (meV) &$\delta_\mathrm{above}$ (meV) & $\delta_\mathrm{below}$ (meV) & $W$ (meV) & $\lambda$\\
         \hline
         \hline
         {\fontfamily{cmtt}\selectfont Param\_one} & $12$  & $33$ & $1.30^\circ$  &  $-21$   &$9.54$   & $25.55$   & $11.04$   & $0.55$\\
         \hline
         {\fontfamily{cmtt}\selectfont Param\_one} & $6$  & $38$ & $1.40^\circ$  &  $-27$   &$21.22$   & $30.13$   & $21.44$   & $0.48$\\
         \hline
         {\fontfamily{cmtt}\selectfont Param\_two} & $12$  & $9$ & $1.30^\circ$  &  $-15$   &$4.29$   & $5.37$   & $17.40$   & $1.14$\\
         \hline
         {\fontfamily{cmtt}\selectfont Param\_two} & $6$  & $17$ & $1.35^\circ$  &  $-24$   &$11.25$   & $12.63$   & $25.64$   & $0.78$\\
         \hline
         {\fontfamily{cmtt}\selectfont Param\_three} & $12$  & $37$ & $1.35^\circ$  &  $-18$   &$11.22$   & $15.27$   & $16.32$   & $0.54$\\
         \hline
         {\fontfamily{cmtt}\selectfont Param\_four} & $12$  & $14$ & $1.40^\circ$  &  $-18$   &$4.45$   & $10.56$   & $19.84$   & $1.01$\\
         \hline
         \hline
    \end{tabular}
    \caption{\textbf{Summary of observations of the mean-field $\mathcal{C}=3$ phase in the parallel configuration.} The first column lists the parameter set used. The second column denotes the dielectric constant $\epsilon_r$. The third column provides the number of data points belonging to such phase and serves as a measure for the area of that phase within the phase diagram. For the data point within that area which minimizes the figure of merit in Eq. \ref{eq: figure of merit}, we display the angle and displacement field where that occurs in columns four and five, the gaps above and below the first conduction band in columns six and seven, the bandwidth of that band in column eight, and the corresponding trace condition in column nine.  }
    \label{tab:C3 parallel}
\end{table}

We emphasize also that ranges of $(\theta,\Delta)$ where nontrivial Chern phases exist after HF renormalization do not often overlap with the ranges of $(\theta,\Delta)$ where Chern bands appear in the non-interacting limit, suggesting the importance of Coulomb interactions.

\section{Numerical Stability Analysis}
\label{sec: Stability Analysis}

\begin{figure}
    \centering
    \includegraphics[width=1\linewidth]{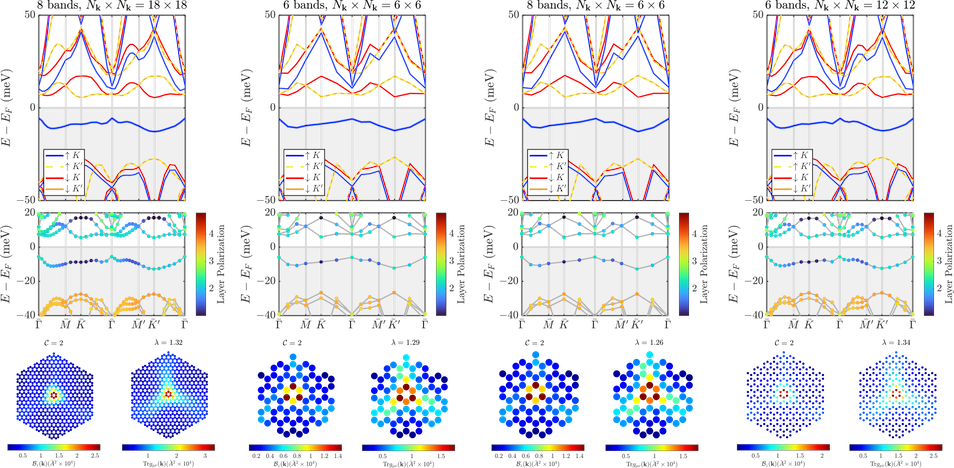}
    \caption{\textbf{Convergence of $\mathcal{C}=2$ phase simulation for the antiparallel configuration.} The state is at $(\theta,\Delta)=(1.20^\circ,-18 \text{ meV})$ with $\epsilon_r=12.$ The different band structures are simulated including different numbers of bands and using different mesh sizes, as indicated in the titles. The corresponding Berry curvature and quantum metric are shown below each band structure. }
    \label{fig:antiparallel_grid_size}
\end{figure}

\begin{figure}
    \centering
    \includegraphics[width=1\linewidth]{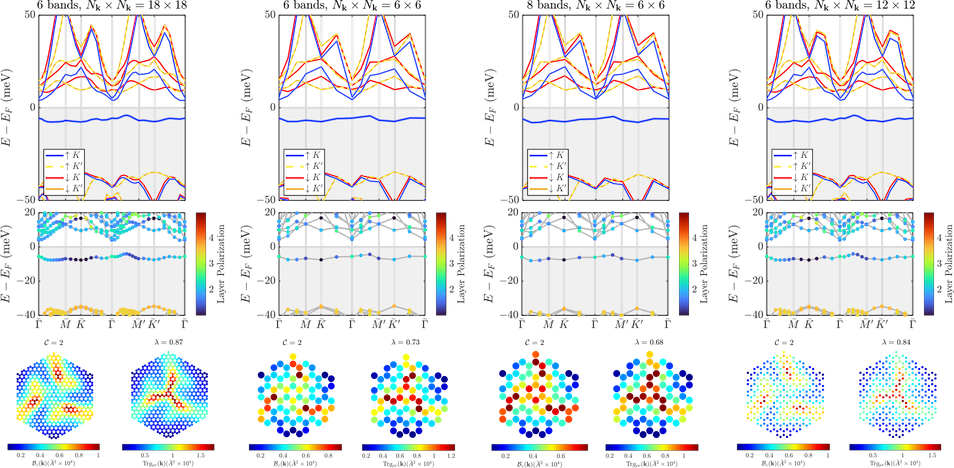}
    \caption{\textbf{Convergence of $\mathcal{C}=2$ phase simulation for the parallel configuration.} The state is at $(\theta,\Delta)=(1.10^\circ,-21 \text{ meV})$ with $\epsilon_r=12.$ The different band structures are simulated including different numbers of bands and using different mesh sizes, as indicated in the titles. The corresponding Berry curvature and quantum metric are shown below each band structure. }
    \label{fig:parallel_grid_size}
\end{figure}

\begin{figure}
    \centering
    \includegraphics[width=1\linewidth]{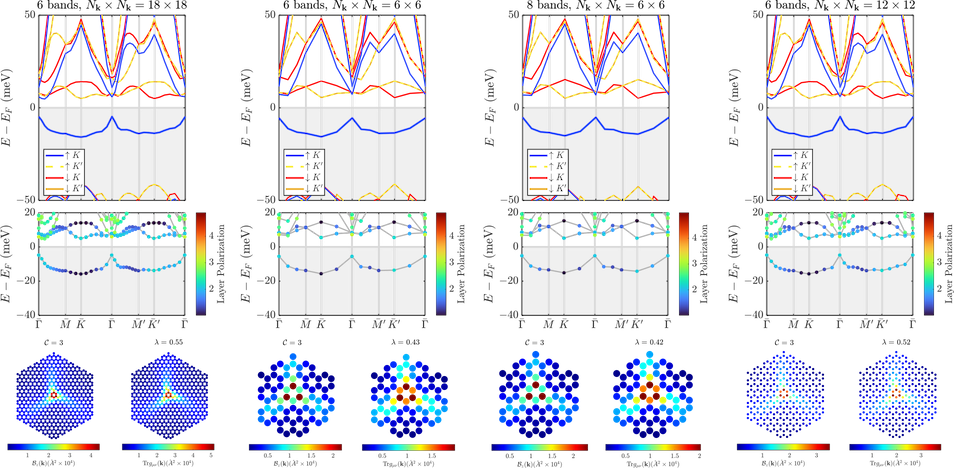}
    \caption{\textbf{Convergence of $\mathcal{C}=3$ phase simulation for the parallel configuration.} The state is at $(\theta,\Delta)=(1.30^\circ,-21 \text{ meV})$ with $\epsilon_r=12.$ The different band structures are simulated including different numbers of bands and using different mesh sizes, as indicated in the titles. The corresponding Berry curvature and quantum metric are shown below each band structure. }
    \label{fig:parallel_grid_size_2}
\end{figure}

We briefly address the question of convergence of our results in this section. For concreteness, we focus only on our principal results presented in the main text. That is, we show that the optimal states for both configurations simulated with {\fontfamily{cmtt}\selectfont Param\_one} and $\epsilon_r=12$ are stable with changes in the number of bands included and the mesh size. In Fig. \ref{fig:antiparallel_grid_size}, we show the optimal $\mathcal{C}=2$ state for the antiparallel configuration. In Figs. \ref{fig:parallel_grid_size} and \ref{fig:parallel_grid_size_2}, we show the optimal $\mathcal{C}=2$ and $\mathcal{C}=3$ states for the parallel configuration respectively. In all three cases, we see that already at $N_\mathbf{k} \times N_\mathbf{k} = 6\times 6,$ we have excellent agreement in the energetics. However, the trace condition violations do deviate from the ``true" values at such a smaller mesh. For a larger mesh of $N_\mathbf{k} \times N_\mathbf{k} = 12\times 12,$ we observe in all three cases that the violations are within $6\%$ of the ``true" values. This analysis adds convincing evidence that our principal results concerning the nontrivial topological states have likely converged numerically.

\section{Phase Diagram Comparison with Layer-\textit{In}dependent Coulomb Potential}

In many previous studies on related twisted and non-twisted systems, the Coulomb potential is often regarded as independent of the layer index, i.e. all of the layers are treated as residing on the same plane. This is probably a good approximation when the number of layers is small (say two or three layers) or when the distance to the gates is much greater than the thickness of the stack, $D/2 \gg N_\ell d, $ where $N_\ell$ is the number of layers. In this work, neither of these assumptions is true: (1) we have a reasonably thick stack (5 layers) and (2) the gates are not set significantly far away since experimentally they can be rather close to the active material. We have therefore considered the full layer dependence of the Coulomb potential in this work. In this section, we compare the results of considering the full layer dependence to those obtained by neglecting that layer index. 

\begin{figure}
    \centering
    \includegraphics[width=1\linewidth]{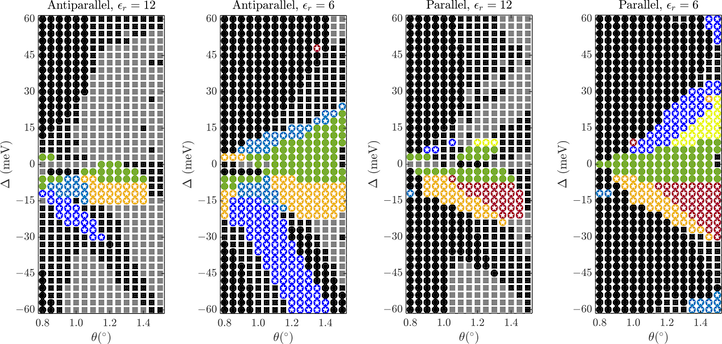}
    \caption{\textbf{Mean-field phase diagrams for the antiparallel and parallel configurations simulated with the layer-independent Coulomb potential.} The results here are simulated with {\fontfamily{cmtt}\selectfont Param\_one} and with $\epsilon_r=12$ or $\epsilon_r = 6.$}
    \label{fig:no_layer_phase_diagram}
\end{figure}

From the practical point of view, neglecting the layer index can lead to a substantial reduction in required computational resources for HF calculations. This is due to the fact that the form factors no longer need to be resolved to individual layers, significantly lowering memory and CPU requirements for the evaluations of the Coulomb matrix elements, which, importantly, are the most intensive routine in our the HF algorithm. Without layer index, Eqs. \eqref{eq: form factors} and \eqref{eq: Coulomb matrix elements} simplify:
\begin{equation}
\label{eq: Coulomb matrix elements 2}
    \begin{split}
        \mathbb{V}^{n_1,n_2,n_3,n_4}_{\xi_1,\xi_2}(\mathbf{k},\mathbf{p},\mathbf{q}) 
        &= \frac{1}{V_\mathrm{sys}} \sum_{\mathbf{Q}} \mathbb{\Lambda}^{n_1,n_3}_{\xi_1}(\mathbf{k},\mathbf{q}+\mathbf{Q}) \mathcal{V}(\mathbf{q}+\mathbf{Q}) \left[\mathbb{\Lambda}^{n_4,n_2}_{\xi_2}(\mathbf{p}-\mathbf{q},\mathbf{q}+\mathbf{Q}) \right]^\dagger,\\
        \mathbb{\Lambda}^{n,n'}_{\xi}(\mathbf{k},\mathbf{q}+\mathbf{Q}) &= \sum_{\mathbf{G},\sigma,\ell} \phi^\dagger_{n,\xi,\sigma,\ell,\mathbf{k}+\mathbf{G}+\mathbf{q}+\mathbf{Q}} \phi_{n',\xi,\sigma,\ell,\mathbf{k}+\mathbf{G}} = \bra{u_{n,\xi,\mathbf{k}+\mathbf{q}}} e^{i \mathbf{Q} \cdot \hat{\mathbf{r}}}  \ket{u_{n',\xi,\mathbf{k}}},\\
        \mathcal{V}(\mathbf{q}) &= \frac{e^2}{2\epsilon_0\epsilon_r |\mathbf{q}|} \tanh \left(D|\mathbf{q}|/2 \right).
    \end{split}
\end{equation}
As illustrated, there is no longer a need to calculate $\mathbb{\Lambda}^{n,n'}_{\xi,\ell}$ since the only quantities which are needed to calculate $\mathbb{V}^{n_1,n_2,n_3,n_4}_{\xi_1,\xi_2}$ are $\mathbb{\Lambda}^{n,n'}_{\xi} = \sum_\ell \mathbb{\Lambda}^{n,n'}_{\xi,\ell}.$ This simple change might have important qualitative implications for the phase diagram, which we now examine. The phase diagrams simulated using {\fontfamily{cmtt}\selectfont Param\_one} with $\epsilon_r=12$ and $\epsilon_r=6$ for both antiparallel and parallel configurations are shown in Fig. \ref{fig:no_layer_phase_diagram}. Qualitatively, the phase diagrams in Fig. \ref{fig:no_layer_phase_diagram} are similar to the phase diagrams obtained using the layer-dependent potential. For instance, there is a robust $\mathcal{C}=2$ phase in the antiparallel configuration for both values of the dielectric constant. There is also a robust $\mathcal{C}=2$ phase and a robust $\mathcal{C}=3$ phase for the parallel configuration, again, for both values of the dielectric constant. To further understand the difference between using the layer-dependent potential and layer-independent potential, we tabulate properties of the optimal states in Table \ref{tab:comparison layer dependence}. In all cases, we observe that the optimal states are pushed to lower values of $\Delta$ when layer dependence is included. This is perfectly sensible because layer screening  reduces the externally-applied displacement field. For the antiparallel configuration, the optimal states appear quite similar with and without layer dependence. For the parallel configuration,  $\lambda$ is reduced without layer dependence. Thus, it is important to include layer dependence in order to accurately capture quantum geometry as closeness to ideal quantum geometry is an important aspect of our conclusions. This alone suggests the necessity of using the layer-dependent Coulomb potential energy for realistic modeling. We leave further analysis of the difference between the two potential energy functions to future studies.

\begin{table}[]
    \centering
    \begin{tabular}{||c|c|c|c|c|c|c|c|c| c|c||}
        \hline\hline
         Configuration & Layer dependence? & $\epsilon_r$ & $\mathcal{C}$   & \# of points & $\theta$ (degrees)  &  $\Delta$ (meV) &$\delta_\mathrm{above}$ (meV) & $\delta_\mathrm{below}$ (meV) & $W$ (meV) & $\lambda$\\
         \hline
         \hline
         Antiparallel & No & $12$ & $2$ & $23$ & $1.20^\circ$  &  $-12$   &$11.64$   & $12.53$   & $7.45$   & $1.32$\\
         \hline
         Antiparallel & Yes & $12$ & $2$ & $26$ & $1.20^\circ$  &  $-18$   &$11.13$   & $14.62$   & $7.25$   & $1.32$\\
         \hline
         Antiparallel & No & $6$ & $2$ & $48$ & $1.25^\circ$  &  $-15$   &$27.37$   & $16.14$   & $14.04$   & $1.27$\\
         \hline
         Antiparallel & Yes & $6$ & $2$ & $39$ & $1.25^\circ$  &  $-24$   &$26.94$   & $17.37$   & $12.36$   & $1.24$\\
         \hline
         Parallel & No & $12$ & $2$ & $25$ & $1.10^\circ$  &  $-15$   &$8.83$   & $25.49$   & $4.39$   & $0.75$\\
         \hline
         Parallel & Yes & $12$ & $2$ & $22$ & $1.10^\circ$  &  $-21$   &$7.85$   & $27.04$   & $3.75$   & $0.87$\\
         \hline
         Parallel & No & $6$ & $2$ & $29$ & $1.15^\circ$  &  $-18$   &$18.29$   & $35.32$   & $7.69$   & $0.59$\\
         \hline
         Parallel & Yes & $6$ & $2$ & $16$ & $1.15^\circ$  &  $-30$   &$16.33$   & $40.88$   & $6.51$   & $0.74$\\
         \hline
         Parallel & No & $12$ & $3$ & $27$ & $1.25^\circ$  &  $-12$   &$11.17$   & $14.56$   & $12.56$   & $0.45$\\
         \hline
         Parallel & Yes & $12$ & $3$ & $33$ & $1.30^\circ$  &  $-21$   &$9.54$   & $25.55$   & $11.04$   & $0.55$\\
         \hline
         Parallel & No & $6$ & $3$ & $36$ & $1.35^\circ$  &  $-18$   &$21.77$   & $27.96$   & $18.96$   & $0.36$\\
         \hline
         Parallel & Yes & $6$ & $3$ & $38$ & $1.40^\circ$  &  $-27$   &$21.22$   & $30.13$   & $21.44$   & $0.48$\\
         \hline         \hline
    \end{tabular}
    \caption{\textbf{Comparison of the optimal states obtained with layer-dependent and layer-independent Coulomb potential energy.}  Column one indicates the configuration. Column two indicates whether a layer-independent or layer-dependent Coulomb potential energy is used. Column three denotes the dielectric constant. Columns four and five indicate the Chern phase and its number of points in phase space. For the data point within that area which minimizes the figure of merit in Eq. \ref{eq: figure of merit}, we display the angle and displacement field where that occurs in columns six and seven, the gaps above and below the first conduction band in columns eight and nine, the bandwidth of that band in column ten, and the corresponding trace condition in column eleven.   }
    \label{tab:comparison layer dependence}
\end{table}

\section{Numerical Evaluation of Quantum Geometric Tensor}
\label{sec: Numerical Evaluation of Quantum Geometric Tensor}

The gauge-invariant geometric tensor for Bloch states $\ket{u_\mathbf{k}}$ from a single isolated band is \cite{pati1991relation,cheng2010quantum}
\begin{equation}
    \mathcal{Q}_{\mu \nu}(\mathbf{k}) = \bra{\partial_\mu u_\mathbf{k}}\ket{\partial_\nu u_\mathbf{k}} - \bra{\partial_\mu u_\mathbf{k}}\ket{u_\mathbf{k}}\bra{u_\mathbf{k}}\ket{\partial_\nu u_\mathbf{k}},
\end{equation}
where all partial derivatives are taken with respect to $\mathbf{k}.$ Recalling that $\bra{u_\mathbf{k}}\ket{\partial_\nu u_\mathbf{k}}$ is purely imaginary, we notice that the second factor in the geometric tensor is purely real. We therefore define the Berry curvature and quantum metric as
\begin{subequations}
\begin{align}
    \mathcal{F}_{\mu\nu}(\mathbf{k}) &= -i \left( \bra{\partial_\mu u_\mathbf{k}}\ket{\partial_\nu u_\mathbf{k}} - \bra{\partial_\nu u_\mathbf{k}}\ket{\partial_\mu u_\mathbf{k}} \right) = 2 \mathrm{Im} \mathcal{Q}_{\mu \nu}(\mathbf{k}),      \\
    g_{\mu\nu}(\mathbf{k}) &= \frac{\bra{\partial_\mu u_\mathbf{k}}\ket{\partial_\nu u_\mathbf{k}} + \bra{\partial_\nu u_\mathbf{k}}\ket{\partial_\mu u_\mathbf{k}}}{2} - \bra{\partial_\mu u_\mathbf{k}}\ket{u_\mathbf{k}}\bra{u_\mathbf{k}}\ket{\partial_\nu u_\mathbf{k}} =  \mathrm{Re} \mathcal{Q}_{\mu \nu}(\mathbf{k}).
\end{align}
\end{subequations}
To facilitate numerical calculation, let us consider finite differences of the form
\begin{equation}
    \bra{u_{\mathbf{k}}}\ket{u_{\mathbf{k}+d\mathbf{k}}} \approx 1 + \bra{u_{\mathbf{k}}} \ket{\partial_\mu u_\mathbf{k}} dk^\mu + \frac{1}{2}
    \bra{u_{\mathbf{k}}} \ket{\partial_\mu \partial_\nu u_\mathbf{k}} dk^\mu  dk^\nu + ...,
\end{equation}
where we have expanded $\ket{u_{\mathbf{k}+d\mathbf{k}}}$ to second order in $d\mathbf{k}$ and the contribution of the normalization, if any, is neglected since it is beyond second order. Taking the absolute value and only keeping up to second order, we find
\begin{equation}
    \begin{split}
        \left|\bra{u_{\mathbf{k}}}\ket{u_{\mathbf{k}+d\mathbf{k}}} \right| &\approx \sqrt{\left(1+\frac{1}{2}
    \mathrm{Re}\bra{u_{\mathbf{k}}} \ket{\partial_\mu \partial_\nu u_\mathbf{k}} dk^\mu  dk^\nu\right)^2 + \left( -i \bra{u_{\mathbf{k}}} \ket{\partial_\mu u_\mathbf{k}} dk^\mu + \frac{1}{2}
    \mathrm{Im}\bra{u_{\mathbf{k}}} \ket{\partial_\mu \partial_\nu u_\mathbf{k}} dk^\mu  dk^\nu\right)^2 } \\
    &\approx 1 + \frac{1}{2}\left[\mathrm{Re}\bra{u_{\mathbf{k}}} \ket{\partial_\mu \partial_\nu u_\mathbf{k}}  + \bra{\partial_\mu u_{\mathbf{k}}} \ket{ u_\mathbf{k}}\bra{u_{\mathbf{k}}} \ket{\partial_\nu u_\mathbf{k}} \right]dk^\mu dk^\nu+... \\
    \end{split}
\end{equation}
We can simplify noting that $\mathrm{Re}\bra{u_{\mathbf{k}}} \ket{\partial_\mu \partial_\nu u_\mathbf{k}} = - \mathrm{Re}\bra{\partial_\mu u_{\mathbf{k}}} \ket{\partial_\nu u_\mathbf{k}}$ because $\bra{u_{\mathbf{k}}} \ket{\partial_\mu \partial_\nu u_\mathbf{k}}+\bra{\partial_\mu u_{\mathbf{k}}} \ket{\partial_\nu u_\mathbf{k}}$ is purely imaginary \cite{Mera2021Engineering}: 
\begin{equation}
\label{eq: numerical quantum metric}
     \left|\bra{u_{\mathbf{k}}}\ket{u_{\mathbf{k}+d\mathbf{k}}}\right| \approx 1 - \frac{1}{2} g_{\mu \nu} dk^\mu dk^\nu + ... \rightarrow g_{\mu \nu} dk^\mu dk^\nu \approx 2-2\left|\bra{u_{\mathbf{k}}}\ket{u_{\mathbf{k}+d\mathbf{k}}}\right| \approx -2\mathrm{Re}\ln \left[\bra{u_{\mathbf{k}}}\ket{u_{\mathbf{k}+d\mathbf{k}}}\right]
\end{equation}
For purposes of  numerical evaluation, let us identify Eq. \eqref{eq: numerical quantum metric} with the quantum metric half-way between $\mathbf{k}$ and $\mathbf{k}+d\mathbf{k}$: $g_{\mu\nu}(\mathbf{k}+d\mathbf{k}/2)$. On a grid, we have wave functions at discrete points $\mathbf{k} = n_1 \mathbf{g}_1 + n_2 \mathbf{g}_2,$ where $\mathbf{g}_i = \mathbf{G}_i/N_\mathbf{k}$ and $N_\mathbf{k}^2$ is the number of points sampled in the moir\'{e} Brillouin zone. We compute the quantum metric and Berry curvature using two different \textit{counterclockwise} triangular loops: (1) $\mathbf{k}\rightarrow \mathbf{k} + \mathbf{g}_1 \rightarrow \mathbf{k} + \mathbf{g}_1 + \mathbf{g}_2 \rightarrow \mathbf{k}$ and (2) $\mathbf{k}\rightarrow \mathbf{k} + \mathbf{g}_1 + \mathbf{g}_2 \rightarrow \mathbf{k}  + \mathbf{g}_2 \rightarrow \mathbf{k}$. For the first triangular loop, the quantum metric is approximately
\begin{equation}
\label{eq: QM1}
    \begin{split}
        &-2\mathrm{Re}\ln \left[\bra{u_{\mathbf{k}}}\ket{u_{\mathbf{k}+\mathbf{g}_1}}\bra{u_{\mathbf{k}+\mathbf{g}_1}}\ket{u_{\mathbf{k}+\mathbf{g}_1+\mathbf{g}_2}}\bra{u_{\mathbf{k}+\mathbf{g}_1+\mathbf{g}_2}}\ket{u_\mathbf{k}} \right] \\
        &\approx   g_{xx}(\mathbf{k}+\mathbf{g}_1/2) \frac{\kappa^2}{4} + 2g_{xy}(\mathbf{k}+\mathbf{g}_1/2) \frac{\sqrt{3}\kappa^2}{4} +g_{yy}(\mathbf{k}+\mathbf{g}_1/2) \frac{3\kappa^2}{4} +g_{xx}(\mathbf{k}+\mathbf{g}_1 + \mathbf{g}_2/2) \kappa^2 \\
        &+g_{xx}(\mathbf{k}+\mathbf{g}_1/2+\mathbf{g}_2/2) \frac{\kappa^2}{4} - 2g_{xy}(\mathbf{k}+\mathbf{g}_1/2+\mathbf{g}_2/2) \frac{\sqrt{3}\kappa^2}{4} +g_{yy}(\mathbf{k}+\mathbf{g}_1/2+\mathbf{g}_2/2) \frac{3\kappa^2}{4} \\
        &\approx \left[ g_{xx}(\mathbf{k} +(2\mathbf{g}_1+\mathbf{g}_2)/3)  + g_{yy}(\mathbf{k} +(2\mathbf{g}_1+\mathbf{g}_2)/3) \right] \frac{3\kappa^2}{2} \\
    \end{split}
\end{equation}
 For the second triangular loop, the quantum metric is approximately
\begin{equation}
\label{eq: QM2}
    \begin{split}
        &-2\mathrm{Re}\ln \left[\bra{u_{\mathbf{k}}}\ket{u_{\mathbf{k}+\mathbf{g}_1+\mathbf{g}_2}}\bra{u_{\mathbf{k}+\mathbf{g}_1+\mathbf{g}_2}}\ket{u_{\mathbf{k}+\mathbf{g}_2}}\bra{u_{\mathbf{k}+\mathbf{g}_2}}\ket{u_\mathbf{k}} \right] \\
        &\approx   g_{xx}(\mathbf{k}+\mathbf{g}_1/2+\mathbf{g}_2/2) \frac{\kappa^2}{4} - 2g_{xy}(\mathbf{k}+\mathbf{g}_1/2+\mathbf{g}_2/2) \frac{\sqrt{3}\kappa^2}{4} +g_{yy}(\mathbf{k}+\mathbf{g}_1/2+\mathbf{g}_2/2) \frac{3\kappa^2}{4} \\
        &+g_{xx}(\mathbf{k}+\mathbf{g}_1/2+\mathbf{g}_2) \frac{\kappa^2}{4} + 2g_{xy}(\mathbf{k}+\mathbf{g}_1/2+\mathbf{g}_2) \frac{\sqrt{3}\kappa^2}{4} +g_{yy}(\mathbf{k}+\mathbf{g}_1/2+\mathbf{g}_2) \frac{3\kappa^2}{4} +g_{xx}(\mathbf{k} + \mathbf{g}_2/2) \kappa^2 \\
        &\approx \left[ g_{xx}(\mathbf{k} +(\mathbf{g}_1+2\mathbf{g}_2)/3)  + g_{yy}(\mathbf{k} +(\mathbf{g}_1+2\mathbf{g}_2)/3) \right] \frac{3\kappa^2}{2} \\
    \end{split}
\end{equation}
Here, $\kappa = |\mathbf{g}_i|.$ In the above derivations, we have assumed that the off-diagonal elements of $g_{\mu\nu}$ approximately cancel since the loops close. Eqs. \eqref{eq: QM1} and \eqref{eq: QM2} are essentially the same expressions in Ref. \cite{Mera2021Engineering}. There is a slightly different way (but similar in spirit) to numerically calculate the trace of the quantum metric presented in Ref. \cite{Marzari1997Maximally}. For the Berry curvature, we can rewrite $\mathcal{B}_z(\mathbf{k}) = \mathcal{F}_{xy}(\mathbf{k}) = \hat{z} \cdot \left[\nabla \times \left( - i \bra{u_\mathbf{k}}\nabla_\mathbf{k} \ket{u_\mathbf{k}} \right)\right].$ So, the integral of the Berry curvature over a triangular area can be turned into a loop integral around the boundary of that area via Stokes' theorem (for the single-band case we are currently assuming). Thus, we have
\begin{equation}
\begin{split}
    \mathcal{B}_z(\mathbf{k} +(2\mathbf{g}_1+\mathbf{g}_2)/3) \frac{\sqrt{3}\kappa^2}{4} &= \mathrm{Im}\ln \left[\bra{u_{\mathbf{k}}}\ket{u_{\mathbf{k}+\mathbf{g}_1}}\bra{u_{\mathbf{k}+\mathbf{g}_1}}\ket{u_{\mathbf{k}+\mathbf{g}_1+\mathbf{g}_2}}\bra{u_{\mathbf{k}+\mathbf{g}_1+\mathbf{g}_2}}\ket{u_\mathbf{k}} \right], \\
    \mathcal{B}_z(\mathbf{k} +(\mathbf{g}_1+2\mathbf{g}_2)/3) \frac{\sqrt{3}\kappa^2}{4} &=  \mathrm{Im}\ln \left[\bra{u_{\mathbf{k}}}\ket{u_{\mathbf{k}+\mathbf{g}_1+\mathbf{g}_2}}\bra{u_{\mathbf{k}+\mathbf{g}_1+\mathbf{g}_2}}\ket{u_{\mathbf{k}+\mathbf{g}_2}}\bra{u_{\mathbf{k}+\mathbf{g}_2}}\ket{u_\mathbf{k}} \right].   
\end{split}
\end{equation}
We can compactify the notation by defining $\triangle(\mathbf{k}_*)$ as the gauge-invariant argument of the logarithm around a counterclockwise loop centered at $\mathbf{k}_*$. Then, we have
\begin{subequations}
\label{eq: numerical calculation of quantum geometric tensor}
    \begin{align}
        \mathrm{Tr} g_{\mu\nu}(\mathbf{k}_*) &= - \frac{4}{3\kappa^2} \mathrm{Re} \ln\triangle(\mathbf{k}_*), \\
        \mathcal{B}_z(\mathbf{k}_*) &=  \frac{4}{\sqrt{3}\kappa^2} \mathrm{Im} \ln\triangle(\mathbf{k}_*).
    \end{align}
\end{subequations}
This shows that both the quantum metric and Berry curvature by just taking real and imaginary parts of $\triangle(\mathbf{k}_*)$ (and then multiplying by non-universal factors that depend on the grid geometry). It is worth emphasizing in closing that the expressions derived Eq. \eqref{eq: numerical calculation of quantum geometric tensor} are manifestly gauge-invariant. Also, it is clear from Eq. \eqref{eq: numerical calculation of quantum geometric tensor} that the trace of the quantum metric is positive semidefinite. This is because $\mathrm{Re} \ln \triangle(\mathbf{k}_*) = \ln |\triangle(\mathbf{k}_*)| \leq 0$ since $0 \leq |\triangle(\mathbf{k}_*)| \leq 1.$ For the Berry curvature, we choose the branch of the logarithm such that $-\pi \leq \mathrm{Im} \ln \triangle(\mathbf{k}_*) < \pi.$

The primary interest of this study is to explore the topology of the first conduction band in cases where it is well-isolated spectrally from the rest of the energy spectrum. For this purpose, the above prescription to calculate the Berry curvature and corresponding Chern number of the first conduction band is sufficient. However, there is a subtlety that may be relevant experimentally. The Chern number of the first conduction band is not necessarily equal to the Hall conductivity (in units of $e^2/h$) of the gap right above it since the latter is equal to the sum of all the Chern numbers (in units of $e^2/h$) of the occupied bands. In an experiment, what is measured is the Hall conductance (or equivalently, Hall resistance) associated with a gap, and not directly the Chern number associated with a band. So, it is meaningful to calculate such a Hall conductance. When computing the Chern numbers of all the other bands to obain the Hall conductivity of a gap, a numerical instability is occasionally observed when the remote bands are close together in energy. To mitigate this effect, we instead calculate the non-Abelian Berry curvature whose trace over bands once integrated over the mBZ gives the \textit{composite} Chern number. This calculation has the advantage that it is well defined even if there are degeneracies in the occupied manifold as long as the occupied and unoccupied states are separated from each other by a sufficiently large energy gap, which we always assume for robust insulating states. In this multiband formalism, the non-Abelian Berry connection is \cite{Wilczek1984Appearance}
\begin{equation}
    \boldsymbol{\mathcal{A}}^{n,m}(\mathbf{k}) = -i \bra{u_{n,\mathbf{k}}}\nabla_\mathbf{k}\ket{u_{m,\mathbf{k}}},
\end{equation}
where $m,n$ are band indices of the occupied bands. This is now a matrix with the number of rows equaling the number of occupied bands.  The non-Abelian Berry curvature matrix is defined using a covariant  derivative $\mathfrak{D}_\mu = \partial_\mu - \mathcal{A}_\mu$:
\begin{equation}
    \mathcal{B}_z(\mathbf{k}) =- \left[\mathfrak{D}_x,\mathfrak{D}_y \right] = \partial_x \mathcal{A}_y - \partial_y\mathcal{A}_x - \left[ \mathcal{A}_x,\mathcal{A}_y \right].
\end{equation}
The non-Abelian Berry curvature differs from the Abelian Berry curvature in the nontrivial commutator $\left[\mathcal{A}_x,\mathcal{A}_y\right],$ which clearly vanishes in the Abelian case. The composite Chern number is 
\begin{equation}
    \mathcal{C}_\mathrm{tot} = \frac{1}{2\pi} \int d^2 \mathbf{k} \mathrm{Tr} \left[\mathcal{B}_z(\mathbf{k}) \right],
\end{equation}
where the trace is over all occupied bands. To compute this quantity numerically, we use the method described in Ref. \cite{fukui2005chern}. We can use the same formulas as before with the replacement:
\begin{equation}
\begin{split}
        \ket{u_\mathbf{k}} &\mapsto \ket{U_\mathbf{k}} = \begin{pmatrix}
        \ket{u_{1,\mathbf{k}}} & \ket{u_{2,\mathbf{k}}} & \ket{u_{3,\mathbf{k}}} & ... & \ket{u_{n,\mathbf{k}}}
        \end{pmatrix}, \\
        \bra{u_\mathbf{k}} &\mapsto \bra{U_\mathbf{k}} = \begin{pmatrix}
        \bra{u_{1,\mathbf{k}}} & \bra{u_{2,\mathbf{k}}} & \bra{u_{3,\mathbf{k}}} & ... & \bra{u_{n,\mathbf{k}}}
        \end{pmatrix}^T.
\end{split}
\end{equation}
Then, we have
\begin{subequations}
\begin{align}
    \mathcal{B}_z(\mathbf{k} +(2\mathbf{g}_1+\mathbf{g}_2)/3) \frac{\sqrt{3}\kappa^2}{4} &= \mathrm{Im} \ln \left[\bra{U_{\mathbf{k}}}\ket{U_{\mathbf{k}+\mathbf{g}_1}}\bra{U_{\mathbf{k}+\mathbf{g}_1}}\ket{U_{\mathbf{k}+\mathbf{g}_1+\mathbf{g}_2}}\bra{U_{\mathbf{k}+\mathbf{g}_1+\mathbf{g}_2}}\ket{U_\mathbf{k}} \right], \\
    \mathcal{B}_z(\mathbf{k} +(\mathbf{g}_1+2\mathbf{g}_2)/3)\frac{\sqrt{3}\kappa^2}{4} &=  \mathrm{Im} \ln  \left[\bra{U_{\mathbf{k}}}\ket{U_{\mathbf{k}+\mathbf{g}_1+\mathbf{g}_2}}\bra{U_{\mathbf{k}+\mathbf{g}_1+\mathbf{g}_2}}\ket{U_{\mathbf{k}+\mathbf{g}_2}}\bra{U_{\mathbf{k}+\mathbf{g}_2}}\ket{U_\mathbf{k}} \right].   
\end{align}
\end{subequations}
Under suitable conditions, tracing over the Berry curvature can be replaced by taking determinants inside the logarithm. We end by emphasizing again that this method is only reliable when the occupied bands are reasonably separated spectrally from the unoccupied bands. Otherwise, this multi-band method suffers the same problems as the single-band method does.

\section{Descriptions of Accompanying .MP4 Movies}

This manuscript is accompanied by several .mp4 files that animate the evolution of band structures using different sets of parameters. In this section, we provide descriptions for all of these .mp4 files.
\begin{enumerate}
    \item {\fontfamily{cmtt}\selectfont noninteracting\_band\_structure\_antiparallel\_with\_relaxation.mp4}: This movie shows the evolution of the non-interacting band structure in the \textcolor{blue}{\textit{antiparallel configuration}} for $\nu = +$ as $\Delta \in \left[-60,60 \right]$ meV and $\theta \in \left[ 0.8^\circ,1.5^\circ \right]$ are varied. Each energy state is color-coded by its layer polarization as defined in Eq. \eqref{eq: layer polarization}. The Chern number of the first conduction band is indicated in the title. We have used parameters in {\fontfamily{cmtt}\selectfont Param\_one} which \textcolor{blue}{\textit{includes some lattice relaxation}} due to the difference in $\gamma_\mathrm{AA}$ and $\gamma_\mathrm{AB}.$
    \item {\fontfamily{cmtt}\selectfont noninteracting\_band\_structure\_antiparallel\_no\_relaxation.mp4}: This movie shows the evolution of the non-interacting band structure in the \textcolor{blue}{\textit{antiparallel configuration}} for $\nu = +$ as $\Delta \in \left[-60,60 \right]$ meV and $\theta \in \left[ 0.8^\circ,1.5^\circ \right]$ are varied. Each energy state is color-coded by its layer polarization as defined in Eq. \eqref{eq: layer polarization}. The Chern number of the first conduction band is indicated in the title. We have used parameters in {\fontfamily{cmtt}\selectfont Param\_two} which \textcolor{blue}{\textit{does not include any lattice relaxation}} since $\gamma_\mathrm{AA}=\gamma_\mathrm{AB}.$
    \item {\fontfamily{cmtt}\selectfont noninteracting\_band\_structure\_parallel\_with\_relaxation.mp4}: This movie shows the evolution of the non-interacting band structure in the \textcolor{blue}{\textit{parallel configuration}} for $\nu = +$ as $\Delta \in \left[-60,60 \right]$ meV and $\theta \in \left[ 0.8^\circ,1.5^\circ \right]$ are varied. Each energy state is color-coded by its layer polarization as defined in Eq. \eqref{eq: layer polarization}. The Chern number of the first conduction band is indicated in the title. We have used parameters in {\fontfamily{cmtt}\selectfont Param\_one} which \textcolor{blue}{\textit{includes some lattice relaxation}} due to the difference in $\gamma_\mathrm{AA}$ and $\gamma_\mathrm{AB}.$
    \item {\fontfamily{cmtt}\selectfont noninteracting\_band\_structure\_parallel\_no\_relaxation.mp4}: This movie shows the evolution of the non-interacting band structure in the \textcolor{blue}{\textit{parallel configuration}} for $\nu = +$ as $\Delta \in \left[-60,60 \right]$ meV and $\theta \in \left[ 0.8^\circ,1.5^\circ \right]$ are varied. Each energy state is color-coded by its layer polarization as defined in Eq. \eqref{eq: layer polarization}. The Chern number of the first conduction band is indicated in the title. We have used parameters in {\fontfamily{cmtt}\selectfont Param\_two} which \textcolor{blue}{\textit{does not include any lattice relaxation}} since $\gamma_\mathrm{AA}=\gamma_\mathrm{AB}.$
    \item {\fontfamily{cmtt}\selectfont noninteracting\_band\_structure\_antiparallel\_reduced\_vF\_with\_relaxation.mp4}: This movie shows the evolution of the non-interacting band structure in the \textcolor{blue}{\textit{antiparallel configuration}} for $\nu = +$ as $\Delta \in \left[-60,60 \right]$ meV and $\theta \in \left[ 0.8^\circ,1.5^\circ \right]$ are varied. Each energy state is color-coded by its layer polarization as defined in Eq. \eqref{eq: layer polarization}. The Chern number of the first conduction band is indicated in the title. We have used parameters in {\fontfamily{cmtt}\selectfont Param\_three} which \textcolor{blue}{\textit{includes some lattice relaxation}} due to the difference in $\gamma_\mathrm{AA}$ and $\gamma_\mathrm{AB}.$
    \item {\fontfamily{cmtt}\selectfont noninteracting\_band\_structure\_antiparallel\_reduced\_vF\_no\_relaxation.mp4}: This movie shows the evolution of the non-interacting band structure in the \textcolor{blue}{\textit{antiparallel configuration}} for $\nu = +$ as $\Delta \in \left[-60,60 \right]$ meV and $\theta \in \left[ 0.8^\circ,1.5^\circ \right]$ are varied. Each energy state is color-coded by its layer polarization as defined in Eq. \eqref{eq: layer polarization}. The Chern number of the first conduction band is indicated in the title. We have used parameters in {\fontfamily{cmtt}\selectfont Param\_four} which \textcolor{blue}{\textit{does not include any lattice relaxation}} since $\gamma_\mathrm{AA}=\gamma_\mathrm{AB}.$
    \item {\fontfamily{cmtt}\selectfont noninteracting\_band\_structure\_parallel\_reduced\_vF\_with\_relaxation.mp4}: This movie shows the evolution of the non-interacting band structure in the \textcolor{blue}{\textit{parallel configuration}} for $\nu = +$ as $\Delta \in \left[-60,60 \right]$ meV and $\theta \in \left[ 0.8^\circ,1.5^\circ \right]$ are varied. Each energy state is color-coded by its layer polarization as defined in Eq. \eqref{eq: layer polarization}. The Chern number of the first conduction band is indicated in the title. We have used parameters in {\fontfamily{cmtt}\selectfont Param\_three} which \textcolor{blue}{\textit{includes some lattice relaxation}} due to the difference in $\gamma_\mathrm{AA}$ and $\gamma_\mathrm{AB}.$
    \item {\fontfamily{cmtt}\selectfont noninteracting\_band\_structure\_parallel\_reduced\_vF\_no\_relaxation.mp4}: This movie shows the evolution of the non-interacting band structure in the \textcolor{blue}{\textit{parallel configuration}} for $\nu = +$ as $\Delta \in \left[-60,60 \right]$ meV and $\theta \in \left[ 0.8^\circ,1.5^\circ \right]$ are varied. Each energy state is color-coded by its layer polarization as defined in Eq. \eqref{eq: layer polarization}. The Chern number of the first conduction band is indicated in the title. We have used parameters in {\fontfamily{cmtt}\selectfont Param\_four} which \textcolor{blue}{\textit{does not include any lattice relaxation}} since $\gamma_\mathrm{AA}=\gamma_\mathrm{AB}.$
\end{enumerate}

In all of these band structures, the presented Chern numbers are calculated numerically by summing over the local Berry curvature computed with Eq. \eqref{eq: numerical calculation of quantum geometric tensor} using a finite grid of $N_\mathbf{k}\times N_\mathbf{k} = 18\times 18$ points. This method is quite reliable except in cases where the \textit{direct} gaps between bands become vanishingly small, such as when a band inversion occurs. To mitigate this, one has to increase the number of mesh points. However, that is quite expensive, especially for HF calculations. 
Thus, Chern numbers at a phase boundary should be taken with some skepticism. With that said, the Chern numbers for states deep inside a phase is very reliable. None of our conclusion relies on data points at a phase boundary. So this small numerical instability has no material effect on our study.

\bibliography{References.bib}

\end{document}